\documentclass[%
superscriptaddress,
preprintnumbers,
 amsmath,amssymb,
aps,
pra,
]{revtex4-2}

\usepackage{graphicx}% Include figure files
\usepackage{dcolumn}% Align table columns on decimal point
\usepackage{bm}% bold math
\usepackage{xspace}
\usepackage{siunitx}
\usepackage{amsmath}
\usepackage{amssymb}
\usepackage{mathtools}
\usepackage{mathrsfs}
\usepackage{comment}
\usepackage{tabularx}
\usepackage{orcidlink}
\usepackage{hyperref}% add hypertext capabilities
\hypersetup{
    colorlinks=true,
    citecolor=blue,
    linkcolor=blue,
    urlcolor=blue,
}
\newcommand{\fulltoday}{\number\day\space \ifcase\month\or
    January\or February\or March\or April\or May\or June\or
    July\or August\or September\or October\or November\or December\fi
    \space\number\year
    }

\date{\fulltoday}

\def\vec#1{\bm{\mathrm{#1}}}    %vector
\def\tsr#1{\mathsf{#1}}     %tensor 
\def\corot#1{\mathring{#1}} %丸が上につく．corotationの記号
\def\etal{~\textit{et al. }}
\def\ILC{\textit{I}}
\def\NLC{\textit{N}}
\def\NsLC{\textit{N}$^\star$\hspace{-0.1em}}
\def\NDLC{\textit{N}\hspace{-0.1em}$_{\text{D}}$\hspace{-0.1em}}

\def\SmCsLC{smectic-\textit{C}$^\star$\hspace{-0.1em}}
\def\EricksenLeslie{EL model\xspace}
\def\ImuraOkano{IO model\xspace}
\def\Reciproc{Onsager's reciprocal relation\xspace}
\def\OnsagerVariational{Onsager's variational principle\xspace}
\def\Rayleighian{\mathscr{R}}

\begin{document}

\preprint{main text}

\title{Symmetry and Thermodynamic Bounds on Cross-Coupling Transport\\ in Chiral Liquid Crystals}%

\newcommand{\FacultyWaseda}{
    \affiliation{%
        Faculty of Science and Engineering, Waseda University, 3-4-1 Okubo, Shinjuku, Tokyo, 169-8555, Japan
    }
}

\newcommand{\GradWaseda}{
    \affiliation{%
        Graduate School of Science and Engineering, Waseda University, TWIns, 2-2 Wakamatsu-cho, Shinjuku, Tokyo, 162-8480, Japan
    }
}

\newcommand{\CompResOrg}{
   \affiliation{%
        Comprehensive Research Organization, Waseda University, TWIns, 2-2 Wakamatsu-cho, Shinjuku, Tokyo, 162-8480, Japan
    }
}

\newcommand{\SKCM}{
    \affiliation{%
        International Institute for Sustainability with Knotted Chiral Meta Matter (WPI-SKCM$^2$),
        Hiroshima University, 1-3-1 Kagamiyama, Higashi-Hiroshima, Hiroshima 739-8526, Japan
    }
}

\author{Shunsuke~Takano\orcidlink{0009-0002-9428-4851}}
    %\email{shunsuke.t-8395@akane.waseda.jp}
    \SKCM\GradWaseda
 
\author{Takuya~Nakanishi\orcidlink{0000-0002-1172-718X}}%
    \CompResOrg

\author{Kenta~Nakagawa\orcidlink{0000-0002-0577-537X}}
    \CompResOrg
    
\author{Toru~Asahi\orcidlink{0000-0003-4925-8259}}
    \email{tasahi@waseda.jp}
    \CompResOrg\FacultyWaseda

\begin{abstract}
    %\textcolor{black}{Full paper in Physical Review E}
    We reformulate the Leslie effects that describe the dynamic cross-couplings in chiral liquid crystals driven by the transport of heat, electric charge, and mass. The Ericksen--Leslie model is extended in the linear response framework by representing nematic order with the Q-tensor. Subsequently, the thermodynamic uncertainty relation is applied to identify the upper bounds of the Leslie cross-coupling coefficients. We reveal that the cross-coupling coefficients are dependent on the scalar order parameter and vanish in the isotropic phase. In addition, the chirality of the phase allows torque induced by a transport current parallel to the director. The mutual signs of the Leslie thermohydrodynamic and thermomechanical coefficients are likely to be opposite in calamitic liquid crystals, as suggested by recent experimental observations. Our model is applicable to the thermal, chemical, and electrical Leslie effects. The present arguments suggest that a common underlying principle may govern both the Leslie effects and the thermal Edelstein effect in chiral solid crystals attributed to chiral phonons.
    %They should consist of one paragraph and be completely self-contained.
    %They should concisely summarize the subjects, conclusions, and results of the manuscript.
    %For experimental papers, specify the quantities measured and objects studied and clearly describe the experimental conditions.
    %Avoid coined words and unexplained acronyms.
    %They should have no displayed equations or tables.
    %Do not use numbered references (incorporate source listings directly into the abstract text itself—see APS Journals Style Guide)
    %Length should be about 5% of the article and less than 500 words.Abstract
    %\textcolor{black}{(161 words less than 500 words)}

    %\url{https://journals.aps.org/authors/style-basics#tables}
\end{abstract}

\maketitle

%\tableofcontents
\section{Introduction}
    True-chirality correlates translation and rotation~\cite{Barron2012chirality}. The Lifshitz invariant and the Dzyaloshinskii--Moriya interaction induce self-assembled twist order in magnets and liquid crystals~\cite{Dzyaloshinsky1958, Moriya1960, de_Gennes, Kishine2005ProgTheorPhysSuppl, Togawa2012PhysRevLett}. The twist in the chiral nematic \mbox{(\NsLC)} phase selectively reflects, transmits, or even amplifies light in a helicity-specific manner~\cite{de_Vries1951ActaCryst, Dreher1971MolCrystLiqCryst, Vellaichamy2025NatPhoton}. %Helicity designates the circularly polarized eigenstates of photons and Weyl fermions~\cite{LandauLifshitzQuantumElectro1980}. 
    Since Lipkin's Zilch interacts with the electric toroidal monopole, photons become spin-polarized when emitted from chiral media~\cite{Lipkin1964JMathPhys,Tang2010PhysRevLett,Kusunose2024ApplPhysLett}. Chiral materials also make electrons spin-polarized: an electric current in noncentrosymmetric metals induces magnetization (the Edelstein effect)~\cite{Edelstein1990SolidStateCommun}, and further spin polarization is achieved in electrons transmitted from DNA (chirality-induced spin selectivity: CISS)~\cite{Gohler2011Science}. CISS is attributed to transfer of materials' chirality to electrons' helicity, and observed also in chiral metals and a superconductor~\cite{Inui2020PhysRevLett, Nakajima2023Nature}. Phonons carry an angular momentum~\cite{Kishine2020PhysRevLett}. Such chiral phonons were detected in chiral crystals~\cite{Ishito2023NatPhys, Ueda2023Nature}, allowing a heat current to induce spin accumulation in \ensuremath{\alpha}-quartz~\cite{Ohe2024prl} and mechanical rotation of whole crystals (thermal Edelstein effect)~\cite{Hamada2018PhysRevLett}. However, the rotation of chiral crystals and topological superconductors~\cite{Nomura2012} remains purely theoretical due to experimental unavailability. 

\begin{figure}
        \includegraphics[width=17.0cm, keepaspectratio]{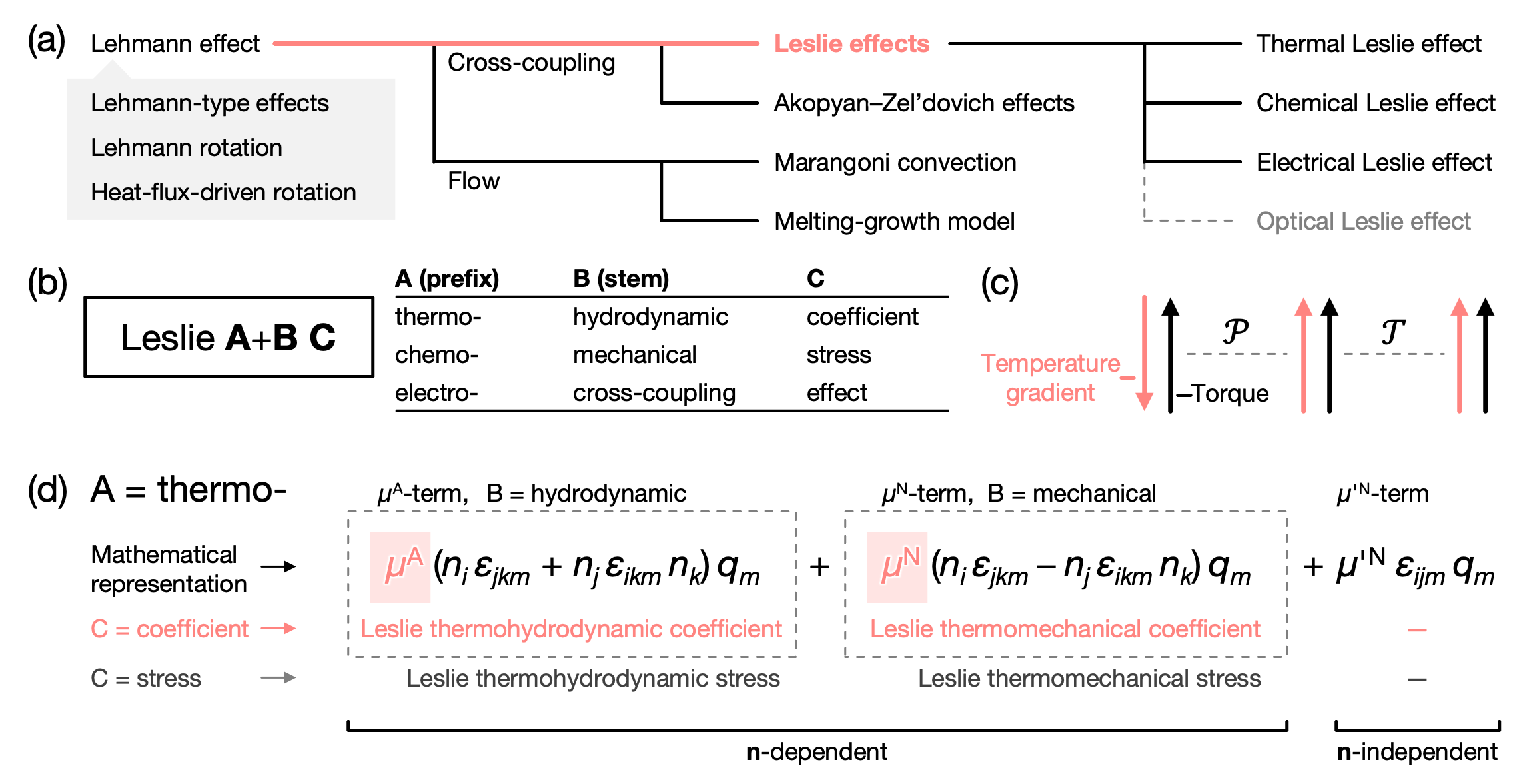}
        \caption{
            \label{fig:terminology} Mechanisms of the Lehmann effect.
            (a) Classification of theoretical models. Synonyms of the Lehmann effect are presented in the gray callout. The thermal Leslie effect and related extensions by de~Gennes~\cite{de_Gennes} are also included.
            (b) Stylistic format ``Leslie A+B C'' for denoting the Leslie effects in this article, where A, B, and C represent allowable terms for slots. 
            (c) Symmetry of collinear temperature gradient and torque under parity $\mathcal{P}$ (spatial inversion) and time reversal $\mathcal{T}$.
            (d) Terms of the stress tensors in the thermal Leslie effect. No specific designation has been assigned to the $\mu'^{\text{N}}$-term derived in this article.
        }
    \end{figure}

    Chiral liquid crystals exhibit an analogue of the thermal Edelstein effect, namely continuous rotation under a temperature gradient~\cite{Lehmann1900, Oswald2008PRL, Oswald2009EPJE, Oswald2012epl, Yoshioka2014, Yamamoto2015, Yoshioka2018, Bono2019, Oswald2019softmatter, Bono2020, Nishiyama2021SoftMatter, Takano2023, Yoshioka2024}. This Lehmann effect is classically described by the Leslie effects~\cite{Leslie1968II} (Fig.~\ref{fig:terminology}). The Leslie effects are classified into thermal, chemical and electrical types but formulated in the unified framework~\cite{de_Gennes}. Since the chirality of the phase always correlates mechanical motion with any transports, the Leslie effects are ubiquitously manifested in liquid crystals of chiral molecules: the thermal Leslie effect has been observed in \NsLC~droplets~\cite{Oswald2008PRL, Oswald2009EPJE, Oswald2012epl, Yoshioka2014, Yamamoto2015, Bono2019, Bono2020, Nishiyama2021SoftMatter, Takano2023} and slabs~\cite{Eber1982, Dequidt2007epl, Dequidt2008epje, Oswald2008EPL, Oswald2008PRE, Oswald2014epl}; the chemical Leslie effect has been investigated in \NsLC~droplets~\cite{Bono2017softmatter, Darmon2016Topo, Yoshioka2018} and \SmCsLC~free-standing films~\cite{Seki2011, Bunel2023nonsingular, Bunel2023singular} or Langmuir monolayers~\cite{Tabe2003}. Contradictory behaviors were observed in \NsLC~droplets surrounded by the isotropic (\ILC) phase~\cite{Oswald2019rev}, prompting the formulation of complementary hypotheses. The Akopyan--Zel’dovich (AZ) effects describe that the distortions of the director field $\vec{n}$ make heat conduction and mechanical motion cross-coupled~\cite{Akopyan1984, Brand2018RheolActa}. The AZ effects are observed in achiral nematic (\NLC) droplets with twisted director field~\cite{Ignes-Mullol2016prl}, and possibly induced by mass diffusion and electrical currents, too~\cite{Brand2013pre}. Hydrodynamic flow can produce apparent rotations since translation and rotation become indistinguishable in a twisted director configuration. This explains rotation in partially miscible emulsions~\cite{Oswald2019softmatter} and in the \NsLC~phase coexisting with the \ILC~phase~\cite{Oswald2018pre}. These complementary hypotheses each describe the Lehmann effect under specific experimental conditions; however, the Leslie effects themselves have not been updated. The upper limit of the cross-coupling coefficients remains unclear. The energy conversion efficiency is of concern for applying the Lehmann effect to soft actuators~\cite{Bono2018} or waste heat recycling, as well as of interest as heat engines. The interrelations between the Leslie cross-coupling coefficients also remain unknown, which was shed in light by recent studies~\cite{Bunel2023nonsingular, Bunel2023singular}. Furthermore, the conventional Leslie effects overlook a term that symmetry allows.

    The Leslie effects comprise $\vec{n}$-dependent and -independent terms [Fig.~\ref{fig:terminology}(d)]. The latter is unexamined~\cite{de_Gennes} but symmetry allows it. When the temperature gradient (polar vector) and torque (axial vector) are collinear, they alternate between parallel and antiparallel under parity $\mathcal{P}$ while remaining invariant under time reversal $\mathcal{T}$ [Fig.~\ref{fig:terminology}(c)]. Their configuration is true-chiral~\cite{Barron2012chirality} and independent of $\vec{n}$. Thus, the $\vec{n}$-independent Leslie effects are possible in chiral phases. 
    So (1) a heat current parallel to $\vec{n}$ ($\infty$-fold axis) would induce torque. 
    In \ensuremath{\alpha}-quartz, a heat current along the $c$-axis induces spin polarization~\cite{Ohe2024prl}. Anisotropy could not be necessary for cross-coupling.
    (2) Chirality allows torque to be induced in the \ILC~phase since $\vec{n}$ is not involved in the latter term. However, the \ILC~phases are thought not to rotate~\cite{de_Gennes, Oswald2021}, and no experimental observations verified this. The answer to this second question may lie beyond symmetry, probably in thermodynamics.
    In summary, the Leslie effects are a still evolving model with open questions, and revision is essential, which is guided by up-to-date understanding of orientation order, chirality, and more.

    Here, we reformulate the Leslie effects by fully incorporating \NLC~order through the Q‑tensor $\tsr{Q}$ and by making the model consistent with the law of entropy increase. This reformulation addresses two aspects: (1) what symmetry permits and (2) what thermodynamics restricts. (1) The Curie principle states that the transport coefficients reflect the symmetries of the phase~\cite{Curie1894JPhysTheorAppl,deGrootMazur1984}. \NLC~order belongs to a dihedral point group $D_\infty$ or $D_2$, which have $C_2'$, $\pi$ rotation around an axis perpendicular to the main axis. Under $C_2'$, $\tsr{Q}$ remains invariant unlike $\vec{n}$, resulting in $-\vec{n}$. So, $\tsr{Q}$ is suitable for representing \NLC~order~\cite{de_Gennes}. Moreover, $\tsr{Q}$ contains the scalar order parameter $S$ and distinguishes the \ILC~from \NLC~order. (2) Since the Leslie effects are irreversible, their energy conversion efficiency never exceeds the Carnot efficiency. Consequently, symmetry-allowed terms would be forbidden should the non-negativity of entropy production be violated. Two different cross-couplings, i.e. Leslie thermohydrodynamic and thermomechanical effects, must satisfy this principle simultaneously, which potentially impose sign- or magnitude- relationships among the cross-coupling coefficients. Our reformulated model must comprehensively consider \NLC~order by $\tsr{Q}$ and the limitation due to thermodynamics.

    This article is organized as follows.
    In Sec.~\ref{sec:linear_response}, the Ericksen--Leslie (EL) model is extended to relate the irreversible stress with the transport current by following linear response theory.
    In Sec.~\ref{sec:thermodynamic_bounds}, thermodynamic bounds of the cross-coupling coefficients are determined by applying the thermodynamic uncertainty relation (TUR). 
    In Sec.~\ref{sec:validation}, our model is validated by comparing with experimental observations.
    In Sec.~\ref{sec:torque}, the presence of torque is investigated, and its sign is discussed further.
    In Sec.~\ref{sec:conclusion}, a conclusion is provided.
    In Appendices, alternative derivations and notation, as well as the computed stress for the experimentally observed director fields, are shown.

    %\url{https://blog.wordvice.jp/how-to-draft-a-compelling-introduction-for-your-journal-article/}

    %http://www.ams.eng.osaka-u.ac.jp/user/ishihara/?p=1597

%==========================================================================
\section{Linear Response}\label{sec:linear_response}
    This section formulates the hydrodynamics incorporating the Leslie effects. Here, we briefly introduce the framework of linear response theory, which we follow in subsequent subsections. This assumption of linear response is supported by experimental evidence that demonstrates linear cross-coupling across diverse affinities. In \NsLC~droplets, the rotation speed is proportional to the temperature gradient~\cite{Oswald2008PRL,Takano2023}. The Leslie thermomechanical torque also increases linearly with the temperature gradient~\cite{Oswald2009EPJE, Oswald2012epje}. In the chemical Leslie effect, the precession frequency of Langmuir monolayers and flows in \SmCsLC~free-standing films vary linearly with the difference in vapor pressure across the layers~\cite{Tabe2003,Bunel2023singular}.
    
    The linear momentum is conserved locally, and the Navier--Stokes equation holds:
    \begin{equation}
        \rho \dot{v}_i - \nabla_j \sigma_{ji} = 0
        ,\hspace{1em}
        \sigma_{ij} = -p \delta_{ij} + \sigma^{\text{elast}}_{ij} + \sigma'_{ij}
        ,
        \label{eq:Linear_Navier_Stokes}
    \end{equation}
    where $\dot{a}$ represents the material derivative of an arbitrary quantity $a$, $\rho$ the mass density, $\vec{v}$ the hydrodynamic flow, $\tsr{\sigma}$ the stress, and $p$ the pressure. The reversible stress $\tsr{\sigma}^{\text{elast}}$ mainly originates from the orientation elasticity (the Ericksen stress), and $\tsr{\sigma}'$ is the irreversible stress. The determination of $\tsr{\sigma}'$ is the main objective of this section, and we applied linear response theory. 
    The order of indices of the stress tensor follows de~Gennes's traditional notation~\cite{de_Gennes}. This notation is different from contemporary notation styles but allows for a clear expression of our equations.
    
    In an irreversible process, the entropy density $s$ is not conserved:
    \begin{equation}
        \dot{s} + \nabla_i\left(\frac{q_i}{T}\right) = \frac{2R}{T}
        ,
    \end{equation}
    where $\vec{q}$ denotes the heat current, $T$ temperature, and $R$ the dissipation function~\cite{Brand2018RheolActa}. 
    The local entropy production is generally caused by an affinity $F_\mu$ and the conjugate irreversible current $j_\mu$: 
    \begin{equation}
        \label{eq:2R_jmuFmu}
        2 R = j_\mu F_\mu
        .
    \end{equation}
    The Einstein summation convention is adopted for repeated indices. Each affinity may induce the conjugate current and nonconjugate currents. The phenomenological equations describe the linear relations between the affinities and the currents:
    \begin{equation}
        \label{eq:jmu_LmunuFnu}
        \langle j_\mu \rangle = L_{\mu\nu} F_\nu
        ,
    \end{equation}
    where $L_{\mu \nu}$ is the transport coefficient, which is an element of the transport (Onsager) matrix $\tsr{L}$.
    Actual responses are possibly slow compared with the observer's time scale, and general responses that include them are formulated with a convolution. The present phenomenological equations in Eq.~\eqref{eq:jmu_LmunuFnu} represent the limiting case in which the response is sufficiently fast. The bracket $\langle a \rangle$ denotes the stochastic average of $a$. A current $j_\mu$ is decomposed into its average $\langle j_\mu \rangle$ and stochastically fluctuating part $\delta j_\mu$, as
    \begin{equation}
        \label{eq:jmu_jmu_deltajmu}
        j_\mu = \langle j_\mu \rangle + \delta j_\mu
        .
    \end{equation}
    The transport coefficient $L_{\mu \nu}$ represents the response speed, and the fluctuation--dissipation theorem (FDT) indicates that $L_{\mu \nu}$ originates from fluctuations of currents of interest~\cite{Pleiner1996}: 
    \begin{equation}
        \label{eq:fdt_deltaj_deltaj}
        \langle \delta j_\mu (\vec{r}, t)\, \delta j_\nu (\vec{0}, 0) \rangle = k_{\text{B}} T (L_{\mu\nu} + L_{\nu\mu}) \delta(\vec{r}) \delta (t)
        ,
    \end{equation}
    where $k_{\text{B}}$ is the Boltzmann constant, and $\delta(\bullet)$ the Dirac delta function.

\subsection{Affinity and current}
    %\textcolor{black}{EL, Okano formalism. EL theory and choice of hydrodynamic variables. Phenomenological equation. Heat conduction and advection is distinguished in hydrodynamics.}
    The \EricksenLeslie~\cite{Ericksen1961,Leslie1968I,Leslie1992} is a standard model of nematodynamics. The \EricksenLeslie selects the director $\vec{n}$ for representing \NLC~order. However, $\vec{n}$ only represents the local averages of molecular orientation, and its magnitude is disregarded. The ideal order parameter for \NLC~order is the Q-tensor $\tsr{Q}$. The \EricksenLeslie was extended by Imura and Okano by selecting $\tsr{Q}$ as the order parameter~\cite{Imura1972}. In this section, we follow the Imura--Okano (IO) model to reformulate the Leslie effects using $\tsr{Q}$. The \ImuraOkano incorporates the hydrodynamic variables below:
    \begin{subequations}
    \begin{align}
        \sigma^{\mathrm{sym}}_{ij}
        &=  \frac{1}{2}
        \left(
            \sigma'_{ij} + \sigma'_{ji}
        \right)
        ,\\
        \sigma^{\mathrm{ant}}_{ij}
        &=  \frac{1}{2}
        \left(
            \sigma'_{ij} - \sigma'_{ji}
        \right)
        ,\\
        A_{ij}
        &=  \frac{1}{2}
        \left(
            \nabla_i v_j + \nabla_j v_i
        \right)
        ,\\
        \corot{n}_i
        &=  \dot{n}_i - \frac{1}{2}
        \left(
            \nabla_j v_i - \nabla_i v_j
        \right)
        n_j
        ,\\
        N_{ij}
        &=  \corot{n}_i n_j - \corot{n}_j n_i
        .
    \end{align}
    \end{subequations}
    The change in $\vec{n}$ relative to the backflow, denoted as $\corot{\vec{n}}$ in our notation, is conventionally represented by $\vec{N}$~\cite{de_Gennes}. The change in $\vec{n}$, rather than the change in $\tsr{Q}$, is still used as a hydrodynamic variable in the \ImuraOkano~\cite{Imura1972}. The change in $\tsr{Q}$ can be treated as a hydrodynamic variable by adopting \OnsagerVariational (see Appendix~\ref{sec:variational}).

    Within our notation on stress tensor indices, the dissipation function $R$ is represented as
    \begin{equation}
    \label{eq:2R_sigma_A_sigma_N_T_q}
        2R = \sigma^{\mathrm{sym}}_{ij} A_{ij}
        + \sigma^{\mathrm{ant}}_{ij} N_{ij}
        + \left(-T^{-1}\nabla_i T\right) q_i
        ,
    \end{equation}
    where $\vec{q}$ is a heat current. The dissipation due to shear and rotation friction is represented by $\tsr{A}$ and $\tsr{N}$, respectively~\cite{Imura1972}, and the dissipation due to heat conduction is indicated by $\vec{q}$~\cite{Leslie1968II,de_Gennes}. In contemporary notation style, the term for the antisymmetric part has the opposite sign. 

    The choice of affinities and currents is arbitrary, provided that the conjugate pair of the variables adheres to Eq.~\eqref{eq:2R_sigma_A_sigma_N_T_q}. We selected the combination of the variables as in Table~\ref{tab:choice_of_variables}. The stresses, $\tsr{\sigma}^{\mathrm{sym}}$ and $\tsr{\sigma}^{\mathrm{ant}}$, and the flow gradients, $\tsr{A}$ and $\tsr{N}$, were chosen as currents and affinities, respectively. This choice follows both the \EricksenLeslie and the \ImuraOkano. For heat conduction, the heat current $\vec{q}$ was chosen as an affinity so that all the affinities $\tsr{A}, \tsr{N}, \vec{q}$ in our selection become $\mathcal{T}$-odd (see Sec.~\ref{sec:time_reversal} for details of time reversal symmetry). Consequently, all currents $\tsr{\sigma}^{\mathrm{sym}}, \tsr{\sigma}^{\mathrm{ant}}, -T^{-1} \vec{\nabla} T$ possess the same time reversal symmetry ($\mathcal{T}$-even) among them. Our allocation of $\vec{q}$ and $-T^{-1} \vec{\nabla} T$ is different from that by de~Gennes~\cite{de_Gennes} however the resulting equations are consistent (see Appendix~\ref{sec:some_other_notations}).
    
    \begin{table}[h]
        \caption{\label{tab:choice_of_variables}%
            Choice of hydrodynamic variables.
        }
        \begin{ruledtabular}
            \begin{tabular}{ccccd}
                   &\multicolumn{1}{c}{Shear flow}    &\multicolumn{1}{c}{Rotation}    &\multicolumn{1}{c}{Heat conduction}    &\epsilon^{\mathcal{T}}_X
                \\
                \hline
                \mbox{Current}    &$\sigma^{\mathrm{sym}}_{ij}$  &$\sigma^{\mathrm{ant}}_{ij}$  &$-T^{-1}\nabla_i T$  &+1\footnotemark[1]\\
                Affinity    &$A_{ij}$  &$N_{ij}$  &$q_i$  &-1\footnotemark[1]\\
            \end{tabular}
        \end{ruledtabular}
        \footnotetext[1]{According to Ref.~\cite{de_Gennes}}
    \end{table}

    Eq.~\eqref{eq:jmu_LmunuFnu} shows that an affinity drives conjugate and nonconjugate currents. Our selection of the hydrodynamic variables, as described in Table~\ref{tab:choice_of_variables}, enables the formulation of the phenomenological equations, which are given by
    \begin{subequations}
        \begin{align}
            \sigma^{\mathrm{sym}}_{ij} 
            &= L^{\mathrm{AA}}_{ijmn} A_{mn} 
            + L^{\mathrm{AN}}_{ijmn} N_{mn}
            + L^{\mathrm{AQ}}_{ijm} q_m
            ,\\
            \sigma^{\mathrm{ant}}_{ij} 
            &= L^{\mathrm{NA}}_{ijmn} A_{mn} 
            + L^{\mathrm{NN}}_{ijmn} N_{mn}
            + L^{\mathrm{NQ}}_{ijm} q_m
            ,\\
            -T^{-1}\nabla_i T
            &= L^{\mathrm{QA}}_{imn} A_{mn} 
            + L^{\mathrm{QN}}_{imn} N_{mn}
            + L^{\mathrm{QQ}}_{im} q_m
            .
        \end{align}
    \end{subequations}
    The transport coefficients, denoted by $L^{\bullet\bullet}_{\bullet\bullet}$, constitute the transport matrix $\tsr{L}$. The viscosity is represented by $L^{\mathrm{AA}}_{ijmn}, L^{\mathrm{AN}}_{ijmn}, L^{\mathrm{NA}}_{ijmn}, L^{\mathrm{NN}}_{ijmn}$, as in the original \ImuraOkano~\cite{Imura1972}, and the pure heat conduction by $L^{\mathrm{QQ}}_{im}$. The remaining coefficients signify the thermal Leslie effect.

\subsection{Anisotropy}
    %\textcolor{black}{Expansion with $Q$-tensor. Director and scalar order parameter. Incompressibility}
    Anisotropy differentiates liquid crystals from isotropic liquids. The uniaxial and biaxial \NLC~phases exhibit the symmetry of the point group $D_\infty$ and $D_2$, respectively, provided that inversion has been omitted so far. The complete information on \NLC~order is contained in $\tsr{Q}$. The \ImuraOkano has shown that the symmetry of the irreversible stress $\tsr{\sigma}'$ is determined by $\tsr{Q}$~\cite{Imura1972}, and the transport coefficients are polynomials in $\tsr{Q}$. The resulting expansions are shown in Eq.~\eqref{eq:Imura_Okano_transport}. 
    The transport coefficients undergo changes or remain unchanged under the exchange of indices (Table~\ref{tab:symmetry}). For example, $L^{\text{AA}}_{ijmn}$ is invariant under the exchange $i \leftrightarrow j,\ m \leftrightarrow n$: 
    \begin{equation}
        L^{\text{AA}}_{ijmn} = L^{\text{AA}}_{jimn} = L^{\text{AA}}_{jinm}= L^{\text{AA}}_{ijnm}
        .
    \end{equation}
    This invariance (variance) is attributed to that of the corresponding current $\sigma^{\text{sym}}_{ij}$ and the affinity $A_{mn}$. These conjugate variables $\sigma^{\text{sym}}_{ij}$ and $A_{mn}$ are invariant under $i \leftrightarrow j$ and $m \leftrightarrow n$, respectively. Other transport coefficients also have their own variance and/or invariance against the indices.
    We considered this variance and/or invariance to expand transport coefficients for the Leslie effects.

    The heat conductivity is a polynomial in $\tsr{Q}$:
    \begin{equation}
    \label{eq:LQQ_Q}
        L^{\text{QQ}}_{im} (\tsr{Q})
        =
        \rho_0 \delta_{im} + \rho_1 Q_{im}
        .
    \end{equation}
    Higher-order terms are possible but not dominant.

    In addition, for the Leslie effects, \NLC~order must be considered by $\tsr{Q}$. The transport coefficients are expanded as
    \begin{subequations}
    \label{eq:LAQ_Q_LNQ_Q}
    \begin{align}
        L^{\text{AQ}}_{ijm} (\tsr{Q})
        &=
        +\mu_{12}^{\text{A}}\left(\epsilon_{ikm}Q_{kj} + \epsilon_{jkm} Q_{ki}\right)  \nonumber
        \\&\hspace{1.1em}
        +\mu_{22}^{\text{A}}\left(\epsilon_{ikm}Q_{kl}Q_{lj} + \epsilon_{jkm}Q_{kl}Q_{li}\right)
        +\mu_{24}^{\text{A}}\left(\epsilon_{ikl}Q_{kj}Q_{lm} + \epsilon_{jkl}Q_{ki}Q_{lm}\right)
        ,\\
        L^{\text{NQ}}_{ijm} (\tsr{Q})
        &=
        + \mu_{01}^{\text{N}} \epsilon_{ijm}
        + \mu_{11}^{\text{N}} \epsilon_{ijk} Q_{km}
        + \mu_{12}^{\text{N}} \left(\epsilon_{ikm} Q_{kj} - \epsilon_{jkm} Q_{ki}\right)    \nonumber
        \\&\hspace{1.1em}
        + \mu_{21}^{\text{N}} \epsilon_{ijk} Q_{kl} Q_{lm}
        + \mu_{22}^{\text{N}} \left(\epsilon_{ikm} Q_{kl} Q_{lj} - \epsilon_{jkm} Q_{kl} Q_{li} \right) 
        + \mu_{23}^{\text{N}} \epsilon_{klm} Q_{ki} Q_{lj}  \nonumber
        \\&\hspace{1.1em}
        + \mu_{24}^{\text{N}} \left(\epsilon_{ikl} Q_{kj} Q_{lm} - \epsilon_{jkl} Q_{ki} Q_{lm} \right)
        + \mu_{25}^{\text{N}} \epsilon_{ijm} Q_{kl} Q_{lk}
        ,
    \end{align}
    \end{subequations}
    where $\epsilon_{ijk}$ is the Levi-Civita symbol defined in Eq.~\eqref{eq:def_levi_civita}. The Leslie cross-coupling coefficients $\mu^{\text{A}}_{\bullet \bullet}, \mu^{\text{N}}_{\bullet \bullet}$ are intrinsic to the material and almost independent of temperature.
    This expansion is consistent with the fact that $L^{\text{AQ}}_{ijm}$ and $L^{\text{NQ}}_{ijm}$ are symmetric and antisymmetric against $i \leftrightarrow j$, respectively.
    The above transport coefficients do not depend on $\vec{\nabla} \tsr{Q}$, unlike the AZ effects~\cite{Akopyan1984}. The AZ effects are confirmed by experiments~\cite{Ignes-Mullol2016prl} but the twisted director field is not necessary for rotation, as demonstrated in systems without twist, such as Langmuir monolayers~\cite{Tabe2003} and compensated mixtures~\cite{Dequidt2008epje}. The Leslie effects are attributed to the local symmetry of the phase, and occur in any chiral liquid crystal phase composed of chiral molecules.

    In the uniaxial \NLC~phase, $\tsr{Q}$ is decomposed into the director $\vec{n}$ and the scalar order parameter $S$:
    \begin{equation}
    \label{eq:QSn}
        Q_{ij} = \frac{3}{2} S \left( n_i n_j - \frac{1}{3} \delta_{ij} \right)
        .
    \end{equation}
    The nondegenerate eigenvalue of $\tsr{Q}$ is $S$, which represents the magnitude of \NLC~order. The corresponding (unit) eigenvector $\vec{n}$ represents the mean orientation direction. 
    Using this relation in Eq.~\eqref{eq:QSn}, the transport coefficients in Eqs.~\eqref{eq:Imura_Okano_transport},~\eqref{eq:LQQ_Q}~and~\eqref{eq:LAQ_Q_LNQ_Q} is represented as functions of $\vec{n},\ S$.
    Within our traditional notation for the stress tensor indices, expressions of viscous stress become
    \begin{equation}
    \label{eq:stress_hydro}
        \sigma^{\text{H}}_{ij}
		=
		\alpha_1 n_i n_j n_m n_n A_{mn}
		+\alpha_2 n_i \corot{n}_j + \alpha_3 n_j \corot{n}_i
		+\alpha_4 A_{ij}
		+\alpha_5 n_i n_m A_{mj} + \alpha_6 n_j n_m A_{mi}
		,
    \end{equation}
    where $\bullet^{\text{H}}$ represents the hydrodynamic part (without the Leslie effects) as in de~Gennes's textbook~\cite{de_Gennes}. The Leslie viscosity coefficients $\alpha_i\ (i = 1,2,\dots,6)$ depend on $S$ [see Eq.~\eqref{eq:Imura_Okano_Leslie_viscosity}].

    For the Leslie effects, we obtain
    \begin{subequations}
    \label{eq:LAQ_LNQ_Sn}
    \begin{align}
        L^{\text{AQ}}_{ijm}(S,\vec{n})
        &=
        \mu^{\text{A}}
        \left(
            n_i \epsilon_{jkm} n_k
            + n_j \epsilon_{ikm} n_k
        \right)
        ,\\
        L^{\text{NQ}}_{ijm}(S,\vec{n})
        &=
        \mu^{\text{N}}
        \left(
            n_i \epsilon_{jkm} n_k
            - n_j \epsilon_{ikm} n_k
        \right)
        +
        \mu'^{\text{N}}
        \epsilon_{ijm}
        ,\\
        \mu^{\text{A}}
        &=
        \frac{3}{2} \mu^{\text{A}}_{12} S
        + \left( \frac{3}{4} \mu^{\text{A}}_{22} - \frac{3}{4} \mu^{\text{A}}_{24} \right) S^2
        ,\\
        \mu^{\text{N}}
        &=
        \left( \frac{3}{2} \mu^{\text{N}}_{11} - \frac{3}{2} \mu^{\text{N}}_{12}\right) S
        + \left( \textcolor{black}{\frac{3}{4}} \mu^{\text{N}}_{21} - \frac{3}{4} \mu^{\text{N}}_{22} + \frac{3}{4} \mu^{\text{N}}_{23} - \frac{3}{4} \mu^{\text{N}}_{24} \right) S^2
        ,\\
        \mu'^{\text{N}}
        &=
        \mu^{\text{N}}_{01}
        + \left( \mu^{\text{N}}_{11} \textcolor{black}{- \mu^{\text{N}}_{12}} \right) S
        + \left( \textcolor{black}{ \mu^{\text{N}}_{21}} \textcolor{black}{+ \frac{1}{2} \mu^{\text{N}}_{22}} + \frac{1}{4} \mu^{\text{N}}_{23} \textcolor{black}{- \mu^{\text{N}}_{24}} + \frac{3}{2} \mu^{\text{N}}_{25} \right) S^2
        .
    \end{align}
    \end{subequations}
    The thermal Leslie effect is characterized by three coefficients $\mu^{\text{A}},  \mu^{\text{N}},  \mu'^{\text{N}}$, which are pseudoscalars (see Sec.~\ref{sec:time_reversal}) that depend on $S$. Leslie proposed the thermomechanical effect (the $\mu^{\text{N}}$-term), and de~Gennes added the thermohydrodynamic effect (the $\mu^{\text{A}}$-term). The $\mu'^{\text{N}}$-term represents the $\vec{n}$-independent effect that induces the torque from the heat current [Fig.~\ref{fig:terminology}(d)], which has not yet been derived within the scalar--vector model. 

    The anisotropy of pure heat conduction takes the form 
    \begin{subequations}
    \begin{align}
        L^{\text{QQ}}_{im}(S, \vec{n})
        &=
        \rho_\perp \delta_{im} + \left(\rho_\| - \rho_\perp \right)n_i n_m
        ,\\
        \rho_\perp
        &=
        \rho_0 - \frac{1}{2}\rho_1 S
        ,\\
        \rho_\|
        &=
        \rho_0 + \rho_1 S
        ,
    \end{align}
    \end{subequations}
    where coefficients $\rho_\|$ and $\rho_\perp$ represent the thermal resistivity parallel and perpendicular to the director, respectively.

    \begin{table}[h]
        \caption{\label{tab:symmetry}%
            Some symmetry of variables and transport coefficients. Symbols $+$ and $-$ represent sign-preserved and -reversed transformations, respectively. A blank space indicates that the operation shown by the label cannot be applied.
        }
        \begin{ruledtabular}
            \begin{tabular}{ccccc}
                &$i \leftrightarrow j$    &$m \leftrightarrow n$    &$\mathcal{T}$    &$\mathcal{P}$
                \\
                \hline
                $\nabla_i$  &   &   &$+$    &$-$
                \\
                $v_i$   &   &   &$-$    &$-$
                \\
                $n_i$   &   &   &$+$    &$-$
                \\
                $S$ &   &   &$+$    &$+$
                \\
                $Q_{ij}$    &   &   &$+$    &$+$
                \\
                $T$   &   &   &$+$    &$+$
                \\
                \hline
                $\sigma^{\mathrm{sym}}_{ij}$    &$+$  &&$+$ &$+$
                \\
                $\sigma^{\mathrm{ant}}_{ij}$    &$-$  &&$+$ &$+$
                \\
                $-T^{-1}\nabla_i T$ &&&$+$  &$-$
                \\
                $A_{ij}$    &$+$    &&$-$   &$+$
                \\
                $N_{ij}$    &$-$    &&$-$   &$+$
                \\
                $q_i$   &&&$-$  &$-$
                \\
                \hline
                $L^{\text{AA}}_{ijmn}$  &$+$    &$+$    &$+$    &$+$
                \\
                $L^{\text{NN}}_{ijmn}$  &$-$    &$-$    &$+$    &$+$
                \\
                $L^{\text{AN}}_{ijmn}$  &$+$    &$-$    &$+$    &$+$
                \\
                $L^{\text{NA}}_{ijmn}$  &$-$    &$+$    &$+$    &$+$
                \\
                $L^{\text{QQ}}_{im}$    &&&$+$  &$+$
                \\
                $L^{\text{AQ}}_{ijm}$   &$+$    &&$+$   &$-$
                \\
                $L^{\text{NQ}}_{ijn}$  &$-$ &&$+$   &$-$
                \\
                $L^{\text{QA}}_{imn}$   &&$+$   &$+$    &$-$
                \\
                $L^{\text{QN}}_{imn}$   &&$-$   &$+$    &$-$
            \end{tabular}
        \end{ruledtabular}
    \end{table}

\subsection{Time reversal and parity}\label{sec:time_reversal}
    %\textcolor{black}{Green--Kubo formula. Fluctuation--Dissipation Theorem. Onsager reciprocal relation. Although this is re-formulated with \OnsagerVariational later, choice of variants are crucial.}
    Two discrete symmetries---time reversal $\mathcal{T}$ and parity $\mathcal{P}$---were examined. These symmetries impose constraints on the degrees of freedom of the transport matrix. As a consequence, the Leslie effects are related to their inverse processes, and their inherent chirality is elucidated. 
    
    With an arbitrary quantity $X$ as a function of time $t$ and position $\vec{r}$, $X$ transforms under time reversal $\mathcal{T}$ and parity $\mathcal{P}$ as
    \begin{subequations}
    \begin{align}
        \mathcal{T}&: X(t,\vec{r}) \mapsto \mathcal{T}X(t,\vec{r}) = \epsilon^{\mathcal{T}}_X X(-t,\vec{r})
        ,\\
        \mathcal{P}&: X(t,\vec{r}) \mapsto \mathcal{P}X(t,\vec{r}) = \epsilon^{\mathcal{P}}_X X(t,-\vec{r})
        ,
    \end{align}
    \end{subequations}
    where $\epsilon^{\mathcal{T}}_X$ and $\epsilon^{\mathcal{P}}_X$ take values of $\pm 1$, and $X$ is said to be $\mathcal{T}$-even if $\epsilon^{\mathcal{T}}_X = +1$, and $\mathcal{T}$-odd if $\epsilon^{\mathcal{T}}_X = -1$. The same terminology is also applied to $\mathcal{P}$.
    In dissipative cross-couplings, irreversible currents are required to have the same time reversal symmetry, as this is a necessary condition for ensuring non-negative entropy production~\cite{Brand2018RheolActa}. The variables we adopted satisfy this requirement (Table~\ref{tab:choice_of_variables}).
    
    The transport coefficients in Eq.~\eqref{eq:jmu_LmunuFnu} associated with the direct and inverse processes are related by \Reciproc~\cite{Onsager1931I,Onsager1931II} given by
    \begin{equation}
    \label{eq:Onsager_Reciprocal_Relations}
        L_{\mu\nu} = \epsilon^{\mathcal{T}}_\mu \epsilon^{\mathcal{T}}_\nu L_{\nu\mu}
        ,
    \end{equation}
    where $\epsilon^{\mathcal{T}}_\mu$ represents the time reversal symmetry of the current $j_\mu$. The current $j_\mu$ is fluctuating, and its time reversal symmetry in equilibrium leads to these relations.
    In the \EricksenLeslie, the Parodi relation 
    \begin{equation}
        \label{eq:parodi}
        \alpha_2 + \alpha_3 =\alpha_6 - \alpha_5
    \end{equation}
    is attributed to \Reciproc~\cite{Parodi1970}.
    Within the \ImuraOkano, the Parodi relation is included in
    \begin{subequations}
    \begin{align}
        L^{\text{AN}}_{ijmn} (S, \vec{n}) = L^{\text{NA}}_{mnij} (S, \vec{n})
        &=
        -\frac{1}{2} \frac{\alpha_2 + \alpha_3}{2} \left( n_i n_m \delta_{jn} + n_j n_m \delta_{in} - n_j n_n \delta_{im} - n_i n_n \delta_{jm} \right)
        \label{eq:LAN_LNA}\\
        &=
        \frac{1}{2 k_{\text{B}} T} \int d\vec{r}\,dt\, \langle \delta \sigma^{\text{sym}}_{ij} (\vec{r}, t) \, \delta \sigma^{\text{ant}}_{mn} (\vec{0}, 0) \rangle
        \label{eq:kubo_LAN_LNA}
        .
    \end{align}
    \end{subequations}
    Eq.~\eqref{eq:LAN_LNA} shows \Reciproc and the explicit form of the transport matrices. For the Leslie effects, the given direct process in Eq.~\eqref{eq:LAQ_LNQ_Sn} specifies the inverse process:
    \begin{subequations}
    \begin{align}
        L^{\text{AQ}}_{ijm} (S, \vec{n}) = L^{\text{QA}}_{mij} (S, \vec{n})
        &=
        \mu^{\text{A}} (n_i \epsilon_{jkm} n_k + n_j \epsilon_{ikm} n_k )
        \\
        L^{\text{NQ}}_{ijm} (S, \vec{n}) = L^{\text{QN}}_{mij} (S, \vec{n})
        &=
        \mu^{\text{N}} (n_i \epsilon_{jkm} n_k - n_j \epsilon_{ikm} n_k ) + \mu'^{\text{N}} \epsilon_{ijm}
        .
    \end{align}
    \end{subequations}

    Eq.~\eqref{eq:kubo_LAN_LNA} originates from the Green--Kubo formula. The Green--Kubo formula demonstrates that the response function is related to the correlation function~\cite{Kubo1957}, indicating that the response function is determined by the fluctuations of the relevant variables. The formula also leads to the FDT. We specifically adopted the FDT to identify the symmetry of response functions based on variables with known parity and time reversal symmetries. This approach is particularly useful in hydrodynamic systems where dissipation plays a crucial role~\cite{Pleiner1996, LandauLifshitzStat1980}.
    Within our choice of variables where all currents are $\mathcal{T}$-even (Table~\ref{tab:choice_of_variables}), \Reciproc $L_{\mu\nu} = L_{\nu\mu}$ implies that the transport matrix is determined by the currents:
    \begin{equation}
        L_{\mu\nu} = L_{\nu\mu} = \frac{1}{2 k_{\text{B}} T} \int d\vec{r}\,dt\, \langle \delta j_\mu (\vec{r}, t) \, \delta j_\nu (\vec{0}, 0) \rangle
        .
    \end{equation}
    The Green–Kubo formula is now applied to the Leslie effects: 
    \begin{subequations}
        \begin{align}
            L^{\text{AQ}}_{ijm} = L^{\text{QA}}_{mij}
            &=
            \frac{1}{2 k_{\text{B}} T} \int d\vec{r}\, dt\,
            \langle \delta \sigma^{\text{sym}}_{ij} (\vec{r}, t) \, \delta (-T^{-1} \nabla_m T) (\vec{0}, 0) \rangle
            ,\\
            L^{\text{NQ}}_{ijm} = L^{\text{QN}}_{mij} 
            &=
            \frac{1}{2 k_{\text{B}} T} \int d\vec{r}\, dt\,
            \langle \delta \sigma^{\text{ant}}_{ij} (\vec{r}, t) \, \delta (-T^{-1} \nabla_m T) (\vec{0}, 0) \rangle
            \label{eq:LNQ_LQN_FDT}
            .
        \end{align}
    \end{subequations}
    The resulting representation is consistent with the one derived and employed in molecular dynamics (MD) simulation~\cite{Sarman1999}. Since the stress tensors $\sigma^{\text{sym}}_{ij}, \sigma^{\text{ant}}_{ij}$ are $\mathcal{P}$-even and the temperature gradient $-T^{-1}\nabla_i T$ is $\mathcal{P}$-odd, the FDT indicates $L^{\text{AQ}}_{ijm} = L^{\text{QA}}_{mij}$ and $L^{\text{NQ}}_{ijm} = L^{\text{QN}}_{mij}$ are $\mathcal{P}$-odd (Table~\ref{tab:symmetry}).
    Consequently, the cross-coupling coefficients $\mu^{\text{A}}_{\bullet\bullet}, \mu^{\text{N}}_{\bullet\bullet}$  and thus $\mu^{\text{A}}, \mu^{\text{N}}, \mu'^{\text{N}}$ are pseudoscalars ($\mathcal{P}$-odd).
    An analogous analysis focusing on time reversal symmetry indicates that these coefficients are $\mathcal{T}$-even.
    Thus, $\mu^{\text{A}}, \mu^{\text{N}}, \mu'^{\text{N}}$ are true-chiral (pseudo)scalars~\cite{Barron2012chirality}, which predicts that the cross-coupling coefficients depend on the chirality of molecules. In bulk \NsLC~phase of mixtures, the Leslie thermomechanical coefficient is proportional to the concentration of chiral molecules (\textit{R})-octan-2-yl 4-((4-(hexyloxy)benzoyl)oxy)benzoate~\cite{Oswald2014epl}. The rotation directions are opposite between two enantiomers~\cite{Takano2023}. MD simulation showed that the cross-coupling coefficient is proportional to the chirality of the intermolecular potential or the fraction of chiral molecules~\cite{Sarman2016}. Also in the chemical Leslie effect, the directions of precession are opposite between mirror image molecules~\cite{Tabe2003}. The Leslie effects are intrinsically chiral phenomena, attributed to the local chirality of the liquid crystal phase, especially to the chirality of the constitutive molecules.

    The molecular origin is indicated by the representation with the correlation functions. The Leslie thermomechanical effect, in particular, originates from the cross-coupling of translational heat transport and rotational mechanical motion, formulated as the correlation between the temperature gradient and the antisymmetric stress tensor in Eq.~\eqref{eq:LNQ_LQN_FDT}. The cross-coupling coefficients $\mu^{\text{A}}, \mu^{\text{N}}, \mu'^{\text{N}}$ are true-chiral, namely they transform in the same manner as the electric toroidal monopole $G_0$. MD simulation for liquid water elucidated that molecular rotational degrees of freedom primarily contribute to heat transport~\cite{Ohara1999JChemPhys}. In crystals, chiral phonons carry orbital angular momentum. Their energies are shifted respectively to the handedness of the lattice, and the transported angular momentum is polarized in chiral crystals~\cite{Hamada2018PhysRevLett, Ohe2024prl}. In liquid crystals, chiral molecules preferentially rotate in a specific direction, dictated by the chirality of the phase or its surrounding environment. Consequently, translational heat transport may induce rotational motion. Molecular chirality ($G_0$) allows the finite Lifshitz invariant $\vec{n}\cdot(\vec{\nabla}\times\vec{n})$, which also transforms as $G_0$. Thus, the \NsLC~and \SmCsLC~phases have spontaneous twist. Since the inner product of the temperature gradient and torque transforms as $G_0$, the thermal Leslie effect is possible in the chiral phases. Note that the smectic-\textit{A} phase of chiral molecules is free of twist or transits into the twist-grain-boundary phase since the layered order suppresses the twist~\cite{de_Gennes1972}. Chirality permits the manifestation of diverse behaviors; however, the emergence of them can be constrained by additional factors. We discuss a constraint by thermodynamics in the following Sec.~\ref{sec:thermodynamic_bounds}.

    The foregoing discussion completes the reformulation of the Leslie effects within the framework of linear response theory. The degrees of freedom of the transport matrix were determined by the order parameter that describes the local symmetry of the phase and the discrete symmetries of the relevant variables. The dependence of the transport coefficients on the magnitude of orientational order was formulated explicitly. Time reversal symmetry led to the equivalence between forward and inverse processes, while parity reaffirmed that the Leslie effects are intrinsic to chiral liquid crystal phases. The outcome is encapsulated in the following two constitutive equations:
    \begin{subequations}
    \label{eq:constitutive}
        \begin{align}
            \sigma'_{ij} 
            &=
            \alpha_1 n_i n_j n_m n_n A_{mn}
            + \alpha_2 n_i \corot{n}_j
            + \alpha_3 n_j \corot{n}_i
            + \alpha_4 A_{ij}
            + \alpha_5 n_i n_m A_{mj}
            + \alpha_6 n_j n_m A_{mi}
            \nonumber\\
            &\hspace{1em}+
            \left[
                \mu^{\text{A}}
                \left(
                    n_i \epsilon_{jkm} n_k 
                    + n_j \epsilon_{ikm} n_k
                \right)
                + \mu^{\text{N}}
                \left(
                    n_i \epsilon_{jkm} n_k 
                    - n_j \epsilon_{ikm} n_k
                \right)
                + \mu'^{\text{N}}
                \epsilon_{ijm}
            \right]
            q_m
            ,   \label{eq:constitutive_sigma}\\
            -T^{-1}\nabla_i T
            &=
            \left[
                \rho_\perp \delta_{im}
                + (\rho_\parallel - \rho_\perp) n_i n_m
            \right]
            q_m
            - 2\mu^{\text{A}}
            \epsilon_{ikn} n_k n_m
            A_{mn}
            + 2 \left( \mu^{\text{N}} - \mu'^{\text{N}}\right)
            \epsilon_{ikm} n_k 
            \corot{n}_m
            \label{eq:constitutive_-T-1nablaT}
            .
        \end{align}
    \end{subequations}
    In the conventional argument, the temperature gradient $-T^{-1}\vec{\nabla} T$ is frequently chosen as an affinity~\cite{Leslie1968II, de_Gennes}. We present Eq.~\eqref{eq:constitutive2}, the equivalent constitutive equations to Eq.~\eqref{eq:constitutive}.

%==========================================================================
\section{Thermodynamic bounds}\label{sec:thermodynamic_bounds}
    %\textcolor{yellow}{bounds to coefficients. thermodynamic uncertainty relation. quantum. second law. }
    Linear response theory permits the Leslie effects that convert energy. However, the efficiency of such processes is fundamentally limited by the second law of thermodynamics given by
    \begin{equation}
        \label{eq:2RT_0}
        \frac{2R}{T} \ge 0
        .
    \end{equation}
    This inequality places bounds on the coefficients in terms of the corresponding conjugate transport coefficients. %In this section, we refine our theory of the Leslie effect by deriving thermodynamic constraints on the coupling coefficients, thereby ensuring consistency with the non-negativity of entropy production.

    The TUR imposes a stronger constraint than the second law. Irreversible processes dissipate energy at a finite rate. The TUR takes this nature into account and provides a more appropriate thermodynamic bound on the transport matrix. We applied the TUR to the Leslie effects to determine thermodynamic bounds on the transport coefficients. Numerical studies~\cite{Li2019natcommun,Manikandan2020prl} have demonstrated the reliability of the TUR in estimating entropy production. To the best of our knowledge, this section represents the first attempt to apply the TUR to actual systems of condensed matter.

\subsection{Application of thermodynamic uncertainty relation}
    %\textcolor{black}{positive semi-definite. all eigenvalues are positive. block-diagonalised matrix}
    The TUR states that entropy production is universally bounded from below by the fluctuations of the current. The TUR in the linear response regime
    \begin{equation}
        \label{eq:TUR_linear}
        \frac{2R}{T}
        \ge
        \frac{2 \langle j_\lambda \rangle^2}{D_\lambda}
        %= \frac{2 \left( \sum_\mu \lambda_\mu j_\mu \right)^2}{2 \sum_{\mu\nu} L_{\mu\nu} \lambda_\mu \lambda_\nu}
    \end{equation}
    was proven via the triangle inequality~\cite{Barato2015}. Here, $j_\lambda \coloneqq \lambda_\mu j_\mu$ is a linear combination of currents weighted by real coefficients $\lambda_\mu$, and $D_\lambda \coloneqq 2 L_{\mu\nu} \lambda_\mu \lambda_\nu$ the fluctuations of $j_\lambda$ around its steady-state value $\langle j_\lambda \rangle$. Eq.~\eqref{eq:TUR_linear} shows that the entropy production is bounded by the fluctuations of the current relative to $\langle j_\lambda \rangle$. The general TUR is proven via large deviation theory~\cite{Horowitz2017pre}, and a weaker inequality applicable to $\mathcal{T}$-odd observables is derived solely from the fluctuation theorem, namely time reversal symmetry~\cite{Hasegawa2019prl}. 

    The second law in Eq.~\eqref{eq:2RT_0} constitutes both a necessary and sufficient condition for the TUR in our system, which is formulated within the framework of linear response theory and a specific choice of variables. The variables were chosen such that all currents are $\mathcal{T}$-even (Table~\ref{tab:symmetry}), and the transport matrix becomes symmetric due to \Reciproc: $L_{\mu\nu} = L_{\nu \mu}$. In the linear response regime, the second law takes the form of
    \begin{equation}
        \frac{2R}{T}
        =
        \frac{1}{T} \sum_{\mu\nu} L_{\mu\nu} F_\mu F_\nu \ge 0
        .
    \end{equation}
    This inequality implies that the symmetric $\tsr{L}$ is positive semidefinite. 
    The application of the Cauchy--Schwarz inequality $(\vec{F}^{\text{T}} \tsr{L} \vec{F}) (\vec{\lambda}^{\text{T}} \tsr{L} \vec{\lambda}) \ge |\vec{F}^{\text{T}} \tsr{L} \vec{\lambda}|^2$ with the vector of real coefficients $\vec{\lambda} \coloneqq (\lambda_1,\lambda_2,\dots)$ and affinity $\vec{F} \coloneqq (F_1, F_2, \dots)$ yields
    \begin{equation}
        \left( L_{\mu\nu} \lambda_\mu \lambda_\nu \right)
        \left( L_{\mu\nu} F_\mu F_\nu \right)
        \ge
        \left( L_{\mu\nu} \lambda_\mu F_\nu \right)^2
        .
    \end{equation}
    The positive semidefiniteness of $\tsr{L}$ ensures that $D_\lambda /2 = L_{\mu\nu} \lambda_\mu \lambda_\nu \ge 0$, leading to
    \begin{equation}
        L_{\mu\nu} F_\mu F_\nu
        \ge
        \frac{\left( L_{\mu\nu} \lambda_\mu F_\nu \right)^2}{L_{\mu\nu} \lambda_\mu \lambda_\nu}
        ,
    \end{equation}
    which is equivalent to the TUR Eq.~\eqref{eq:TUR_linear} since the assumption of linear response in Eq.~\eqref{eq:jmu_LmunuFnu} leads to
    \begin{equation}
        L_{\mu\nu} \lambda_\mu F_\nu = \lambda_\mu \langle j_\mu \rangle = \langle j_\lambda \rangle
        .
    \end{equation}
    Thus, the TUR has been derived from the second law within the assumption that is applicable to our system~\cite{Macieszczak2018prl}.

    In our system, the TUR is equivalent to the second law. This equivalence underscores that the positive semidefiniteness of the transport matrix $\tsr{L}$ constitutes a refined thermodynamic constraint. The positive semidefiniteness of $\tsr{L}$ is a necessary and sufficient condition for all its eigenvalues to be non-negative real numbers. The non-negativity of the eigenvalues constitutes ten thermodynamic bounds on the transport coefficients:
    \begin{subequations}
        \begin{align}
            \alpha_4
            \ge 0
            &, \label{eq:bound_alpha_4}\\
            \alpha_3 - \alpha_2
            \ge 0
            &, \label{eq:bound_alpha_3-alpha_2}\\
            2\alpha_4 + (\alpha_5 + \alpha_6)
            \ge 0
            &, \label{eq:bound_2alpha_4}\\
            2 \alpha_1 + 3 \alpha_4 + 2 (\alpha_5 + \alpha_6)
            \ge 0
            &, \label{eq:bound_2alpha_1}\\
            \rho_\perp
            \ge 0
            &, \label{eq:bound_rho_perp}\\
            \rho_\parallel
            \ge 0
            &, \label{eq:bound_rho_parallel}\\
            X^2 + Y^2 + Z^2 + 2 X Y Z
            \le 1
            &, \label{eq:bound_X2+Y2+Z2-2XYZ}\\
            |X|
            \le 1
            &, \label{eq:bound_|X|}\\
            |Y|
            \le 1
            &, \label{eq:bound_|Y|}\\
            |Z|
            \le 1
            &, \label{eq:bound_|Z|}
        \end{align}
    \end{subequations}
    where
    \begin{subequations}
        \begin{align}
            X
            &\coloneqq
            \frac{2\mu^{\text{A}}}{\sqrt{(2\alpha_4 + (\alpha_5 + \alpha_6))\rho_\perp}}
            , \label{eq:bound_X2muA}\\
            Y
            &\coloneqq
            \frac{2\left(\mu^{\text{N}}-\mu'^{\text{N}}\right)}{\sqrt{(\alpha_3 - \alpha_2)\rho_\perp}}
            , \label{eq:bound_X2muN}\\
            Z
            &\coloneqq
            \frac{\alpha_2 + \alpha_3}{\sqrt{(2\alpha_4 + (\alpha_5 + \alpha_6))(\alpha_3 - \alpha_2)}}
            . \label{eq:bound_Zalpha2alpha3}
        \end{align}
    \end{subequations}
    The details of the derivation are shown in Appendix~\ref{sec:derivation_bounds}. 
    The inequalities above are thermodynamic bounds in a general framework incorporating the Leslie effects. The subsequent three subsections verify the set of these bounds by comparing it with known inequalities argued by previous studies~\cite{Leslie1968II, Leslie1979, BerisEdwards1994}. The conventional inequalities were actually derived in specific limiting cases.

\subsection{Isotropic limit}\label{sec:isotropic_limit}
    The \ILC~phase is considered. This system is formed by chiral molecules and is chiral but devoid of anisotropy. Brand\etal mentioned that reversible cross-coupling is forbidden in isotropic fluids~\cite{Brand2018RheolActa}. We extend their mention to include the irreversible Leslie effects. 

    The \ILC~phase is characterized as the zero scalar order parameter
    \begin{equation}
        S = 0
        .
    \end{equation}
    An isotropic, incompressible fluid has a single viscosity coefficient. Eq.~\eqref{eq:Imura_Okano_Leslie_viscosity} shows that all coefficients other than $\alpha_4$ vanish in the \ILC~phase:
    \begin{subequations}
    \begin{align}
        \alpha_1 = \alpha_2 = \alpha_3 = \alpha_5 = \alpha_6 &= 0
        ,\\
        \alpha_4 &= \eta
        .
    \end{align}
    \end{subequations}
    With $\eta \ge 0$, the thermodynamic bounds on the viscosity in Eqs.~\eqref{eq:bound_alpha_4}--\eqref{eq:bound_2alpha_1} are satisfied. 
    The isotropic viscosity $\alpha_4$ remains nonzero in the \ILC~phase. Despite its name ``isotropic'', $\alpha_4$ has the $S$-dependence in the \NsLC~phase, as is described in several models~\cite{Imura1972,BerisEdwards1994}. The $S$-dependence of $\alpha_4$ is observed in the calamitic phase~\cite{Luis1983MolCrystLiqCryst}, the discotic phase~\cite{Grecov2003molcrystliqcryst} and polymer solutions~\cite{Doi1983faraday}. The anisotropy of the phase influences even on the isotropic components of the response. The Leslie viscosity coefficients except $\alpha_4$ appear only in an anisotropic phase. Eq.~\eqref{eq:Imura_Okano_Leslie_viscosity} shows that the presence of $\alpha_5$ and $\alpha_6$ in the \NsLC~phase corresponds to the $S$-dependence of $\alpha_4$. The coefficients, which are absent in the \ILC~phase and only appears in the anisotropic phase, inevitably depend on $S$.

    The anisotropy of heat conduction also vanishes in the \ILC~phase:
    \begin{equation}
        \rho_\perp = \rho_\parallel = \rho_0
        .
    \end{equation}
    Eqs.~\eqref{eq:bound_rho_perp}~and~\eqref{eq:bound_rho_parallel} are satisfied since $\rho_0 > 0$.
    
    Most of the cross-coupling coefficients vanish due to their $S$-dependence in Eq.~\eqref{eq:LAQ_LNQ_Sn}:
    \begin{subequations}
    \begin{align}
        \mu^{\text{A}}
        =
        \mu^{\text{N}}
        &=
        0,\\
        \mu'^{\text{N}}
        &=
        \mu^{\text{N}}_{01}
        ,
    \end{align}
    \end{subequations}
    but the chirality of the phase allows $\mu^{\text{A}}_{12}, \mu^{\text{A}}_{22}, \mu^{\text{A}}_{24}, \mu^{\text{N}}_{11}, \mu^{\text{N}}_{12}, \mu^{\text{N}}_{21}, \mu^{\text{N}}_{22}, \mu^{\text{N}}_{23}, \mu^{\text{N}}_{24}, \mu^{\text{N}}_{25}$ to be nonzero. However, a thermodynamic bound in Eq.~\eqref{eq:bound_|Y|} and $\alpha_2 - \alpha_3 = 0$ require
    \begin{equation}
        \mu^{\text{N}}_{01} = 0
        .
    \end{equation}
    Symmetry does not prohibit $\mu'^{\text{N}}$, but thermodynamics forbids it in the \ILC~phase. From a perspective of symmetry, the $\mu'^{\text{N}}$-term is possible in the \ILC~phase since rotation and heat current form a chiral configuration without the anisotropy of the phase [Fig.~\ref{fig:terminology}(c)]. However, the twist viscosity $\gamma_1 = \alpha_3 - \alpha_2$ is absent in the \ILC~phase, and thermodynamics forbids the cross-coupling of antisymmetric strain and heat conduction. Note that the vanishing of the cross-coupling coefficients $\mu^{\text{A}}, \mu^{\text{N}}, \mu'^{\text{N}}$ in the \ILC~phase originates from the $S$-dependence, which is consistent with their emergence in the \NsLC~phase.

\subsection{Achiral limit}
    The achiral \NLC~phase is considered. The system is locally invariant under $\mathcal{P}$. Twist in the director field is acceptable but it never arises the Leslie effects. In this achiral limit, one obtains the \EricksenLeslie and pure heat conduction without the Leslie effects. Since chirality of the phase is essential for the Leslie effects (see subsection~\ref{sec:time_reversal}), chiral coefficients vanish:
    \begin{equation}
        \mu^{\text{A}} = \mu^{\text{N}} = \mu'^{\text{N}} = 0
        .
    \end{equation}
    This relation is equivalent to
    \begin{equation}
        X = Y = 0
        ,
    \end{equation}
    and thus Eqs.~\eqref{eq:bound_|X|}~and~\eqref{eq:bound_|Y|} are omitted. The remaining Eqs.\eqref{eq:bound_alpha_4}--\eqref{eq:bound_2alpha_1}, \eqref{eq:bound_X2+Y2+Z2-2XYZ}, and \eqref{eq:bound_|Z|} are bounds on the Leslie viscosity coefficients. These bounds are consistent with those originally derived by Leslie~\cite{Leslie1979, BerisEdwards1994} from the second law. Since the TUR and the second law are equivalent under linear response with a proper choice of variables, this agreement is to be expected. For pure but anisotropic heat conduction, Eqs.~\eqref{eq:bound_rho_perp}~and~\eqref{eq:bound_rho_parallel} are obvious as the heat current always proceeds from the hot side to the cold side. In the achiral limit, Eqs.~\eqref{eq:bound_X2+Y2+Z2-2XYZ}~and~\eqref{eq:bound_|Z|} are equivalent.

    Thus, our model and thermodynamic bounds reproduce the well-known transport behavior of achiral liquid crystals.

\subsection{Leslie limit}
    The \NsLC~phase that has chirality and anisotropy is considered. Our thermodynamic bounds Eqs.~\eqref{eq:bound_alpha_4}--\eqref{eq:bound_|Z|} should be compared with the inequalities derived by Leslie~\cite{Leslie1968II}:
    \begin{subequations}
        \begin{eqnarray}
            \lambda_1
            &&\le 0
            , \label{eq:Leslie_bound_lambda_1}\\
            \kappa_1
            &&\le 0
            , \label{eq:Leslie_bound_kappa_1}\\
            \kappa_1 + \kappa_2
            &&\le 0
            , \label{eq:Leslie_bound_kappa_1+kappa_2}\\
            \frac{4\lambda_1\kappa_1}{T}
            &&\ge
            \left(\frac{\kappa_3}{T} - \lambda_3 + T \frac{d}{dT} \frac{\alpha}{T}\right)^2
            \label{eq:Leslie_bound_4lambda_1kappa_1}
        \end{eqnarray}
    \end{subequations}
    in Leslie's notation (except for the director and some more quantities).
    With the irreversible stress $\hat{\sigma}'_{ij}$, the molecular field $\hat{g}'_i$, and the heat current $q_i$, Leslie presented the constitutive equations
    \begin{subequations}
        \begin{align}
            \hat{\sigma}'_{ij}
            &=
            \mu_2 n_i N_j + \mu_3 n_j N_i + \mu_7 n_i \epsilon_{jkm} n_k \nabla_m T + \mu_8 n_j \epsilon_{ikm} n_k \nabla_m T
            ,   \label{eq:Leslie_bound_hatsigma}\\
            \hat{g}'_i
            &=
            \lambda_1 N_i + \lambda_3\epsilon_{ikm} n_k \nabla_m T
            ,   \label{eq:Leslie_bound_hatg}\\
            q_i
            &=
            \kappa_1 \nabla_i T + \kappa_2 n_i n_m \nabla_m T + \kappa_3 \epsilon_{ikm} n_k N_m
            ,   \label{eq:Leslie_bound_q}\\
            \lambda_1
            &=
            \mu_2 - \mu_3
            ,\\
            \lambda_3
            &=
            \mu_7 - \mu_8
            .
        \end{align}
    \end{subequations}
    
    See Eq.~\eqref{eq:leslie_notation} for the correspondence of notations between Leslie's and ours.
    These constitutive equations consider the Leslie thermomechanical effect, while omitting the Leslie thermohydrodynamic effect as well as most of the symmetric viscosity.
    Keeping $\hat{\sigma}'_{ij} = \sigma'_{ij}$ in mind, the comparison of Eqs.~\eqref{eq:constitutive2_sigma} and \eqref{eq:Leslie_bound_hatsigma} as well as \eqref{eq:constitutive2_q} and \eqref{eq:Leslie_bound_q} tells us the correspondence. 
    The application of Eqs.~\eqref{eq:bound_alpha_3-alpha_2} and \eqref{eq:bound_|Y|} shows
    \begin{equation}
        \lambda_1
        = 
        - \left( \alpha_3 - \alpha_2 \right)
        \left(
            1 - \frac{\left(2\mu^{\text{N}}\right)^2}{(\alpha_3 - \alpha_2) \rho_\perp}   
        \right)
        \le 0
        ,
    \end{equation}
    which is the same as Eq.~\eqref{eq:Leslie_bound_lambda_1}. 
    Eqs.~\eqref{eq:Leslie_bound_kappa_1} and \eqref{eq:Leslie_bound_kappa_1+kappa_2} are non-positivity of heat conductivities perpendicular and paralel to the director, respectively.
    The application of Eqs.~\eqref{eq:bound_rho_perp} and \eqref{eq:bound_rho_parallel} shows
    \begin{align}
        \kappa_1
        &=
        -\frac{1}{\rho_\perp T}
        <
        0
        ,\\
        \kappa_1 + \kappa_2
        &=
        -\frac{1}{\rho_\parallel T}
        <
        0
        .
    \end{align}
    The application of Eqs.~\eqref{eq:bound_alpha_3-alpha_2}, \eqref{eq:bound_rho_perp} and \eqref{eq:bound_|Y|} leads to
    \begin{equation}
        \frac{4\lambda_1\kappa_1}{T}
        - \left(
            \frac{\kappa_3}{T}
            - \lambda_3
            + \frac{1}{T} \frac{d}{dT} \frac{\alpha}{T}
        \right)^2
        =
        \frac{4}{\rho_\perp T^2}
        (\alpha_3 - \alpha_2)
        \left( 1 - \frac{\left(2 \mu^{\text{N}}\right)^2}{(\alpha_3 - \alpha_2) \rho_\perp} \right)^2
        \ge 0
        ,
    \end{equation}
    which is equivalent to Eq.~\eqref{eq:Leslie_bound_4lambda_1kappa_1}. From the up-to-date understanding~\cite{de_Gennes}, $\alpha=0$ is known.

    The Leslie limit is thus recovered. Our model is more general in that it takes into account the $\mu^{\text{A}}$- and $\mu'^{\text{N}}$-terms (as well as all the Leslie viscosity).

%==========================================================================
\section{Comparison with known experimental observations}\label{sec:validation}
    We tested our model in several situations. Our model is consistent with all the situations considered in the following.

\subsection{Leslie thermomechanical coefficient compared with known values}
    %\textcolor{black}{range of $|Y|$, impose hydro, discuss slope in MD, step/descrete function?}
    Eq.~\eqref{eq:bound_|Y|} shows the upper bound of the Leslie thermomechanical coefficient. This inequality is validated by comparing it with the results of experiments. Previous studies often adopted the constitutive equations in Eqs.~\eqref{eq:constitutive2_sigma}~and~\eqref{eq:constitutive2_q} and measured $\nu = 2 \tilde{\mu}^{\text{N}}$, where $\nu$ is the frequent notation of the Leslie thermomechanical coefficient in these studies~\cite{Dequidt2007epl, Dequidt2008epje, Dequidt2008epje, Oswald2008PRE, Oswald2008PRL, Oswald2008EPL, Oswald2012epl, Oswald2014epl, Bono2019, Bono2019, Nishiyama2021SoftMatter, Takano2023}. As shown in Table~\ref{tab:nu_exp}, the first measurement was reported with $|\nu| < 2\times 10^{-6}\,\si{kg.s^{-2}.K^{-1}}$ for a mixture of 4$'$-alkyl-4-cyanobiphenyl~($n$CB) and cholesteryl chloride~(CC)~\cite{Eber1982}. Subsequent studies demonstrated the distribution of $|\nu|$ in the ranges of $10^{-8}\text{--}10^{-6}\, \si{kg.s^{-2}.K^{-1}}$ by employing several experimental techniques. The value and sign of $\nu$ are known to depend on the composition of the sample and the temperature. However, thermodynamics imposes bounds on $\nu$ as shown in Eq.~\eqref{eq:bound_|Y|}, which holds beyond the details of experimental systems.

    The cross-coupling coefficient $\nu$ is bounded by the twist viscosity $\gamma_1$ and the heat conductivity $\sigma_\perp = {\rho_\perp}^{-1}$ (and temperature). Both $\gamma_1$ and $\sigma_\perp$ depend on the material and the temperature. We computed $Y$ by applying the appropriate values of them for each experimental condition. Unfortunately, at least one of them is not shown in the previous publications. 
    In Ref.~\cite{Oswald2008PRL}, $\nu \sim 1.5\times10^{-6}\, \si{kg.s^{-2}.K^{-1}}$ was obtained from the rotational speed of \NsLC~droplets dispersed in the \ILC~phase~(Table~\ref{tab:nu_exp}). A mixture of 4$'$-octyloxy-4-cyanobiphenyl~(8OCB) and CC was used as the liquid crystal sample, and $\gamma_1 \sim 0.015\, \si{Pa.s}$ was estimated at the phase transition temperature. However, the thermal conductivity $\sigma_\perp$ is approximately one-seventh that of glass~\cite{Oswald2008PRL}. The same group mentioned that the heat conductivity of glass is $\sim 1\, \si{W.m^{-1}.K^{-1}}$~\cite{Oswald2014epl}, so we calculated $\sigma_\perp \sim 0.14\, \si{W.m^{-1}.K^{-1}}$ for that liquid crystal sample, which is comparable to the typical heat conductivity of $n$CB perpendicular to the director $\sim 0.13\, \si{W.m^{-1}.K^{-1}}$~\cite{Marinelli1998PRE}. Since the experiment was carried out in the coexistence of the \NsLC~and \ILC~phases (\NsLC--\ILC) in $66\text{--}66.6\si{\degreeCelsius}$, we adopted this temperature to estimate $|Y| \sim 6\times 10^{-4}$.
    For the other experimental conditions, we also estimated $|Y|$ by supplementing $\gamma_1,\ \sigma_\perp$ based on the materials investigated in each study. We postulated a wide range of potential values for $\gamma_1$ and $\sigma_\perp$ to ensure that their true values would fall within that range. The range is at most $10$ times larger and smaller than the representative values, which appear as error bars in $|Y|$ across the digits. 

    \begin{figure}
        \includegraphics[width=8.5cm, keepaspectratio]{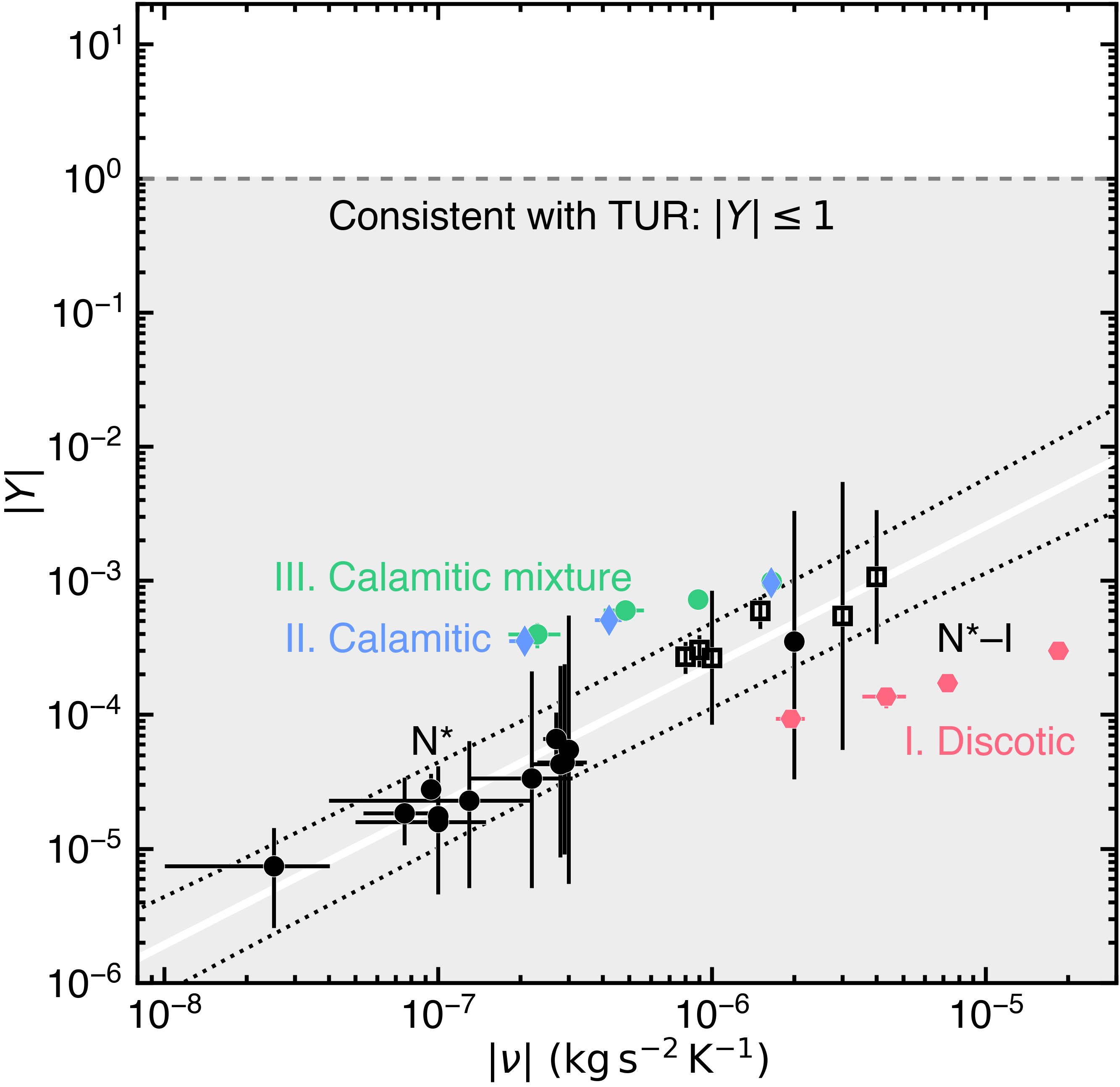}
        \caption{
            \label{fig:nuY_exp} 
            Consistency of the thermodynamic bound with the experiments. 
            The black plots show the experimentally obtained Leslie thermomechanical coefficient $\nu$: the closed circles in the \NsLC~phase and the open squares in the \NsLC--\ILC~coexistence state. We calculated $|Y|$ from $\nu$ and supplemented $\gamma_1$ and $\sigma_\perp$ (and temperature) as shown in Table~\ref{tab:nu_exp}. The white line represents the regression curve $\log_{10}{|Y|} = 1.0\, \log_{10}{|\nu|} + 2.6$ for all the experiments, and the black dotted lines represent the $95\%$ prediction interval. The colored plots show the estimation (Table~\ref{tab:nu_MDsim}) by MD simulation in Ref.~\cite{Sarman2016}. The red hexagons, the blue diamonds, and the green circles are system I (disklike molecules), II (rodlike molecules), and III (mixture of chiral and achiral rodlike molecules), respectively. The gray dashed line is the upper limit of $|Y|$ imposed by the TUR, and the gray region is consistent with the thermodynamics as shown in Eq.~\eqref{eq:bound_|Y|}. 
        }
    \end{figure}

    \begingroup
    \setlength{\tabcolsep}{0pt}
    \begin{table}[h]
        \caption{\label{tab:nu_exp}%
            The Leslie thermomechanical coefficient $\nu$ obtained by experiments. For the twist viscosity $\gamma_1$, the heat conductivity perpendicular to the director $\sigma_\perp$, the temperature $T$, and $|Y|$, the values or estimated ranges that they can take are indicated within the bracket $[\bullet]$, with the representative values indicated its outside. The data are organized chronologically, with the most recent publication at the bottom.
            %~\cite{KlemanLavrentovich2003}.
            %Numbers in columns Three--Five are aligned with the ``d'' column specifier (requires the \texttt{dcolumn} package). Non-numeric entries (those entries without a ``.'') in a ``d'' column are aligned on the decimal point. Use the ``D'' specifier for more complex layouts. 
        }
        \begin{ruledtabular}
            \begin{tabular}{
            d	d	D{,}{\times 10}{3}		d	D{,}{\text{--}}{5} @{\hspace{1em}}		d	D{,}{\text{--}}{5} @{\hspace{1em}}		d	D{,}{\text{--}}{5} @{\hspace{1em}}		d	D{,}{\text{--}}{3}	D{,}{\times 10}{3} @{\hspace{1em}}	l @{\hspace{0.5em}}	r	}
\multicolumn{3}{c}{	$ |\nu| $ }			& \multicolumn{2}{c}{	$ \gamma_1 $ }		& \multicolumn{2}{c}{	$ \sigma_\perp $ }		& \multicolumn{2}{c}{	$ T $ }		& \multicolumn{3}{c}{	$ |Y|$ }		&Phase	&Ref.	\\
\multicolumn{3}{c}{	$ ( \si{kg.s^{-2}.K^{-1}} ) $ }			& \multicolumn{2}{c}{	$ ( \si{Pa.s} ) $ }		& \multicolumn{2}{c}{	$ ( \si{W.m^{-1}.K^{-1}} ) $ }		& \multicolumn{2}{c}{	$ ( \si{\degreeCelsius} ) $ }		& \multicolumn{3}{c}{	}		&	&	\\
																		\hline
2	& 	& ,^{-6}		& 0.1	& [0.01 , 1]	\footnotemark[2]	& 0.1	& [0.01 , 1]	\footnotemark[8]	& 36	& [4.3 , 23.1]		& 3.5	& [0.3 , 33.1]	& ,^{-4} 	& \NsLC	& \cite{Eber1982}	\\
2.8	&  \pm 0.6	& ,^{-7}		& 0.1	& [0.01 , 1]	\footnotemark[3]	& 0.14	& [0.071 , 0.21]	\footnotemark[9]	& 59	& 		& 4.3	& [0.9 , 23.2]	& ,^{-5} 	& \NsLC	& \cite{Dequidt2007epl}	\\
2.9	&  \pm 0.6	& ,^{-7}		& 0.1	& [0.01 , 1]	\footnotemark[3]	& 0.14	& [0.071 , 0.21]	\footnotemark[9]	& 59	& 		& 4.4	& [0.9 , 23.9]	& ,^{-5} 	& \NsLC	& \cite{Dequidt2008epje}	\\
2.2	&  \pm 0.9	& ,^{-7}		& 0.1	& [0.01 , 1]	\footnotemark[3]	& 0.14	& [0.071 , 0.21]	\footnotemark[9]	& 59	& 		& 3.4	& [0.5 , 21.1]	& ,^{-5} 	& \NsLC	& \cite{Dequidt2008epje}	\\
1.3	&  \pm 0.9	& ,^{-7}		& 0.075	& [0.055 , 0.095]		& 0.14	& [0.071 , 0.21]	\footnotemark[9]	& 57.5	& 		& 2.3	& [0.5 , 6.4]	& ,^{-5} 	& \NsLC	& \cite{Oswald2008PRE}	\\
1.5	& 	& ,^{-6}		& 0.015	& 		& 0.14	& [0.071 , 0.21]	\footnotemark[9]	& 66.3	& [66 , 66.6]	\footnotemark[10]	& 6	& [4.4 , 7.6]	& ,^{-4} 	& \NsLC --\ILC	& \cite{Oswald2008PRL}	\\
1	& 	& ,^{-7}		& 0.075	& 		& 0.14	& [0.071 , 0.21]	\footnotemark[9]	& 57.8	& [57.7 , 57.9]		& 1.8	& [1.4 , 2.5]	& ,^{-5} 	& \NsLC	& \cite{Oswald2008EPL}	\\
1	&  \pm 0.5	& ,^{-7}		& 0.09	& [0.06 , 0.18]	\footnotemark[4]	& 0.14	& [0.071 , 0.21]	\footnotemark[9]	& 51.5	& 		& 1.6	& [0.5 , 4.1]	& ,^{-5} 	& \NsLC	& \cite{Oswald2012epl}	\\
9	& 	& ,^{-7}		& 0.02	& 		& 0.14	& [0.071 , 0.21]	\footnotemark[9]	& 53.8	& [53.5 , 54]	\footnotemark[11]	& 3	& [2.3 , 3.9]	& ,^{-4} 	& \NsLC --\ILC	& \cite{Oswald2012epl}	\\
8	& 	& ,^{-7}		& 0.02	& 		& 0.14	& [0.071 , 0.21]	\footnotemark[9]	& 53.8	& [53.5 , 54]	\footnotemark[11]	& 2.7	& [2.0 , 3.5]	& ,^{-4} 	& \NsLC --\ILC	& \cite{Oswald2012epl}	\\
27.0	&  \pm 2.8	& ,^{-8}	\footnotemark[1]	& 0.04	& [0.02 , 0.06]	\footnotemark[5]	& 0.13	& 		& 38.8	& [35.8 , 41.8]	\footnotemark[12]	& 6.6	& [4.8 , 10.4]	& ,^{-5} 	& \NsLC	& \cite{Oswald2014epl}	\\
7.5	&  \pm 2.2	& ,^{-8}	\footnotemark[1]	& 0.04	& [0.02 , 0.06]	\footnotemark[5]	& 0.13	& 		& 38.8	& [35.8 , 41.8]	\footnotemark[12]	& 1.8	& [1.1 , 3.4]	& ,^{-5} 	& \NsLC	& \cite{Oswald2014epl}	\\
9.4	&  \pm 0.8	& ,^{-8}	\footnotemark[1]	& 0.028	& [0.019 , 0.036]	\footnotemark[6]	& 0.13	& 		& 41.6	& [39.1 , 44.1]	\footnotemark[13]	& 2.8	& [2.2 , 3.6]	& ,^{-5} 	& \NsLC	& \cite{Oswald2014epl}	\\
2.5	&  \pm 1.5	& ,^{-8}	\footnotemark[1]	& 0.028	& [0.019 , 0.036]	\footnotemark[6]	& 0.13	& 		& 41.6	& [39.1 , 44.1]	\footnotemark[13]	& 7.4	& [2.6 , 14.3]	& ,^{-6} 	& \NsLC	& \cite{Oswald2014epl}	\\
																		
4	& 	& ,^{-6}		& 0.05	& 		& 0.1	& [0.01 , 1]	\footnotemark[8]	& 82	& [79 , 85]	\footnotemark[14]	& 1.1	& [0.3 , 3.4]	& ,^{-3} 	& \NsLC --\ILC	& \cite{Bono2019}	\\
1	& 	& ,^{-6}		& 0.05	& 		& 0.1	& [0.01 , 1]	\footnotemark[8]	& 82	& [79 , 85]	\footnotemark[14]	& 2.7	& [0.8 , 8.4]	& ,^{-4} 	& \NsLC --\ILC	& \cite{Bono2019}	\\
3	& 	& ,^{-6}		& 0.1	& [0.01 , 1]	\footnotemark[7]	& 0.1	& [0.01 , 1]	\footnotemark[8]	& 58	& 	\footnotemark[15]	& 5.5	& [0.5 , 54.6]	& ,^{-4} 	& \NsLC --\ILC	& \cite{Nishiyama2021SoftMatter}	\\
3	& 	& ,^{-7}		& 0.1	& [0.01 , 1]	\footnotemark[7]	& 0.1	& [0.01 , 1]	\footnotemark[8]	& 61.6	& 		& 5.5	& [0.5 , 54.9]	& ,^{-5} 	& \NsLC	& \cite{Takano2023}	\\
            \end{tabular}
        \end{ruledtabular}
        \footnotetext[1]{Converted for $1\text{wt}\%$ concentration of chiral dopants from the Leslie thermomechanical power.}
\footnotetext[2]{Roughly estimated based on the viscosity of pure 8CB and cholesteryl acetate, $0.035\, \si{Pa.s}$ at $33.5\si{\degreeCelsius}$ and $0.053\, \si{Pa.s}$ at $105\si{\degreeCelsius}$, respectively~\cite{Pestov2018}.}
\footnotetext[3]{Roughly estimated based on the viscosity of pure 8OCB and cholesteryl acetate, $16\, \si{mm^{-2}.s}$ at $70\si{\degreeCelsius}$ and $0.053\, \si{Pa.s}$ at $105\si{\degreeCelsius}$, respectively~\cite{Pestov2018}.}
\footnotetext[4]{The effective viscosity $\gamma_1^\ast$ at the compensation temperature (solidus temperature $-2\si{\degreeCelsius}$) is $\sim 0.18\, \si{Pa.s}$. Since $\gamma_1$ and the surface viscosity contributed to $\gamma_1^\ast$ in similar proportions, $\gamma_1^\ast$ was estimated as the upper limit of $\gamma_1$ and $\gamma_1^\ast / 3$ as the lower limit.}
\footnotetext[5]{$\gamma_1$ of pure 7CB in $-7\text{--}-1\si{\degreeCelsius}$ from the \NLC--\ILC~transition temperature~\cite{Luis1983MolCrystLiqCryst} was adopted.}
\footnotetext[6]{$\gamma_1$ of pure MBBA in $40.0\text{--}44.0\si{\degreeCelsius}$~\cite{Kneppe1982JChemPhys} was adopted.}
\footnotetext[7]{Roughly estimated based on the viscosity of $n$CB~\cite{Pestov2018}.}
\footnotetext[8]{Roughly estimated based on the heat conductivity of $n$CB~\cite{Marinelli1998PRE}.}
\footnotetext[9]{The heat conductivity of the 8CB + CC mixture is estimated to be ca. one-seventh that of glass ($\sim 1\, \si{W.m^{-1}.K^{-1}}$~\cite{Oswald2014epl})~\cite{Oswald2008PRL}, which is comparable to that of $n$CB and adaptable to 8CB + 8OCB + CC mixture.}
\footnotetext[10]{The experiment was conducted in the coexistence state of the \NsLC~and \ILC~phases. The sample began to melt at $66\si{\degreeCelsius}$ and possessed a freezing range of ca. $0.6\si{\degreeCelsius}$.}
\footnotetext[11]{The experiment was conducted in the coexistence state of the \NsLC~and \ILC~phases. The sample melted in $53.5\text{--}54\si{\degreeCelsius}$.}
\footnotetext[12]{The experiment was conducted $-7\text{--}-1\si{\degreeCelsius}$ below the transition temperature~\cite{Oswald2012epl}, which we assume $42.8\si{\degreeCelsius}$ for pure 7CB\cite{Pestov2018}.}
\footnotetext[13]{The experiment was conducted $-6\text{--}-1\si{\degreeCelsius}$ below the transition temperature~\cite{Oswald2012epl}, which we assume $45.1\si{\degreeCelsius}$ for pure MBBA~\cite{Kneppe1982JChemPhys}.}
\footnotetext[14]{The experiment was conducted in the coexistence state of the \NsLC~and \ILC~phases, whose temperature range is shown in Ref.~\cite{Bono2018}.}
\footnotetext[15]{The \NLC--\ILC~transition temperature of E7~\cite{Pestov2018}.}
    \end{table}
    \endgroup

    \begingroup
    \setlength{\tabcolsep}{0pt}
    \begin{table}[h]
        \caption{\label{tab:nu_MDsim}%
            The Leslie thermomechanical coefficient converted from the results of MD simulations~\cite{Sarman2016}. We postulated $\sigma_0 = 2\times10^{-10}\, \si{m}$ as the size of a molecule.
        }
        \begin{ruledtabular}
            \begin{tabular}{
            l	d	d	 @{\hspace{1em}}	d	d	D{,}{\times 10}{3}	d	d	D{,}{\times 10}{3}	d	d	D{,}{\times 10}{3}	}
System	& p/\sigma_0	& \multicolumn{1}{c}{ $c$, $x_{\text{d}}$	\footnotemark[1] }	& \multicolumn{3}{c}{	$ \tilde{\gamma}_1 \nu \cdot 3 k_{\text{B}} / ( p \sigma_0 ) $ }		& \multicolumn{3}{c}{	$ \tilde{\gamma}_1 \lambda_{zz} $ }		& \multicolumn{3}{c}{	$ |\nu | / \sqrt{ \tilde{\gamma}_1 \lambda_{zz}} $ }		\\
	&	&		& \multicolumn{3}{c}{	$ ( \si{kg.s^{-2}.K^{-1}} ) $ }		& \multicolumn{3}{c}{	}		& \multicolumn{3}{c}{	 }		\\
													\hline
I	& 25	& 0.31		& 1.8	& \pm 0.1	& ,^{-5}	& 1.31	& \pm 0.01	& ,^{3}	& 3.0	& \pm 0.2	& ,^{-4}	\\
disklike	& 37.5	& 0.2		& 7.3	& \pm 0.5	& ,^{-6}	& 1.34	& \pm 0.01	& ,^{3}	& 1.7	& \pm 0.1	& ,^{-4}	\\
	& 50	& 0.145		& 4.3	& \pm 0.8	& ,^{-6}	& 1.34	& \pm 0.01	& ,^{3}	& 1.4	& \pm 0.2	& ,^{-4}	\\
	& 75	& 0.11		& 1.9	& \pm 0.2	& ,^{-6}	& 1.33	& \pm 0.01	& ,^{3}	& 9.3	& \pm 1.1	& ,^{-5}	\\
													\hline
II	& 45	& 0.24		& 1.6	& \pm 0.1	& ,^{-6}	& 5.07	& \pm 0.02	& ,^{1}	& 9.7	& \pm 0.7	& ,^{-4}	\\
rodlike	& 90	& 0.13		& 4.2	& \pm 0.5	& ,^{-7}	& 5.04	& \pm 0.02	& ,^{1}	& 5.1	& \pm 0.6	& ,^{-4}	\\
	& 125	& 0.11		& 2.1	& \pm 0.3	& ,^{-7}	& 4.99	& \pm 0.03	& ,^{1}	& 3.5	& \pm 0.4	& ,^{-4}	\\
													\hline
III	& 45	& 0.684		& 1.7	& \pm 0.1	& ,^{-6}	& 5.01	& \pm 0.02	& ,^{1}	& 9.9	& \pm 0.6	& ,^{-4}	\\
rodlike	& 60	& 0.505		& 8.9	& \pm 0.5	& ,^{-7}	& 4.96	& \pm 0.02	& ,^{1}	& 7.2	& \pm 0.4	& ,^{-4}	\\
mixture	& 90	& 0.3607		& 4.8	& \pm 0.8	& ,^{-7}	& 4.93	& \pm 0.02	& ,^{1}	& 6.0	& \pm 1.0	& ,^{-4}	\\
	& 125	& 0.291		& 2.3	& \pm 0.5	& ,^{-7}	& 4.94	& \pm 0.01	& ,^{1}	& 4.0	& \pm 0.9	& ,^{-4}	\\
            \end{tabular}
        \end{ruledtabular}
        \footnotetext[1]{$c$: parameter characterizing chirality of interactions between two molecules for systems I and II. $x_{\text{d}}$: the number fraction of chiral molecules for for system III.}
    \end{table}
    \endgroup
    
    The distribution of $|Y|$ is shown in Fig.~\ref{fig:nuY_exp}. The estimated $|Y|$ under all known experimental conditions satisfied Eq.~\eqref{eq:bound_|Y|} $|Y| \le 1$. Thus, our thermodynamic bound is valid in that it can reproduce experimental results. In the \NsLC--\ILC~coexistence state, $\nu$ is often higher as $\sim 10^{-6}\, \si{kg.s^{-2}.K^{-1}}$ than in the \NsLC~range as $\sim 10^{-7}\, \si{kg.s^{-2}.K^{-1}}$. Oswald\etal proposed a new model, the \textit{Melting Growth model}, to explain the quantitative discrepancy of $\nu$~\cite{Oswald2019rev}. The higher $\nu$ is observed as the fast rotation of \NsLC~droplets dispersed in the \ILC~phase. In their model, the fast texture rotation is induced by the migration of molecules along the temperature gradient. The Leslie thermomechanical effect is dismissed as negligible in this regime. However, $\nu \sim 2\times10^{-6}\, \si{kg.s^{-2}.K^{-1}}$ was obtained also in the \NsLC~range~\cite{Eber1982}, which is the same order as in the coexistence state. Recent studies by Tabe\etal demonstrated that Leslie's description is still effective even in the coexistence state~\cite{Bono2020, Nishiyama2021SoftMatter}. To consider the ongoing debate, we temporarily assumed that the Leslie thermomechanical coefficient is dominant in the coexistence state. Consequently, $|Y|$ at the coexistence temperature is still consistent with $|Y| \le 1$. Even if we adopt the Melting Growth model and assume that the Leslie thermomechanical effect is renormalized to $\nu$ in the coexistence state, $|Y|$ is smaller than that computed under the assumption mentioned above. In any case, our thermodynamic argument is validated. 

    The experimental plots are aligned on the regression line. The regression equation $\log_{10}{|Y|} = 1.0\pm0.1\, \log_{10}{|\nu|} + 2.6\pm0.8$ indicates that $|Y|$ is almost proportional to $|\nu|$, and consequently the viscosity $\gamma_1$ and the thermal conductivity $\sigma_\perp$ are similar among the liquid crystal samples used in the experiments. The regression coefficients reproduced $\gamma_1 \sim 0.015\, \si{Pa.s},\ \sigma_\perp \sim 0.13\, \si{W.m^{-1}.K^{-1}},\ T \sim 330\, \si{K}$, which are typical for $n$CB. With these constants, the thermodynamic upper limit $|Y| = 1$ is reached when $|\nu| \sim 3.1 \times 10^{-3}\, \si{kg.s^{-2}.K^{-1}}$. The highest $|\nu|$ known so far is $\sim 10^{-6}\, \si{kg.s^{-2}.K^{-1}}$ (see Refs.~\cite{Eber1982, Oswald2008PRL, Bono2019}), leaving room to increase it by $10^3$ times.

    MD simulation also provided $\nu,\ \gamma_1,\ \sigma_\perp$~\cite{Sarman2016}, and we estimated $Y$. Our $\nu$ and $|Y|$ correspond to $\tilde{\gamma}_1 \nu \cdot 3 k_{\text{B}} / (p \sigma_0)$ and $|\nu|/\sqrt{\tilde{\gamma}_1 \lambda_{zz}}$ in the notation of Ref.~\cite{Sarman2016}, respectively. MD simulation considered three systems. In systems I and II, the discotic and calamitic phases were investigated, respectively. The molecular interaction is decomposed into the achiral Gay--Berne potential and the chiral Lennard--Jones potential. The chiral interaction is characterized by the chiral parameter $c$. System III is the calamitic phase of a mixture of achiral and chiral molecules. To calculate $Y$, we set the Boltzmann constant $k_{\text{B}} = 1.38\times10^{-23}\, \si{J.K^{-1}}$, assumed the molecular diameter $\sigma_0 = 2\times10^{-10}\, \si{m}$, referring to the typical width of rodlike liquid crystal molecules, and calculated the pitch length $p = (p/\sigma_0)\cdot \sigma_0$. These assumption corresponds to $300\, \si{K}$. The estimated $|Y|$ is shown in Table~\ref{tab:nu_MDsim} and Fig.~\ref{fig:nuY_exp}. MD simulation is also consistent with the thermodynamics, for both rodlike and disklike molecules. The discotic phase showed a relatively higher $|\nu|$ but $|Y|$ is lower than the calamitics and their mixtures. The low $|Y|$ is attributed to the high viscosity and the high heat conductivity. The discotic phase possessed $25$ times greater $\tilde{\gamma}_1 \lambda_{zz}$ than the calamitic phase (Table~\ref{tab:nu_MDsim}). The high viscosity is experimentally demonstrated in the discotic nematic (\NDLC) phase of 2,3,6,7,10,11-hexakis(4-heptyloxybenzoyloxy)triphenylene~(C$_7$OBzTp)~\cite{Mourey1982MolCrystLiqCryst} and 2,3,7,8,12,13-hexa(\textit{n}-tetradecanoyloxy)truxene~($(\text{C}_{13}\text{H}_{27}\text{COO})_6\text{-TX}$)~\cite{Negita2004pre}. The greater $|\nu|$ in disklike molecules would be attributed to its effective energy transfer. The viscosity and the heat conductivity represent diffusion of momentum and thermal energy, respectively. Since the energy of a molecule is effectively transferred to neighboring molecules, the discotic phase possess the greater $\tilde{\gamma}_1 \lambda_{zz}$. The effective intermolecular energy transfer would lead to high energy conversion via cross-couplings, namely the Leslie effects.

\subsection{Direct and inverse processes}
    When the direct process of the Lehmann effect is defined as the induction of rotation by the temperature gradient, its inverse process is the induction of heat current by rotation. Sato\etal have investigated the inverse Lehmann effect of \NsLC~droplets and compared its cross-coupling coefficient with that of the direct process~\cite{Sato2017JPSJ}. Sato\etal adopted a different choice of variables from ours in Table~\ref{tab:choice_of_variables} but their model is reduced from ours. With the angular velocity of the director rotation $\Omega$, torque exerted on the director $\tau$, they assumed the dissipation function of the form
    \begin{equation}
        R
        =
        \tau \Omega + \left(-T^{-1}\nabla T\right) q
        ,
    \end{equation}
    where $\nabla T$ and $q$ are the $z$-component of the temperature gradient $\vec{\nabla} T$ and the heat current $\vec{q}$, respectively.
    The choice of variables is arbitrary. Sato\etal selected $-T^{-1}\nabla T$ and $\tau$ as affinities, and $\Omega$ and $q$ as currents, leading to the constitutive equations
    \begin{subequations}
    \label{eq:Sato_Lehmann}
    \begin{align}
        \Omega
        &=
        \gamma_1^{-1} \tau + b_{\text{dir}} \left(-T^{-1}\nabla T\right)
        ,\\
        q
        &=
        b_{\text{inv}} \tau + \sigma_{\text{q}}T \left(-T^{-1}\nabla T\right)
        ,
    \end{align}
    \end{subequations}
    where $b_{\text{dir}}$ and $b_{\text{inv}}$ are the cross-coupling coefficients for the direct and inverse processes, respectively. \Reciproc imposes $b_{\text{dir}} = b_{\text{inv}}$.
    Tsori\etal proposed similar equations that describe the chemical Leslie effects~\cite{Tsori2004EPJE}, which was adopted by Tabe\etal to analyze the precession in Langmuir monolayers~\cite{Tabe2003}. Eq.~\eqref{eq:Sato_Lehmann} is reduced from Eqs.~\eqref{eq:constitutive2} by assuming that the director $\vec{n}$ is perpendicular to $\vec{\nabla} T$ and the hydrodynamic flow is absent [see Eq.~\eqref{eq:notation_sato} for the correspondence of the notation]. The Leslie thermohydrodynamic effect, characterized by $\tilde{\mu}'^{\text{N}}$, is omitted. Eqs.~\eqref{eq:bound_alpha_3-alpha_2} and \eqref{eq:bound_|Y|} lead to $\gamma_1\ge0$. Note that $b_{\text{dir}} = b_{\text{inv}}$ can take an arbitrary value, which is consistent with thermodynamic bounds. Sato\etal examined \Reciproc by measuring $b_{\text{dir}}$ and $b_{\text{inv}}$ independently, and reported $b_{\text{dir}} \sim 10^{-3}\,\si{m.s^{-1}}$ and $b_{\text{inv}} \sim 10\,\si{m.s^{-1}}$. Although such a discrepancy between $b_{\text{dir}}$ and $b_{\text{inv}}$ is not consistent with \Reciproc, each value is not forbidden by the thermodynamic bounds. From another point of view, an additional thermodynamic bounds will be considered, which corresponds to the positive semidefinitness of the transport matrix in Eq.~\eqref{eq:Sato_Lehmann}. The bound on the cross-coupling coefficients are given by
    \begin{equation}
        |b_{\text{dir}}|,\ |b_{\text{inv}}|
        \le
        \sqrt{\gamma_1^{-1}\sigma_{\text{q}}T}
        \sim 14\,\si{m.s^{-1}}
        ,
    \end{equation}
    with $\sigma_{\text{q}}\sim 2\times 10^{-2}\,\si{W.K^{-1}.m^{-1}},\ T\sim 3\times 10^2\,\si{K}$, as Sato\etal assumed~\cite{Sato2017JPSJ}. We adopted $\gamma_1\sim 0.03\,\si{Pa.s}$ for $n$CB around the \NLC--\ILC~transition temperature~\cite{Luis1983MolCrystLiqCryst} but the viscosity of the material, RDP-V0639 (DIC), is lower according to a data sheet owned by the authors. The heat conductivity of liquid crystals is typically greater than that Sato\etal assumed. So, the thermodynamic bounds are gentler, and the inverse coefficient $b_{\text{inv}} \sim 10\,\si{m.s^{-1}}$ is still consistent with it.
    Sato\etal concluded that the $10^4$-times discrepancy between $b_{\text{dir}}$ and $b_{\text{inv}}$ was due to hidden artifacts such as hydrodynamic convection and ion transport. 
 
\subsection{Mutual signs of Leslie thermohydrodynamic and thermomechanical coefficients}
    In the Leslie effects, the cross-coupling coefficients $\mu^{\text{A}}, \mu^{\text{N}}, \mu'^{\text{N}}$ are true-chiral, becoming nonzero only in the chiral phases. Their signs are determined by the chirality of the liquid crystal molecules, and thus their signs are opposite in enantiomeric materials. We now focus on the mutual sign relationships among these coefficients within a single material, rather than comparing opposite enantiomers. We demonstrate that the signs need not be identical, and that their relative signs are determined by the shape of molecules.

\begin{figure}
        \includegraphics[width=17.5cm, keepaspectratio]{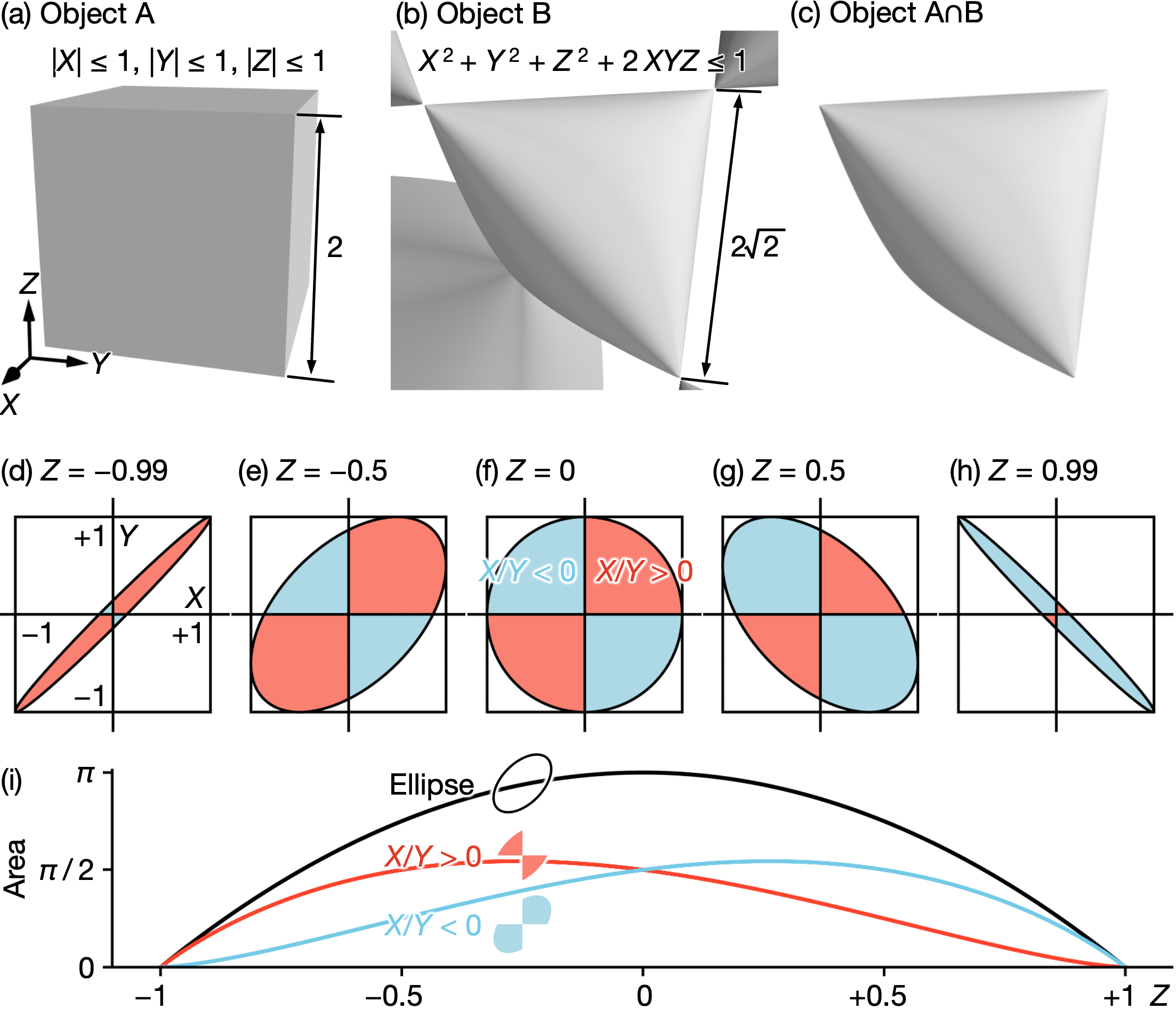}
        \caption{
            \label{fig:ellipses}
            Coefficients $X,Y,Z$ that are consistent with thermodynamics. 
            (a)--(c) Objects locating in $XYZ$-space whose interior is consistent with thermodynamics.
            (a) Object $A$ showing Eqs.~\eqref{eq:bound_|X|}--\eqref{eq:bound_|Z|}.
            (b) Object $B$ showing Eq.~\eqref{eq:bound_X2+Y2+Z2-2XYZ}.
            (c) The intersection $A \cap B$ showing Eqs.~\eqref{eq:bound_|X|}--\eqref{eq:bound_X2+Y2+Z2-2XYZ}, all inequalities to be concerned. Objects $A,\ B$ and $A \cap B$ are centered at the origin ($X=Y=Z=0$). 
            (d)--(h) The cross sections of $A \cap B$ at a fixed $Z$. Probable $X,Y$ is included in the ellipse. In the ellipse, the regions where $X/Y > 0$ and $X/Y < 0$ are highlighted in orange and blue, respectively. On the perimeter of the ellipse, the Carnot efficiency is attained. The black square circumscribing the ellipse represents the perimeter of $|X| \le 1 \land |Y| \le 1$.
            (i) Area of the ellipse and its parts as a function of $Z$. The black line is the area of entire ellipse, the orange and blue lines are areas of $X/Y > 0$ and $X/Y < 0$ parts, respectively.
        }
    \end{figure}

    Typical calamitic liquid crystals, including 5CB and \textit{N}-(4-methoxybenzylidene)-4-butylaniline~(MBBA), show $\alpha_2 + \alpha_3 < 0$, which is equivalent to $Z<0$~(Table~\ref{tab:Z_is_negative}). Thermodynamic bounds Eqs.~\eqref{eq:bound_X2+Y2+Z2-2XYZ}--\eqref{eq:bound_|Z|} suggest that $X$ and $Y$ are likely to have opposite signs when $Z < 0$. Here we identify a thermodynamically possible and preferred range of $X$ and $Y$. 
    Eqs.~\eqref{eq:bound_|X|}--\eqref{eq:bound_|Z|} constitute a cube with edge length $2$ in $XYZ$-space~[Fig.~\ref{fig:ellipses}(a)], and the cube is denoted by $A$. Let $B$ be the object defined in Eq.~\eqref{eq:bound_X2+Y2+Z2-2XYZ}, a rounded regular tetrahedron together with four cones extending from its vertices. Object $B$ is invariant under the action of the tetrahedral group $T_{\text{d}}$.  Eqs.~\eqref{eq:bound_X2+Y2+Z2-2XYZ}--\eqref{eq:bound_|Z|} represent the intersection of the objects $A$ and $B$, and it is a rounded tetrahedron with edge length $2\sqrt{2}$. With $Z$ intrinsic to the material, we fix its value and explore the resulting $XY$-region. The cross-section of the rounded tetrahedron $A \cap B$ at a fixed $Z$ is ellipse [Fig.~\ref{fig:ellipses}(d)--\ref{fig:ellipses}(h)], whose interior determines the thermodynamically possible range of $X$ and $Y$ for that material. The perimeter of the ellipse corresponds to the Carnot cycle. Since the Leslie effects occur such that the dissipation rate is minimized~\cite{Sarman2025PCCP}, one expects that a point $(X,Y)$ is likely to approach the perimeter to reduce dissipation. When $Z < 0$, higher efficiency is attained in the region of $X/Y<0$ for fixed $|X|$ and $|Y|$, namely $X$ and $Y$ prefer to be opposite in sign. This preference is due to the hydrodynamic coupling, characterized by $Z$, of the irrotational and rotational flows induced by the Leslie thermohydrodynamic and thermomechanical effects, characterized by $X$ and $Y$, respectively. The hydrodynamic coupling of the Leslie effects improves efficiency when $XYZ > 0$, in which the thermohydrodynamic and thermomechanical effects cooperatively suppress dissipation (see Appendix~\ref{sec:derivation_bounds} for details). Given that the liquid crystals investigated so far have small $|X|, |Y| \lesssim 10^{-3}$ (Fig.~\ref{fig:nuY_exp}), the proximity of the perimeter to the origin determines the preferred $X/Y$. A region whose perimeter is close to the origin has a small area. The total area of the ellipse $\pi (1-Z^2)$ is occupied by the area of $X/Y<0$ by
    \begin{equation}
        1 - \frac{\cos^{-1}{(-Z)}}{\pi}
    \end{equation}
    times. The area is useful for representing the proximity of the perimeter of interest to the origin. For $Z < 0$, $X$ and $Y$ are localized in the small but thermodynamically preferred region of $ X/Y <0 $.

    \begin{table}[h]
        \caption{\label{tab:Z_is_negative}%
            $Z$ of calamitic liquid crystals calculated from the Leslie viscosity coefficients presented in Ref.~\cite{KlemanLavrentovich2003}.
        }
        \begin{ruledtabular}
            \begin{tabular}{cdddddd}
                &\alpha_2    &\alpha_3   &\alpha_4   &\alpha_5  &\alpha_6   &Z
                \\
                &\multicolumn{5}{c}{$(\si{m.Pa.s})$}   &\text{--}\footnotemark[1]
                \\\hline
                5CB &-83    &-2 &75 &102    &-27    &-0.63
                \\
                MBBA    &-109   &-1 &83 &80 &-34    &-0.73
                \\
                PBG\footnotemark[2] &-6920  &18 &348    &6610   &-292   &-0.99
                \\
            \end{tabular}
        \end{ruledtabular}
        \footnotetext[1]{$Z$ is dimensionless (unitless).}
        \footnotetext[2]{Lyotropic liquid crystal of poly-\ensuremath{\gamma}-benzyl-glutamate in a solvent mixture of methylene chloride and dioxane}
    \end{table}

    Bunel\etal observed $X/Y < 0$ in the chemical Leslie effect and predicted that $X/Y < 0 $ would hold also in the thermal Leslie effect~\cite{Bunel2023singular,Bunel2023nonsingular}. We will first clarify the correspondence between the chemical and thermal Leslie effects. Self-standing films of the \SmCsLC~phase were formed, and the difference in ethanol vapor pressure was applied across the film to drive the transmembrane permeation of the ethanol molecules~\cite{Bunel2023nonsingular,Bunel2023singular}. This vapor permeation induced rotational mass flow within the films via the chemical Leslie effect. The stress tensor in our notation is
    \begin{eqnarray}
        \label{eq:chemLeslie}
        \sigma'_{ij}
        &=&
        \alpha_1 c_i c_j c_m c_n A_{mn}
        + \alpha_2 c_i \corot{c}_j
        + \alpha_3 c_j \corot{c}_i
        + \alpha_4 A_{ij}
        + \alpha_5 c_i c_m A_{mj}
        + \alpha_6 c_j c_m A_{mi}
        \nonumber\\
        &&%\hspace{1em}
        +
        \left[
            \frac{\mu}{2}(\epsilon_{ikm} c_k c_j + \epsilon_{jkm} c_k c_i)
            - \frac{\nu}{2}(\epsilon_{ikm} c_k c_j - \epsilon_{jkm} c_k c_i)
            + \frac{\nu'}{2}\epsilon_{ijm}
        \right]
        \Delta P\ k_m
        ,
    \end{eqnarray}
    where $\vec{c}$ denotes the $c$-director, $\corot{c}_i = \dot{c}_i - c_j (\nabla_j v_i - \nabla_i v_j)/2$ the corotational time derivative of $\vec{c}$ (denoted by $C_i$ in the original Refs.~\cite{Bunel2023nonsingular, Bunel2023singular}), and $\vec{k}$ the unit vector normal to the smectic layers. The notation of Bunel\etal is provided by exchanging stress tensor's indices $i\leftrightarrow j$. The difference in ethanol vapor pressure $-\Delta P\ \vec{k}$ in the chemical Leslie effect in Eq.~\eqref{eq:chemLeslie} corresponds to the temperature gradient $-T^{-1} \vec{\nabla} T$ in the thermal Leslie effect in Eq.~\eqref{eq:constitutive2_sigma}. Consequently, the coefficients $\mu,\, \nu,\, \nu'$ in the chemical Leslie effect correspond to $2\tilde{\mu}^{\text{A}},\, 2\tilde{\mu}^{\text{N}},\, 2\tilde{\mu}'^{\text{N}}$ in the thermal Leslie effect, respectively. The contributions of $\nu$ and $\nu'$ are indistinguishable as long as we focus on the torque: 
    \begin{subequations}
    \begin{align}
        - \epsilon_{ijk}\sigma'_{jk}
        &=
        \epsilon_{ijk}
        \left[
            (\alpha_3 - \alpha_2)  c_j \corot{c}_k
            + (\alpha_6 - \alpha_5) c_j c_m A_{mk}
        \right]
        + \left( \nu - \nu' \right) \Delta P\ k_i
        .
    \end{align}
    \end{subequations}
    Alternatively, one may regard $\nu'$ as renormalized to $\nu$. 
    
    Bunel\etal reported that $\mu$ and $\nu$ have opposite signs in FELIX~M4851-100 (Merck), the smectic material they investigated. They adopted the Stannarius approximation~\cite{Harth2011softmatter} on the Leslie viscosity coefficients:
    \begin{subequations}
    \begin{align}
        \alpha_4 > 0,&\\
        \alpha_2 = -\alpha_5 = -1,&\\
        \alpha_1 = \alpha_3 = \alpha_6 = 0,&
    \end{align}
    \end{subequations}
    which satisfies $Z<0$. Then, $X/Y<0$ leads to $\mu^{\text{A}}/\left(\mu^{\text{N}} - \mu'^{\text{N}}\right) < 0$, and $\tilde{\mu}^{\text{A}} / \left( \tilde{\mu}^{\text{N}} - \tilde{\mu}'^{\text{N}}\right) < 0$. Since $\nu - \nu'$ corresponds to $+2\tilde{\mu}^{\text{N}} - 2 \tilde{\mu}'^{\text{N}}$, one obtains $\mu/(\nu - \nu' ) < 0$. We can consider that $\nu'$ is renormalized to $\nu$. Bunel\etal mentioned that $\mu / \nu = -2.5$ reproduced the experimental results well, under their assumption of $\alpha_4=1$. This condition corresponds to $X/Y = -2.5/\sqrt{2} < 0$ and $Z=-1/\sqrt{3} < 0$ in our notation. From $Z <0$, we predict that $X/Y < 0 $ is thermodynamically preferred, and this prediction is evidenced by the dashed line of $X/Y = -2.5/\sqrt{2}$ passing through the smaller $X/Y<0$ region. The dashed line intersects the perimeter of the ellipse at two points. At these intersects, FELIX~M4851-100 attains the Carnot efficiency in the chemical Leslie effect.

    \begin{figure}
        \includegraphics[width=8.5cm, keepaspectratio]{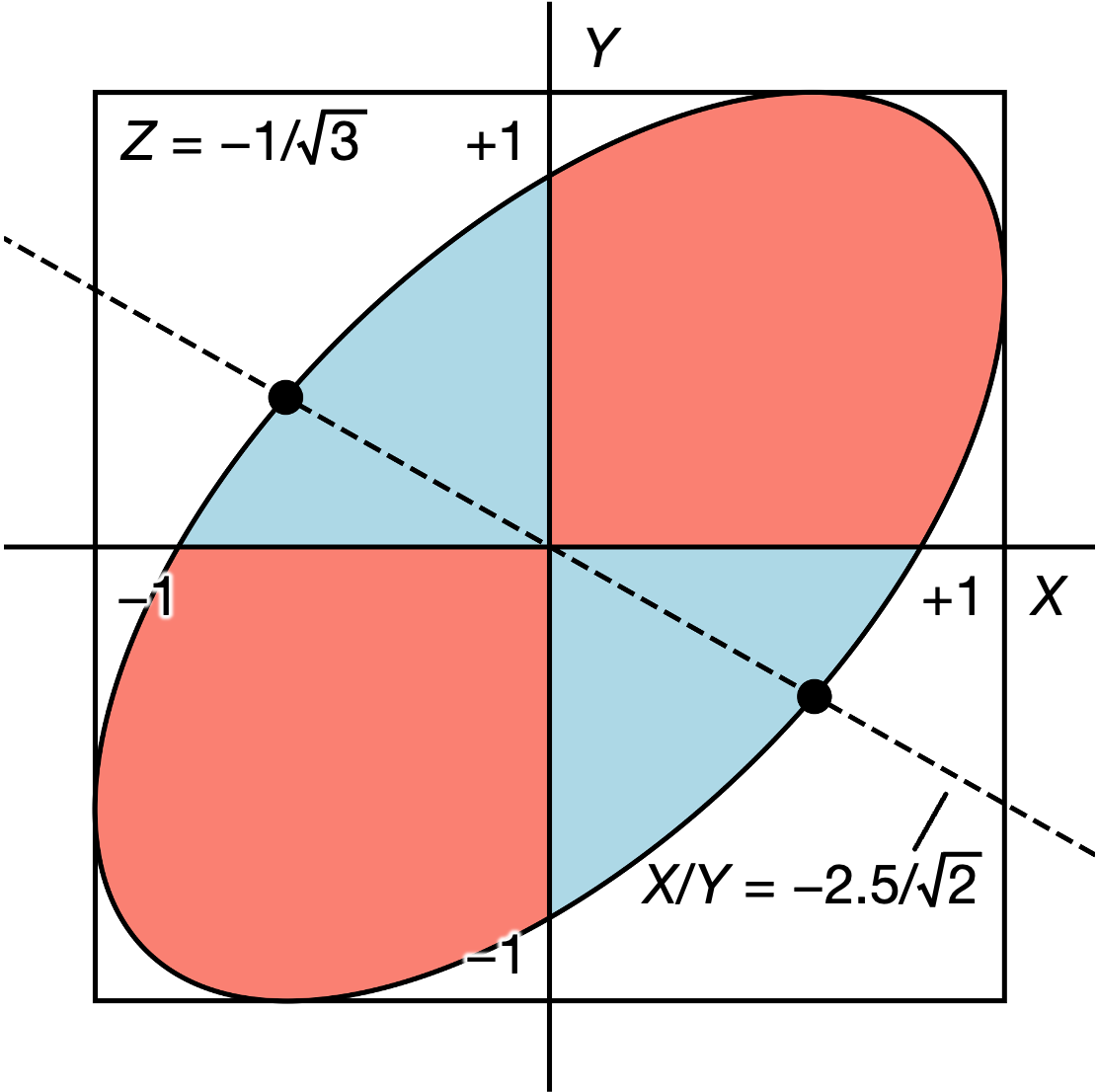}
        \caption{
            \label{fig:ellipse_oswald}
            Thermodynamic bounds for the chemical Leslie effect in \SmCsLC~phase investigated in Refs.~\cite{Bunel2023nonsingular,Bunel2023singular}. The oblique ellipse shows thermodynamically acceptable Leslie chemohydrodynamic and chemomechanical cross-coupling coefficients. The perimeter of the ellipse intersect the dashed line that represents $X/Y = -2.5/\sqrt{2}$ at two points.
        }
    \end{figure}

\begin{figure}
        \includegraphics[width=8.5cm, keepaspectratio]{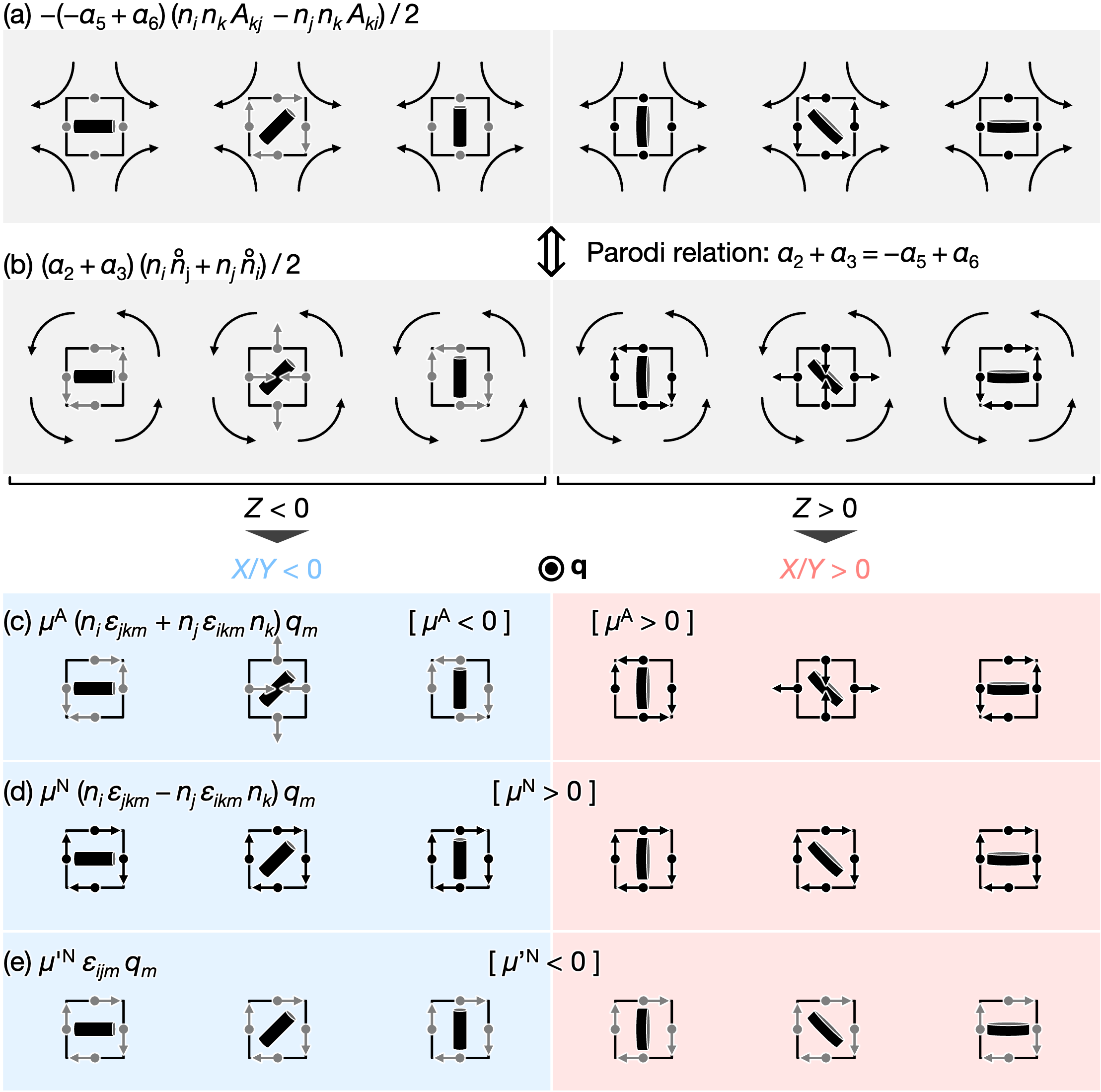}
        \caption{
            \label{fig:corrdevrot} The irreversible stress for rodlike and disklike molecules. 
            (a) The rotational stress induced by an irrotational flow, depending on the shapes and orientations of molecules. The arrows extending from the square represent the forces acting on the face at each arrow’s base. Arrows are black when the coefficient is positive and gray when negative. The streamlines and the director lie in the plane. Each of the three panels on the left and right displays rod‑shaped and disc‐shaped molecules, respectively.
            (b) The deviatoric stress induced by a rotational flow. 
            (c) The deviatoric stress induced by the Leslie effects. The coefficient $\mu^{\text{A}}$ considers a case where it is negative $(X < 0)$ in the rodlike molecule and positive $(X > 0)$ in the disklike molecule as an example. The heat current $\vec{q}$ is directed out of the page.
            Under these conditions, the thermodynamically favored rotational stress (for $Y > 0$) is shown in (d) and (e).
        }
    \end{figure}

    The dominant sign of $X/Y$ coincides with that of $Z$ [Fig.~\ref{fig:ellipses}(d)--\ref{fig:ellipses}(i)], and this correspondence is attributed to the molecular shape. The calamitic phases composed of rodlike molecules typically exhibit $Z < 0$, which favors $X/Y<0$. Here, we discuss how the molecular shape governs the preferred sign of $X/Y$. Stiff polymers are representative examples of rodlike molecules. For the lyotropic \NLC~phase formed by a polymer solution, the Leslie viscosity coefficients have been theoretically investigated:
    \begin{equation}
        \alpha_2 + \alpha_3 = -2S \frac{n_{\text{p}} k_{\text{B}}T}{2\bar{D}_{\text{r}}}
        < 0
        ,
    \end{equation}
    where $n_{\text{p}}$ represents the number of polymers per unit volume, and $\bar{D}_{\text{r}}$ the effective rotational diffusion constant of the polymer~\cite{Doi1983faraday}. For longer polymers, rotation diffusion is suppressed (smaller $\bar{D}_{\text{r}}$), and $\alpha_2 + \alpha_3$ becomes further negative. In the idealized case of infinitely thin and straight rodlike molecules, $(\alpha_2 + \alpha_3) / (\alpha_2 - \alpha_3 )$ is close to $+1$~\cite{Volovik1980jetplett}. Given that a thermodynamic bound $\alpha_3 - \alpha_2 \ge 0$ in Eq.~\eqref{eq:bound_alpha_3-alpha_2}, this implies $\alpha_2 + \alpha_3 < 0$. Thus, the negativity of $Z$ arises from the elongated shape of the liquid crystal molecules. Figure~\ref{fig:corrdevrot}(a) shows the deviatoric stress induced by irrotational flow, which arises specifically in liquid crystals and is absent in isotropic liquids. %The stress depends on the orientation of the molecule and becomes finite when the molecular long axis is misaligned with the principal axis of the irrotational flow. 
    As the flow effectively ``rows'' against the two tips of the rodlike molecule, a clockwise torque is induced, which leads to $- \alpha_5 + \alpha_6 < 0$. Together with the Parodi relation Eq.~\eqref{eq:parodi}, which originates from \Reciproc based on time reversal symmetry, this determines the deviatoric stress induced by rotational flow in Fig.~\ref{fig:corrdevrot}(b). This stress is also specific to liquid crystals, and the induced deviatoric stress has principal axes tilted by $45^\circ$ relative to the molecular long axis.

    In contrast, the discotic phases formed by the disklike molecules show $Z > 0$, leading to $X/Y>0$. The \NDLC~phase has $\alpha_2, \alpha_3 > 0$ as shown by theory~\cite{Volovik1980jetplett}, molecular simulation~\cite{Grecov2003molcrystliqcryst} and experiments~\cite{Negita2004pre}. In the \NDLC~phase, disklike oblate molecules are distributed without long-range positional order. Under an irrotational flow in Fig.~\ref{fig:corrdevrot}(a), the rim of the molecule is rowed to rotate in the anticlockwise direction, which is opposite to the rodlike molecules. Thus, disklike molecules tend to orient with their disk plane parallel to the flow gradient~\cite{Carlsson1982molcrystliqcryst}. The \NDLC~phase of $(\text{C}_{13}\text{H}_{27}\text{COO})_6\text{-TX}$ shows a large positive viscosity estimated as $\alpha_2 = 0.38\, \si{Pa.s}$ at $347.1\si{K}$~\cite{Negita2004pre}, which is opposite in sign and five times greater than that of 5CB. Thus, $\alpha_2>0$ is achieved in the \NDLC~phase and thus $Z>0$ is satisfied, which leads to that $X/Y>0$ is preferred in the chiral \NDLC~phases. %Since the correlation of deviatoric and rotational stress in \NDLC~phase is opposite to that in calamitic \NLC~phase, $\mu^{\text{A}} > 0$ is compatible with $\mu^{\text{N}} < 0,\ \mu'^{\text{N}} > 0$. 
    Experimental verification of the Leslie effects in discotic liquid crystals remains an open challenge, as no previous studies have addressed this phenomenon to the best knowledge of the authors.
    
    Considering the results of MD simulations~\cite{Sarman2016}, $X\ (\propto \mu^{\text{A}})$ changes sign depending on the molecular shape, whereas $Y\ (\propto \mu^{\text{N}} \text{ and } \mu'^{\text{N}})$ remains unaffected by it. Sarman\etal simulated the Leslie thermomechanical coefficient ($\propto Y$) for rodlike and disklike molecules, and showed that its sign is same (Table~\ref{tab:nu_MDsim}) under the fixed sign of the chiral parameter $c$. That suggests that $X$ would change its sign by reflecting the aspect ratio of the molecule. Figures~\ref{fig:corrdevrot}(c)--(e) compare the case of $Y>0$ both for rodlike and disklike molecules. Such a situation coincides with the results obtained by Sarman\etal that examined the effect of the aspect ratio~\cite{Sarman2016}. %Thermodynamics as well as correlation prefers $X<0$ and $X>0$ for rodlike and disklike molecules, respectively. 
    If the shape of molecules could be transformed from one to the other while keeping the handedness of the material fixed, the sign of $X$ would be reversed, whereas the sign of $Y$ would remain unchanged. %signs of $X$ are nega ( and  Molecular shape would influence on thermohydrodynamic effect but not thermomechanical effect. It is believed that prolate or oblate do not affect direction of torque in rotational flow but reverses sign of torque in irrotational flow or deviatoric stress in rotational flow. Thus, molecular shape influences on thermohydrodynamic effect. %Sarman\etal also reported larger viscosity $\alpha_3 - \alpha_2$ for oblate molecules compared with prolate molecules from simulation~\cite{Sarman2016}. 

    Chromonic liquid crystals are another interesting material for investigating the Leslie effects. The \NLC~phase of disodium cromoglycate (DSCG) in an aqueous solution exhibits anomalous viscosity behaviors. The viscosity coefficients of DSCG follow the order~\cite{Zhou2014softmatter}:
    \begin{equation}
    \label{eq:etasplay_etatwist_etabend}
        \eta_{\text{splay}} > \eta_{\text{twist}} > \eta_{\text{bend}} > 0
        ,
    \end{equation}
    namely $Z > 0$ is satisfied since
    \begin{equation}
        \alpha_2 + \alpha_3
        = 
        \frac{\left(\eta_{\text{bend}} - \eta_{\text{splay}}\right){\alpha_3}^2}{\left(\eta_{\text{twist}} - \eta_{\text{splay}}\right) \eta_{\text{bend}}}
    \end{equation}
    holds~(see also Ref.~\cite{de_Gennes}). Therefore, DSCG possibly shows $X/Y > 0$ alike the \NDLC~phase. %DSCG molecules form columnar aggregates with average length $\bar{L}$, and these viscosity coefficients are proportional to ${\bar{L}}^2$, which decrease with a rise of temperature due to molecular dissociation. 
    However, the inequality in Eq.~\eqref{eq:etasplay_etatwist_etabend} would shift to
    \begin{equation}
        \eta_{\text{twist}} >\eta_{\text{splay}} > \eta_{\text{bend}} > 0
    \end{equation}
    in the vicinity of the \NLC--\ILC~transition temperature~\cite{Zhou2014softmatter}, and $Z < 0$ is likely to be obtained. Therefore, the sign-reversal of $X$ is expected with increasing temperature. 
    In contrast, the \NsLC~phase formed by oligo-DNA will show $X/Y < 0$ since the racemic mixture of \textsc{d}- and \textsc{l}-DNA possesses $\eta_{\text{bend}} > \eta_{\text{splay}} > \eta_{\text{twist}} > 0$~\cite{Lucchetti2020acsmacrolett}, which corresponds to $Z<0$. The oligo-DNA molecule must be considered as a stiff rod that is consistent with the microscopic Doi theory~\cite{Doi1983faraday}.
    The preceding discussion of lyotropic liquid crystals is valuable for designing micromotors composed of biocompatible molecules, suitable for medical applications, and driven by the Leslie and AZ effects.

%==========================================================================
\section{Torque induced by Leslie effects}\label{sec:torque}
\subsection{Force along director}
    We confirmed that cross-coupling is prohibited in the \ILC~phase (see Sec.~\ref{sec:isotropic_limit}). However, the Leslie effects are possible under affinity along the isotropic axis of the \NsLC~phase. We assumed that both the heat current $\vec{q}$ and the director $\vec{n}$ are parallel to the $z$-axis:
    \begin{equation}
        \vec{q} = q_z \hat{\vec{z}}
        ,\quad
        \vec{n} = \hat{\vec{z}}
        .
    \end{equation}
    The stress arising from the thermal Leslie effect $\tsr{\sigma}^{\text{L}}$ is given by
    \begin{subequations}
    \begin{align}
        \sigma^{\text{L}}_{ij}
        &=
        \left[
            \mu^{\text{A}} \left( n_i \epsilon_{jkm} n_k + n_j \epsilon_{ikm} n_k \right)
            + \mu^{\text{N}} \left( n_i \epsilon_{jkm} n_k - n_j \epsilon_{ikm} n_k \right)
            + \mu'^{\text{N}} \epsilon_{ijm}
        \right]
        q_m
        \nonumber\\
        &=
        \mu'^{\text{N}} q_z \epsilon_{ijm} \hat{z}_m\\
        \sigma^{\text{L}}_{xy}
        &= - \sigma^{\text{L}}_{yx}
        =
        \mu'^{\text{N}} q_z 
        ,\ \text{otherwise}\ 0
        .
    \end{align}
    \end{subequations}
    Only the antisymmetric part proportional to $\mu'^{\text{N}}$ remains nonzero. The thermohydrodynamic (the $\mu^{\text{A}}$-term) and thermomechanical (the $\mu^{\text{N}}$-term) effects derived from the conventional framework~\cite{de_Gennes} do not appear, whereas the new $\mu'^{\text{N}}$-term predicted by the Q-tensor formulation may emerge. Since this new term has not been discussed previously, validation by experiments and MD simulations is awaited. Nevertheless, the $\mu'^{\text{N}}$-term is also derived systematically from \OnsagerVariational (see Appendix~\ref{sec:variational}). Since the arrangement of $\vec{q}$ and the induced rotational flow above is true-chiral, in Barron's definition~\cite{Barron2012chirality}, the presence of the $\mu'^{\text{N}}$-term is natural in terms of symmetry.

    Thermodynamic bounds prohibit the existence of the $\mu'^{\text{N}}$-term in the \ILC~phase; thus, the presence of anisotropy is a necessary condition for its manifestation. Accordingly, in a chiral liquid crystal phase exhibiting uniaxial anisotropy, $\vec{q}$ can induce a torque. A recent experiment has demonstrated that a heat current along the $c$-axis in \ensuremath{\alpha}-quartz induces a polarization of the angular momentum~\cite{Ohe2024prl}. \ensuremath{\alpha}-Quartz is a chiral condensed phase that crystallizes in noncentrosymmetric $P3_121$ or $P3_221$ and exhibits uniaxial anisotropy. Our identification of the $\mu'^{\text{N}}$-term  suggests that thermally induced chiral cross-couplings are a universal feature of condensed matter systems beyond liquid crystals.

\subsection{Sign reversal of torque}
    The torque originates from the antisymmetric part of the stress tensor, as shown in Eq.~\eqref{eq:Gamma_def_torque}. For arbitrary $\vec{n}$, the torque due to the thermal Leslie effect is
    \begin{subequations}
    \begin{align}
        \Gamma_i^{\text{N}}
        &=
        - \epsilon_{ijk} \, \mu^{\text{N}} \left( n_j \epsilon_{klm} n_l - n_k \epsilon_{jlm} n_l \right) q_m
        %= - 2 \mu^{\text{N}} \epsilon_{ijk} \epsilon_{lmk} n_j n_l  q_m
        %= - 2 \mu^{\text{N}} (\delta_{il} \delta_{jm} - \delta_{im} \delta_{jl}) n_j n_l  q_m
        = 2 \mu^{\text{N}} ( \delta_{im} - n_i n_m ) q_m
        ,\\
        \Gamma_i'^{\text{N}}
        &=
        - \epsilon_{ijk} \, \mu'^{\text{N}} \epsilon_{jkm} q_m
        = - 2 \mu'^{\text{N}} q_i
        .
    \end{align}
    \end{subequations}
    The torques $\vec{\Gamma}^{\text{N}}$ and $\vec{\Gamma}'^{\text{N}}$ are attributed to the Leslie thermomechanical effect (the $\mu^{\text{N}}$-term) and the $\mu'^{\text{N}}$-term, respectively. The Leslie thermohydrodynamic effect (the $\mu^{\text{A}}$-term) drives a deviatoric stress, and thus does not induce torque. Assume that $\mu^{\text{N}}, \mu'^{\text{N}}, \vec{q}$ are fixed for simplicity without loss of generality. The magnitude of $\vec{\Gamma}^{\text{N}}$ depends on $\vec{n}$ but its direction is not reversed. With the identity matrix $\tsr{I} \coloneqq \mathrm{diag} (1,1,1)$, the projection $\tsr{I} - \vec{n} \vec{n}$ extracts the component of $\vec{q}$ perpendicular to $\vec{n}$. Since $(\tsr{I} - \vec{n} \vec{n}) \vec{q}$ is never the opposite to $\vec{q}$, the direction of $\vec{\Gamma}^{\text{N}}$ is confined to one of the hemispheres whose pole is $\vec{q}$. For $\vec{\Gamma}'^{\text{N}}$, its is independent of $\vec{n}$ and both magnitude and direction are constant. Thus, the change in $\vec{n}$ never causes sign-reversal of $\vec{\Gamma}^{\text{N}}$ and $\vec{\Gamma}'^{\text{N}}$.

    Variations of $\vec{n}$ would cause sign-reversal of their sum $\vec{\Gamma}^{\text{N}}+\vec{\Gamma}'^{\text{N}}$. We focus on the component parallel to $\vec{q}$ and choose a coordinate as $\vec{q} = q_z \hat{\vec{z}}$:
    \begin{equation}
        \Gamma_z^{\text{N}} + \Gamma_z'^{\text{N}}
        =
        2 \mu^{\text{N}} \left[ 1 - {n_z}^2 - \frac{\mu'^{\text{N}}}{\mu^{\text{N}}} \right] q_z
        .
    \end{equation}
    The sign-reversal with $n_z$ appears if
    \begin{equation}
        0 < \mu'^{\text{N}} / \mu^{\text{N}} < 1
        .
    \end{equation}
    $\mu'^{\text{N}}$ must have the same sign as $\mu'^{\text{N}}$ and its magnitude must be smaller than that of $\mu^{\text{N}}$. This condition is likely to be satisfied according to the $S$-dependence in Eq.~\eqref{eq:LAQ_LNQ_Sn} as 
    \begin{subequations}
    \begin{align}
        \mu^{\text{N}} 
        &=
        \frac{3}{2} \left( \mu^{\text{N}}_{11} - \mu^{\text{N}}_{12} \right) S + O(S^2)
        ,\\
        \mu'^{\text{N}} 
        &= 
        \left( \mu^{\text{N}}_{11} - \mu^{\text{N}}_{12} \right) S + O(S^2)
        .
    \end{align}
    \end{subequations}
    Up to the order of $S^1$, $\mu^{\text{N}}$ is the same in sign as $\mu'^{\text{N}}$ and $\left|\mu'^{\text{N}}\right| < \left| \mu^{\text{N}}\right|$ holds.
    In chiral liquid crystals with the same composition, if a reversal of the rotation direction is observed in systems with different director fields, one possible mechanism for this reversal is the competition between the $\mu^{\text{N}}$- and $\mu'^{\text{N}}$-terms. This also indicates the presence of the $\mu'^{\text{N}}$-term. In particular, a system with uniform director field is free of the AZ effects, and the $\mu'^{\text{N}}$-term in the Leslie effects is responsible for the reversal. The actual director fields are usually wound, and the effect of the twisted director structure is shown in Appendix~\ref{sec:some_director_fields}.
    %\mu^{\text{N}}$-term and $\mu'^{\text{N}}$-term must be comparable so that the competition between them leads to the reversal. 
    %In other cases, $\mu'^{\text{N}}$ dominates the sign of $\Gamma_z^{\text{N}} + \Gamma_z'^{\text{N}}$ and no reversal emerges. 

    Variations of $S$ would also cause sign-reversal of each $\vec{\Gamma}^{\text{N}}$ and $\vec{\Gamma}'^{\text{N}}$. Both $\mu^{\text{N}}$ and $\mu'^{\text{N}}$ are polynomials in $S$. The $S^1$-term slowly varies than the $S^2$-term, and the net sign of the polynomial may change with increasing $S$. Consequently, $\vec{\Gamma}^{\text{N}}$ and $\vec{\Gamma}'^{\text{N}}$ would change in sign $S$-dependently. The sign-reversal of $\mu^{\text{N}}$ and $\mu'^{\text{N}}$ is consistent with the thermodynamic bounds in Eqs.~\eqref{eq:bound_|Y|} and \eqref{eq:X2-2XYZ+Y2=1-Z2}.

%===Conclusion=================================================
\section{Conclusion}\label{sec:conclusion}
    We proposed a modified model for describing the Leslie effects by extending the Ericksen--Leslie model with the Q‑tensor that fully represents the nematic order. Our model reduces to the classical Leslie effects~\cite{Leslie1968II, de_Gennes} in the uniaxial chiral nematic phase and reveals that the Leslie cross-coupling coefficients explicitly depend on the scalar order parameter. By applying the thermodynamic uncertainty relation, we derived thermodynamic bounds that include upper limitations and mutual sign relationships among the Leslie cross-coupling coefficients, in accordance with the principle of entropy increase. The mutual signs of the thermohydrodynamic coefficient $\mu^{\text{A}}$ and the thermomechanical coefficient $ \mu^{\text{N}} - \mu'^{\text{N}} $ tend to coincide with the sign of $\alpha_2 + \alpha_3$. In particular, rodlike molecules ($\alpha_2 + \alpha_3 < 0$) would possess $\mu^{\text{A}} / \left( \mu^{\text{N}} - \mu'^{\text{N}} \right) < 0$, which is consistent with recent experimental observations of the chemical Leslie effect~\cite{Bunel2023singular, Bunel2023nonsingular}. Conversely, we predicted that $\mu^{\text{A}} / \left( \mu^{\text{N}} - \mu'^{\text{N}} \right) >0 $ would be observed in disklike molecules ($\alpha_2 + \alpha_3 > 0$). Since the temperature gradient and the stress are true-chiral even in the absence of orientational order, a Leslie cross-coupling torque may be induced by a heat current parallel to the director. This new effect (the $\mu'^{\text{N}}$-term) is considered by the Q‑tensor formulation and is consistent with the thermodynamic uncertainty relation.

%==========================================================================
%==========================================================================
%==========================================================================
\vskip\baselineskip
\begin{acknowledgments}
	S.T. thanks Y. Tabe and Y. Maruyama for their support at the early stage, J. Lee for the helpful notice on the sign of $\alpha_2 + \alpha_3$.
    %S.T. thanks Y. Tabe and Y. Maruyama for their help and discussions at the early stage. 
	This work was supported by JST SPRING, Grant Number JPMJSP2128; JSPS Grant-in-Aid for JSPS Fellows, Grant Number JP25KJ2180; and Waseda Research Institute for Science and Engineering Grant-in-Aid for Young Scientists (Early Bird) to S.T.
\end{acknowledgments}

%==========================================================================
\appendix
%==========================================================================
\section{Imura--Okano model}\label{sec:Okano_model}
    %\textcolor{black}{original but modified}
    The \EricksenLeslie adopts the director $\vec{n}$ to consider the anisotropy of the viscosity, and the transport coefficients regarding to the viscosity are polynomials in $\vec{n}$. In the \ImuraOkano~\cite{Imura1972}, the transport coefficients are polynomials in the Q-tensor $\tsr{Q}$:
    \begin{subequations}
    \label{eq:Imura_Okano_transport}
    \begin{align}
        L^{\mathrm{AA}}_{ijmn} (\tsr{Q})
        &=
        \frac{\eta}{2} \left(\delta_{im} \delta_{jn} + \delta_{jm} \delta_{in} \right)  \nonumber
        \\&\hspace{1.1em}
        +a_1 \left(Q_{im} \delta_{jn} + Q_{jm} \delta_{in}\right)	\nonumber
        +\frac{a_2}{2} Q_{kl} Q_{lk} \left(\delta_{im} \delta_{jn} + \delta_{jm} \delta_{in}\right)
        +a_3 Q_{ij} Q_{mn}
        \\&\hspace{1.1em}
        +\frac{a_4}{2} \left(Q_{ik} Q_{km} \delta_{jn} + Q_{jk} Q_{km}  \delta_{in} + Q_{ik} Q_{kn} \delta_{jm} + Q_{jk} Q_{kn}  \delta_{im}\right)
        +a_5 Q_{im} Q_{jn}
        ,\\
        L^{\mathrm{AN}}_{ijmn} (\tsr{Q})
        &=
        \textcolor{black}{
        \frac{b_1}{2} \left( Q_{im} \delta_{jn} + Q_{jm} \delta_{in} - Q_{in} \delta_{jm} - Q_{jn} \delta_{im} \right) \nonumber        
        }
        \\&\hspace{1.1em}
        \textcolor{black}{
        +\frac{b_2}{2} \left( Q_{ik} Q_{km} \delta_{jn} + Q_{jk} Q_{km} \delta_{in} - Q_{ik} Q_{kn} \delta_{jm} - Q_{jk} Q_{kn} \delta_{im} \right)
        }
        %-\frac{b_1}{2} \left(Q_{in} \delta_{jm}  + Q_{jn} \delta_{im}  - Q_{im} \delta_{jn}  - Q_{jm} \delta_{in} \right)	\nonumber
        %\\&\hspace{1.1em}
        %-\frac{b_2}{2} \left(Q_{ik} Q_{km} \delta_{jn} + Q_{jk} Q_{kn} \delta_{im} - Q_{ik} Q_{kn} \delta_{jm} - Q_{jk} Q_{km} \delta_{in} \right)
        ,\\
        L^{\mathrm{NA}}_{ijmn} (\tsr{Q})
        &=
        \frac{b_1}{2} \left(Q_{im} \delta_{jn} - Q_{jm} \delta_{in} + Q_{in} \delta_{jm} - Q_{jn} \delta_{im} \right)   \nonumber
        \\&\hspace{1.1em}
        +\frac{b_2}{2} 
        \textcolor{black}{
        \left(Q_{ik} Q_{km} \delta_{jn} - Q_{jk} Q_{km}  \delta_{in} + Q_{ik} Q_{kn} \delta_{jm} - Q_{jk} Q_{kn} \delta_{im} \right)
        }
        %\left(Q_{ik} Q_{km} \delta_{jn} A_{mn} - Q_{jk} Q_{km}  \delta_{in} A_{mn} + Q_{ik} Q_{kn} \delta_{jm} A_{mn} - Q_{jk} Q_{kn} \delta_{im} A_{mn} \right)
        ,\\
        L^{\mathrm{NN}}_{ijmn} (\tsr{Q})
        &=
        \frac{c_1}{2} \left(Q_{im} \delta_{jn} - Q_{jm} \delta_{in} - Q_{in} \delta_{jm} + Q_{jn} \delta_{im} \right)
        +\frac{c_2}{2} \left( Q_{kl} Q_{lk} \delta_{im} \delta_{jn} - Q_{kl} Q_{lk} \delta_{jm} \delta_{in} \right) \nonumber
        \\&\hspace{1.1em}
        +\frac{c_3}{2} \left(Q_{ik} Q_{km} \delta_{jn} + Q_{jk} Q_{km} \delta_{in} - Q_{ik} Q_{kn} \delta_{jm} - Q_{jk} Q_{kn} \delta_{im} \right)
        +\frac{c_4}{2} \left(Q_{im} Q_{jn} - Q_{in} Q_{jm} \right)
        ,
    \end{align}
    \end{subequations}
    where $a_1, a_2, a_3, a_4, a_5, b_1, b_2, b_2, c_1, c_2, c_3, c_4$ denotes the viscosity coefficients that are independent of $\tsr{Q}$ and temperature. The viscosity coefficient for the \ILC~phase $\eta$ may show the Arrhernius-type temperature-dependence but is independent of $\tsr{Q}$. \Reciproc~\cite{Onsager1931I,Onsager1931II} reads $L^{\mathrm{AN}}_{ijmn} = L^{\mathrm{NA}}_{mnij}$, as discussed in Sec.~\ref{sec:time_reversal}. 
    From these transport coefficients, the symmetric and antisymmetric parts of the irreversible stress tensor are obtained: 
    \begin{subequations}
    \begin{align}
        \sigma^{\mathrm{sym}}_{ij}
        &=
        \eta A_{ij}
        +a_1 \left(Q_{ik} A_{kj} + Q_{kj} A_{ik}\right)	\nonumber
		\\
        &\hspace{3.32em}
        +a_2 Q_{kl} Q_{lk} A_{ij}
        +a_3 Q_{ij} Q_{kl} A_{kl}
        +a_4 \left(Q_{il} Q_{lk} A_{kj} + Q_{kl} Q_{lj} A_{ik}\right)
        +a_5 Q_{ik} Q_{lj} A_{kl}	\nonumber
        \\&\hspace{3.32em}
        -b_1 \left(Q_{ik} \left(n_k \corot{n}_j - \corot{n}_k n_j \right) - Q_{kj} \left(n_i \corot{n}_k - \corot{n}_i n_k \right)\right)	\nonumber
        \\&\hspace{3.32em}
        -b_2 \left(Q_{il} Q_{lk} \left(n_j \corot{n}_k - \corot{n}_k n_j \right) - Q_{kl} Q_{lj} \left(n_i \corot{n}_k - \corot{n}_i n_k \right)\right)
        ,\\
        \sigma^{\mathrm{ant}}_{ij}
        &=
        b_1 \left(Q_{ik} A_{kj} - Q_{kj} A_{ik} \right)
        +b_2 \left(Q_{il} Q_{lk} A_{kj} - Q_{kl} Q_{lj} A_{ik} \right)	\nonumber
        \\&\hspace{3.32em}
        -c_1 \left(Q_{ik} \left(n_k \corot{n}_j - \corot{n}_k n_j \right)+Q_{kj} \left(n_i \corot{n}_k - \corot{n}_i n_k \right)\right)
        -c_2 Q_{kl} Q_{lk} \left(n_i \corot{n}_j - \corot{n}_i n_j \right)	\nonumber
        \\&\hspace{3.32em}
        -c_3 \left(Q_{il} Q_{lk} \left(n_k \corot{n}_j - \corot{n}_k n_j \right) + Q_{kl} Q_{lj} \left(n_i \corot{n}_k - \corot{n}_i n_k \right)\right)
        -c_4 Q_{ik} Q_{lj} \left(n_k \corot{n}_l - \corot{n}_k n_l \right)
        .
    \end{align}            
    \end{subequations}
    It should be mentioned that $n_i \corot{n}_j - \corot{n}_i n_i$ in our notation replaced $\Omega_{ij} - \omega_{ij}$ in the original notation. This replacement is indicated in Ref.~\cite{Imura1972} but is not performed. Note that $\Omega_{ij}, \omega_{ij}$ are different from the quantities in Appendix.~\ref{sec:variational}.

    The \ImuraOkano focuses on the uniaxial liquid crystal phase. Since $\tsr{Q}$ is reduced to the scalar order parameter $S$ and the director $\vec{n}$ with Eq.~\eqref{eq:QSn}, the hydrodynamic part of the stress tensor reads Eq.~\eqref{eq:stress_hydro} as in the \EricksenLeslie. The Leslie viscosity coefficients possess the $S$-dependence: 
    \begin{subequations}
        \label{eq:Imura_Okano_Leslie_viscosity}
    \begin{align}
		\alpha_1
		&=
        \frac{9}{4} \left(a_3 + a_5 \right) S^2
		,\\
		\alpha_2
		&=
		-\left(\frac{3}{2} b_1 + \frac{1}{2} c_1 \right) S
		-\left(\frac{3}{4} b_2 + \frac{1}{4} \left(6 c_2 + 5 c_3 - 2 c_4 \right)\right) S^2
		,\\
		\alpha_3
		&=
		-\left(\frac{3}{2} b_1 - \frac{1}{2} c_1 \right) S
		-\left(\frac{3}{4} b_2 - \frac{1}{4} \left(6 c_2 + 5 c_3 - 2 c_4 \right)\right) S^2
		,\\
		\alpha_4
		&=
		\eta - a_1 S + \frac{1}{4} \left(6 a_2 + 2 a_4 + a_5 \right) S^2
		,\\
		\alpha_5
		&=
		\left(\frac{3}{2} a_1 + \frac{3}{2} b_1 \right) S
		+\left(\frac{3}{4} \left(a_4 - a_5 \right) + \frac{3}{4} b_2 \right) S^2
		,\\
		\alpha_6
		&=
		\left(\frac{3}{2} a_1 - \frac{3}{2} b_1 \right)S
		+\left(\frac{3}{4} \left(a_4 - a_5 \right) - \frac{3}{4} b_2 \right) S^2
        .
	\end{align}
    \end{subequations}
    $S$ varies with temperature, and the \ImuraOkano reproduced the temperature dependence of the viscosity of the \NLC~phase~\cite{Imura1972}. 

    A suggestive point of the \ImuraOkano is that the antisymmetric part $\tsr{\sigma}^{\text{ant}}$ is derived from the transport coefficients $\tsr{L}^{\mathrm{NA}}$ and $\tsr{L}^{\mathrm{NN}}$ as polynomials in $\tsr{Q}$. In the \EricksenLeslie, $\tsr{\sigma}^{\text{ant}}$ is manually introduced from the molecular field $\vec{h}$ as
    \begin{equation}
        -\frac{1}{2} \left( n_i h_j - n_j h_i \right)
        ,
    \end{equation}
    and thus the \ImuraOkano has the potential to consider the force on $\vec{n}$ and on the volume element. The antisymmetric stress tensor $\tsr{\sigma}^{\text{ant}}$ is derived from a comprehensive consideration of $D_\infty$-symmetry. Thus, $\tsr{Q}$ enables to consider a stress that is not attributed to the force acting on $\vec{n}$ can be taken into account. Since $\tsr{Q}$ vanishes in the \ILC~phase, $\tsr{\sigma}^{\text{ant}}$ is absent as known (the $\mu'^{\text{N}}$-term is allowed in the \ILC~phase due to symmetry but prohibited by the TUR). We derived the $\mu'^{\text{N}}$-term with the \ImuraOkano, which induces torque on the volume element but is not attributed to the force on $\vec{n}$. The antisymmetric part that includes the hydrodynamic part and the thermal Leslie effect is
    \begin{equation}
        \sigma^{\text{ant}}_{ij}
        =
        - \frac{\alpha_3 - \alpha_2}{2} (n_i \corot{n}_j - n_j \corot{n}_i) 
        - \frac{\alpha_6 - \alpha_5}{2} (n_i n_m A_{mj} - n_j n_m A_{mi} ) 
        + \left[
            \mu^{\text{N}} (n_i \epsilon_{jkm} n_k - n_j \epsilon_{ikm} n_k)
            + \mu'^{\text{N}} \epsilon_{ijm}
        \right]
        q_m
        ,
    \end{equation}
    which could be written with the irreversible part of the molecular field $\vec{h}'$:
    \begin{equation}
        h'_i = (\alpha_3 - \alpha_2 ) \corot{n}_i
        + (\alpha_6 - \alpha_5 ) n_m A_{mi}
        - 2 \mu^{\text{N}} \epsilon_{ikm} n_k q_m %\epsilon_{ijk}\epsilon_{klm} n_j n_l k_m= \delta_{il}\delta_{jm} - \delta_{im}\delta_{jl}
        ,
    \end{equation}
    as
    \begin{equation}
        \sigma^{\text{ant}}_{ij}
        = -\frac{1}{2} \left( n_i h'_j - n_j h'_i \right) + \mu'^{\text{N}} \epsilon_{ijm}q_m
        .
    \end{equation}
    The molecular field $\vec{h}'$ gives the net irreversible torque exerted by $\vec{n}$~\cite{de_Gennes}, given by
    \begin{equation}
        \epsilon_{ijk} n_j h'_k
        =
        \epsilon_{ijk}
        \left[
            (\alpha_3 - \alpha_2)  n_j \corot{n}_k
            + (\alpha_6 - \alpha_5) n_j n_m A_{mk}
            - 2 \mu^{\text{N}} n_j\epsilon_{klm}n_l q_m
        \right]
        .
    \end{equation}
    The torque due to $\vec{h}'$ excludes the $\mu'^{\text{N}}$-term. The torque equivalent to $\tsr{\sigma}^{\text{ant}}$ explicitly has the $\mu'^{\text{N}}$-term:
    \begin{equation}
        -\epsilon_{ijk} \sigma^{\text{ant}}_{jk}
        =
        \epsilon_{ijk} n_j \left[ ( \alpha_3 - \alpha_2 ) \corot{n}_k + ( \alpha_6 - \alpha_5 ) n_m A_{mk} - 2 \mu^{\text{N}} \epsilon_{lmk} n_l q_m  \right] - 2 \mu'^{\text{N}} q_i
        .
    \end{equation}
    Thus, the $\mu'^{\text{N}}$-term is not attributed to the molecular field $\vec{h}'$. One may doubt the presence of the $\mu'^{\text{N}}$-term but chirality, tensorial argument with $\tsr{Q}$, and thermodynamics allows it in liquid crystal phases.

    The Levi-Civita symbol is a rank-$3$ completely antisymmetric tensor defined as
    \begin{equation}
    \label{eq:def_levi_civita}
        \epsilon_{ijk}
        =
        \left\{
        \begin{aligned}
            &+1 & (i,j,k) = (x,y,z), (y,z,x), (z,x,y)
            \\
            &-1 & (i,j,k) = (z,y,x), (y,x,z), (x,z,y)
            \\
            &0  &
            \text{otherwise}.
        \end{aligned}
        \right.
    \end{equation}
    The Levi-Civita symbol is a tensor density and is $\mathcal{P}$-even. 
    The transport coefficients for the Leslie effect are expanded with $\tsr{Q}$ and then written as a function of $S$ and $\vec{n}$. The reduction from $\tsr{Q}$ is done using
    \begin{equation}
        \epsilon_{ijk} n_k  n_l
        = n_i \epsilon_{jkl} n_k - n_j \epsilon_{ikl} n_k + \epsilon_{ijl}
        .
    \end{equation}
    
    %Comparison with coefficients in Beris--Edwards approach.

    \textit{Torque}---We confirm the change in angular momentum due to torque, by following the argument by de~Gennes~\cite{de_Gennes}. Let $V$ be a volumetric region of interest, and $\vec{L}$ be the angular momentum of $V$: 
    \begin{equation}
        L_i
        \coloneqq
        \int_V d^3\vec{r}\ \epsilon_{ijk}x_j(\rho v_k)
        ,
    \end{equation}
    where $\rho$ is the mass density. This angular momentum $\vec{L}$ is attributed to the translational motion of molecules. The angular momentum associated with molecular spinning is possibly added but it is excluded in the \EricksenLeslie, as the inertia of the molecules is negligibly small and thus has an insignificant effect on the total torque~\cite{Leslie1992, Pleiner1996}. The spin angular momentum $\lesssim \hbar N_{\text{A}}$ is also negligible, where $\hbar$ is the Planck constant divided by $2\pi$, and $N_{\text{A}}$ the Avogadro number.
    The Lagrange derivative describes the evolution of $\vec{L}$:
    \begin{equation}
        \dot{L}_i
        =
        \frac{d}{dt} \int_V d^3\vec{r}\ \epsilon_{ijk}x_j(\rho v_k)
        =
        \int_V d^3\vec{r}\ \epsilon_{ijk}x_j(\rho \dot{v}_k)
        .
    \end{equation}
    The application of the Navier--Stokes equation in Eq.~\eqref{eq:Linear_Navier_Stokes} leads to
    \begin{subequations}
    \begin{align}
        \dot{L}_i
        &=
        \int_V d^3\vec{r}\ \epsilon_{ijk} x_j (\nabla_l \sigma_{lk})
        \\
        &=
        \oint_{\partial V} dS_l\ \epsilon_{ijk} x_j \sigma_{lk}
        - \int_V d^3\vec{r}\ \epsilon_{ijk} \sigma_{jk}
        ,
    \end{align}
    \end{subequations}
    where $\partial V$ is the surface of $V$, and $dS_l$ the vector of the area element that is oriented outside $V$. This equation shows that the stress on the surface $\partial V$ (first term) and the antisymmetric part of the stress tensor (second term) drive the rotation of the region $V$ or its internal structures. The second term in the right-side is the torque per volume
    \begin{equation}
        \label{eq:Gamma_def_torque}
        \Gamma_i
        \coloneqq
        - \epsilon_{ijk} \sigma_{jk}
        .
    \end{equation}
    $\vec{\Gamma}$ is simply referred to as ``torque'' if there is no concern of confusion. The antisymmetric part induces the torque.

%==========================================================================
\section{Some other notations}\label{sec:some_other_notations}
    %\textcolor{black}{move to appendix}
    We derived the constitutive equations Eqs.~\eqref{eq:constitutive_sigma}~and~\eqref{eq:constitutive_-T-1nablaT} that describe viscosity, the Leslie effects, and heat conduction. Another expression is obtained by having the operator $\rho_\perp^{-1} \delta_{ij} + (\rho_\parallel^{-1} - \rho_\perp^{-1}) n_i n_j$, the inverse of $\rho_\perp \delta_{ij} + (\rho_\parallel - \rho_\perp) n_i n_j$, act on Eq.~\eqref{eq:constitutive_-T-1nablaT}:
    \begin{subequations}
    \label{eq:constitutive2}
    \begin{align}
            \sigma'_{ij}
            &=
            \tilde{\alpha}_1 n_i n_j n_m n_n A_{mn}
            + \tilde{\alpha}_2 n_i \corot{n}_j
            + \tilde{\alpha}_3 n_j \corot{n}_i
            + \tilde{\alpha}_4 A_{ij}
            + \tilde{\alpha}_5 n_i n_m A_{mj}
            + \tilde{\alpha}_6 n_j n_m A_{mi}
            \nonumber\\&\hspace{1em}
            + \left[
            \tilde{\mu}^{\text{A}} (n_i \epsilon_{jkm} n_k + n_j \epsilon_{ikm} n_k)
            + \tilde{\mu}^{\text{N}} (n_i \epsilon_{jkm} n_k - n_j \epsilon_{ikm} n_k)
            + \tilde{\mu}'^{\text{N}} \epsilon_{ijm} 
            \right] 
            (-T^{-1}\nabla_m T) 
            ,\label{eq:constitutive2_sigma}\\
            q_i
            &=
            \left[
                \sigma_\perp \delta_{im}
                + \left(\sigma_\parallel - \sigma_\perp \right) n_i n_m
            \right]
            (-T^{-1}\nabla_m T)
            + 2\tilde{\mu}^{\text{A}} %\frac{2\mu^{\text{A}}}{\rho_\perp}
            \epsilon_{ikn} n_k n_m
            A_{mn}
            - 2 \left(\tilde{\mu}^{\text{N}} - \tilde{\mu}'^{\text{N}} \right) %\frac{2 \left( - \mu^{\text{N}} + \mu'^{\text{N}}\right)}{\rho_\perp}
            \epsilon_{ikm} n_k 
            \corot{n}_m
            ,\label{eq:constitutive2_q}
            \\
            \tilde{\alpha}_1
            &=
            \alpha_1 
            + \frac{\left(2\mu^{\text{A}}\right)^2}{\rho_\perp}
            ,\\
            \tilde{\alpha}_2
            &=
            \alpha_2
            + \left(
                \mu^{\text{A}}
                + \left(
                    \mu^{\text{N}} - \mu'^{\text{N}}
                \right)
            \right)
            \frac{2\left(\mu^{\text{N}} - \mu'^{\text{N}}\right)}{\rho_\perp}
            ,\\
            \tilde{\alpha}_3
            &=
            \alpha_3 
            + \left(
                \mu^{\text{A}}
                - \left(
                    \mu^{\text{N}} - \mu'^{\text{N}}
                \right)
            \right)
            \frac{2\left(\mu^{\text{N}} - \mu'^{\text{N}}\right)}{\rho_\perp}
            ,\\
            \tilde{\alpha}_4
            &=
            \alpha_4
            ,\\
            \tilde{\alpha}_5
            &=
            \alpha_5
            + \left(
                - \mu^{\text{A}}
                - \left(
                    \mu^{\text{N}} - \mu'^{\text{N}}
                \right)
            \right)
            \frac{2\mu^{\text{A}}}{\rho_\perp}
            ,\\
            \tilde{\alpha}_6
            &=
            \alpha_6
            + \left(
                - \mu^{\text{A}}
                + \left(
                    \mu^{\text{N}} - \mu'^{\text{N}}
                \right)
            \right)
            \frac{2\mu^{\text{A}}}{\rho_\perp}
            ,\\
            \sigma_\perp
            &=
            \frac{1}{\rho_\perp}
            ,\\
            \sigma_\parallel
            &=
            \frac{1}{\rho_\parallel}
            ,\\
            \tilde{\mu}^{\text{A}}
            &=
            \frac{\mu^{\text{A}}}{\rho_\perp}
            ,\\
            \tilde{\mu}^{\text{N}}
            &=
            \frac{\mu^{\text{N}}}{\rho_\perp}
            + \frac{\mu'^{\text{N}}}{\rho_\perp}
            \left(
                \frac{\rho_\perp}{\rho_\parallel} - 1
            \right)
            ,\\
            \tilde{\mu}'^{\text{N}}
            &=
            \frac{\mu'^{\text{N}}}{\rho_\parallel}
            .
    \end{align}  
    \end{subequations}
    This expression is useful in comparing our model with the conventional models. Leslie and de~Gennes~\cite{Leslie1968II, de_Gennes} selected the temperature gradient as an affinity. We selected the heat current $\vec{q}$ so that all affinities are $\mathcal{T}$-odd. The conventional models and our model are equivalent, as Eq.~\eqref{eq:constitutive} and Eq.~\eqref{eq:constitutive2} are interconvertible.

    For reference, the correspondence between our notation and that of Leslie is shown. The leftmost terms represent the notation by Leslie~\cite{Leslie1968II}, and other terms are written in our notation in Eqs.~\eqref{eq:constitutive2} and \eqref{eq:constitutive2}.
    \begin{subequations}
    \label{eq:leslie_notation}
    \begin{align}
        0
        &=
        \tilde{\alpha}_1 = \tilde{\alpha}_4 = \tilde{\alpha}_5 = \tilde{\alpha}_6
        ,\\
        0
        &=
        \tilde{\mu}^{\text{A}} = \tilde{\mu}'^{\text{N}}
        ,\\
        \mu_2
        &=
        \tilde{\alpha}_2
        =
        \alpha_2
        + \frac{2\left(\mu^{\text{N}}\right)^2}{\rho_\perp}
        ,\\
        \mu_3
        &=
        \tilde{\alpha}_3
        =
        \alpha_3
        - \frac{2\left(\mu^{\text{N}}\right)^2}{\rho_\perp}
        ,\\
        \mu_7
        &=
        -\frac{\tilde{\mu}^{\text{N}}}{T}
        =
        -\frac{\mu^{\text{N}}}{\rho_\perp T}
        ,\\
        \mu_8
        &=
        +\frac{\tilde{\mu}^{\text{N}}}{T}
        =
        +\frac{\mu^{\text{N}}}{\rho_\perp T}
        ,\\
        \kappa_1
        &=
        -\frac{\sigma_\perp}{T}
        =
        -\frac{1}{\rho_\perp T}
        ,\\
        \kappa_2
        &=
        -\frac{\sigma_\parallel - \sigma_\perp}{T}
        =
        -\frac{1}{\rho_\parallel T} + \frac{1}{\rho_\perp T}
        ,\\
        \kappa_3
        &=
        2 \tilde{\mu}^{\text{N}}
        =
        2 \frac{\mu^{\text{N}}}{\rho_\perp}
        ,\\
        \lambda_1
        &=
        \tilde{\alpha}_2 - \tilde{\alpha}_3
        =
        \alpha_2 - \alpha_3 + \frac{4 \left(\mu^{\text{N}}\right)^2}{\rho_\perp}
        ,\\
        \lambda_3
        &=
        2 \frac{\tilde{\mu}^{\text{N}}}{T}
        =
        2 \frac{\mu^{\text{N}}}{\rho_\perp T}
        .
    \end{align}
    \end{subequations}

    The constitutive equations in Eq.~\eqref{eq:Sato_Lehmann} by Sato\etal are reduced from Eq.~\eqref{eq:constitutive2}. The correspondence of the notation is
    \begin{subequations}
    \label{eq:notation_sato}
    \begin{align}
        &\dot{n}_i 
        &&=
        \Omega \epsilon_{ijk} \hat{z}_j n_k
        ,&&\\
        &q
        &&=
        q_z,
        \\
        &-T^{-1}\nabla T
        &&=
        -T^{-1}\nabla_z T
        ,&&\\
        &\tau
        &&=
        -\epsilon_{zjk}\sigma'_{jk}
        ,&\\
        &\gamma_1
        &&=
        \tilde{\alpha}_3 - \tilde{\alpha}_2
        &&=
        (\alpha_3 - \alpha_2)
        \left(
            1 -
            \frac{\left(2\mu^{\text{N}}\right)^2}{(\alpha_3 - \alpha_2) \rho_\perp}
        \right)
        ,\\
        &\sigma_{\text{q}} T
        &&=
        \sigma_\perp - 
        \frac{\left(2\tilde{\mu}^{\text{N}}\right)^2}{\tilde{\alpha}_3 - \tilde{\alpha}_2}
        &&=
        \frac{1}{\rho_\perp}
        \left(
            1 - 
            \frac{\left(2\mu^{\text{N}}\right)^2}{(\alpha_3 - \alpha_2) \rho_\perp - \left(2\mu^{\text{N}}\right)^2}
        \right)
        ,\\
        &b_{\text{dir}} = b_{\text{inv}}
        &&=
        -\frac{2\tilde{\mu}^{\text{N}}}{\tilde{\alpha}_3 - \tilde{\alpha}_2}
        &&=
        -\frac{2\mu^{\text{N}}}{(\alpha_3 - \alpha_2) \rho_\perp - \left(2\mu^{\text{N}}\right)^2}
        .
    \end{align}
    \end{subequations}
    The left, middle, and right sides correspond to the notation of Sato\etal\cite{Sato2017JPSJ}, Eq.~\eqref{eq:constitutive}, and Eq.~\eqref{eq:constitutive2}, respectively.

%==========================================================================
\section{Formulation with \OnsagerVariational}\label{sec:variational}
    %\textcolor{yellow}{Onsager's variational principle,  cf.~\cite{Ericksen1991archrationalmechanal,Tovkach2017}. Dynamics of order parameter. Extension to biaxial nematics. }
    \OnsagerVariational is a beneficial framework for describing the dynamics of soft matter~\cite{Doi2011}. In liquid crystals, the time evolution of the velocity field $\vec{v}$, the Q-tensor $\tsr{Q}$, and the heat current $\vec{q}$ is governed by the minimization of the Rayleighian $\Rayleighian [\vec{v}, \dot{\tsr{Q}}, \vec{q}]$, which is a functional of these variables. The physically realized evolution corresponds to a stationary point of $\Rayleighian$:
    \begin{equation}
        \delta \Rayleighian [\vec{v}, \dot{\tsr{Q}}, \vec{q}]
        = \frac{\delta \Rayleighian}{\delta v_i} \delta v_i
        + \frac{\delta \Rayleighian}{\delta \dot{Q}_{ij}} \delta \dot{Q}_{ij}
        + \frac{\delta \Rayleighian}{\delta q_i} \delta q_i
        = 0
        .
    \end{equation}
    This \OnsagerVariational is applicable to the formulation of the Leslie effects, as we see in this Appendix~\ref{sec:variational}. 

    \textit{Rayleighian}---The Rayleighian $\Rayleighian$ consists of the Landau–de~Gennes (LdG) free energy density $f_{\text{LdG}}$ and the dissipation function $R$ of the system:
    \begin{equation}
        \Rayleighian
        = \frac{d}{dt} \int_V d^3\vec{r}\, f_{\text{LdG}}
        + \int_V d^3\vec{r}\, R
        + \int_V d^3\vec{r}\, \left(-p \nabla_i v_i - \Lambda Q_{ij}\delta_{ji} \right)
        ,
    \end{equation}
    where $p$ and $\Lambda$ are the Lagrange multipliers associated with the flow incompressibility and the tracelessness of $\tsr{Q}$, respectively. Liquid crystals respond to the external electric field, and the contribution of the electrostatic interaction can be renormalized into the free energy. 
    
    The irreversible contributions, including friction, heat conduction, and the Leslie effects, are described by the dissipation function $R$. Friction is induced by the flow gradient $\vec{\nabla}\vec{v}$, especially its symmetric part $\tsr{A}$ is responsible in normal isotropic fluids. The friction includes those arising from the vorticity, whose contribution is absent in isotropic fluids. The vorticity $\vec{\omega}$ is associated with the antisymmetric part of the velocity gradient $\tsr{\Omega}$, which is the irreducible 3-dimensional representation of the 3D rotation group $\mathrm{SO}(3)$:
    \begin{align}
        \omega_i
        &=
        \frac{1}{2} \epsilon_{ijk} \nabla_j v_k = \frac{1}{2} \epsilon_{ijk} \Omega_{jk}
        ,\\
        \Omega_{ij} 
        &=
        \frac{1}{2}\left(\nabla_i v_j - \nabla_j v_i \right)
        .
    \end{align}
    The anisotropic molecules may rotate independently of the backflow, leading to friction. This kinetics is described by the rotation of the order parameter relative to the vortex flow. For a variable $a$, let $\corot{a}$ be its change in time relative to flow, which would be considered as the covariant time derivative. The relative change in the order parameters is given by
    \begin{align}
        N_i = \corot{n}_i &= \dot{n}_i + \Omega_{ij} n_j
        ,\\
        \corot{Q}_{ij} &= \dot{Q}_{ij} + \Omega_{ik} Q_{kj} + \Omega_{jk} Q_{ki}
        .
    \end{align}
    The contributions of them are considered by expanding $R$ with $\corot{\vec{n}}$ or $\corot{\tsr{Q}}$. Friction in liquid crystals is complicated due to the anisotropy, and the Leslie viscosity has five degrees of freedom. The dissipation function $R$ accounts for the anisotropy by being expanded with $\tsr{Q}$. We assume that $R$ is expressed as a quadratic expansion in terms of $\tsr{Q}$ and/or $\corot{\tsr{Q}}$:
    \begin{align}
        2R
        &= 
        \zeta_1 \corot{Q}_{ij} \corot{Q}_{ji} 
        + 2 \zeta_2 \corot{Q}_{ij} A_{ji} 
        + 2 \zeta_3 Q_{ij} \corot{Q}_{jk} A_{ki} 
        + 2 \zeta_4 Q_{ij} A_{jk} A_{ki}  
        \nonumber\\
        &\hspace{1em}
        + \zeta_5 Q_{ij} Q_{jk} A_{kl} A_{li} 
        + \zeta_6 Q_{ij} Q_{kl} A_{ij} A_{kl}
        + \zeta_7 Q_{ij} Q_{ji} A_{kl} A_{lk}
        + \zeta_8 A_{ij} A_{ji} 
        \nonumber\\&\hspace{1em}
        + 2 \zeta_{9} (\epsilon_{ikl} Q_{kj} + \epsilon_{jkl} Q_{ki} ) q_l \corot{Q}_{ij} 
        + 2 \zeta_{10} (\epsilon_{ikl} Q_{kj} + \epsilon_{jkl} Q_{ki} ) q_l A_{ij}
        + \zeta_{11} q_i q_i 
        + \zeta_{12} Q_{ij} q_i q_j
        ,
    \end{align}
    where $\zeta_1,\zeta_2,\dots,\zeta_8$ denote the achiral viscosity introduced by Tovkach\etal\cite{Tovkach2017}, $\zeta_9, \zeta_{10}$ the Leslie effects, and $\zeta_{11}, \zeta_{12}$ heat conduction. This form of our $R$ that incorporates the Leslie effects is an extension of the dissipation function in the \EricksenLeslie~\cite{Tovkach2017}. 

    \textit{Evolution of orientational order}---The variational of the LdG free energy is useful to calculate the orientation field in equilibrium. The driving force that induces the change in $\tsr{Q}$ is given by
    \begin{equation}
    \label{eq:LdG_molecular_field}
        \frac{\delta}{\delta Q_{ij}} \int_V d^3\vec{r}\, f_{\text{LdG}}(\tsr{Q}, \nabla \tsr{Q})
        =
        \left(
            \frac{\partial }{\partial Q_{ij}} - \nabla_k \frac{\partial}{\partial (\nabla_k Q_{ij})}
        \right)
        f_{\text{LdG}}
        .
    \end{equation}
    According to \OnsagerVariational, change in $\tsr{Q}$ is determined by
    \begin{equation}
       \frac{\delta \Rayleighian}{\delta \dot{Q}_{ij}}
        =
        \frac{\delta}{\delta \dot{Q}_{ij}} \int_V d^3\vec{r}\, \dot{f}_{\text{LdG}}
        + \frac{\delta}{\delta \dot{Q}_{ij}} \int_V d^3\vec{r}\, R
        - \Lambda \delta_{ij}
        =
        0
        .
    \end{equation}
    The first term is the reversible driving force mentioned above in Eq.~\eqref{eq:LdG_molecular_field}~\cite{Tovkach2017}. 
    The second term describes the irreversible dynamics determined as
    \begin{subequations}
    \begin{align}
        \frac{\delta}{\delta \dot{Q}_{ij}} \int_V d^3\vec{r}\, R 
        &= \frac{\partial R}{\partial \dot{Q}_{ij}}
        \\
        &=
        \zeta_1 \corot{Q}_{ij}
        + \zeta_2 A_{ij}
        + \frac{1}{2} \zeta_3 ( Q_{ik} A_{kj} + Q_{jk} A_{ki} ) 
        + \zeta_9 (\epsilon_{ikm} Q_{kj} + \epsilon_{jkm} Q_{ki}) q_m
        .
    \end{align}
    \end{subequations}
    The terms of $\zeta_1, \zeta_2, \zeta_3$ show friction and the last term of $\zeta_9$ is the Leslie thermomechanical effect. The evolution of $\tsr{Q}$ reduces to the evolution of $\vec{n}$ and the scalar order parameter $S$ in the uniaxial \NLC~phase. The evolution of $\vec{n}$ is dominant in thermotropic liquid crystals. However, in active matter systems, the flow alignment, namely, variations in $S$ is induced by the flow gradient~\cite{Kralj2023prl}, as described by the Beris--Edwards model~\cite{BerisEdwards1994}. Thermomechanical effects result in the rotation of $\vec{n}$ but do not cause flow alignment.

    \textit{Irreversible stress}---The stress is given by the functional derivative of the Rayleighian. Since the LdG free energy is independent of the velocity $\vec{v}$, flow do not induce the reversible stress. The irreversible part of the stress $\tsr{\sigma}'$ is derived from the dissipation function $R$:
    \begin{subequations}
    \begin{align}
        - \nabla_j \sigma'_{ji}
        &\coloneqq
        \frac{\delta}{\delta v_i} \int_V d^3\vec{r}\, R 
        =
        - \nabla_j
        \left[
            \frac{\partial A_{mn}}{\partial (\nabla_j v_i)} \frac{\partial R}{\partial A_{mn}}
            + \frac{\partial \corot{Q}_{mn}}{\partial (\nabla_j v_i)} \frac{\partial R}{\partial \corot{Q}_{mn}}
        \right]
        ,\\
        \sigma'_{ji}
        &=
        \zeta_1 \left( Q_{ik} \corot{Q}_{kj} - Q_{jk} \corot{Q}_{ki} \right)
        + \zeta_2 \corot{Q}_{ij}
        + \zeta_2 ( Q_{ik} A_{kj} - Q_{jk} A_{ki} )
        \\
        &\hspace{1em}
        + \frac{1}{2} \zeta_3 \left( Q_{ik} \corot{Q}_{kj} + Q_{jk} \corot{Q}_{ki} \right)
        + \frac{1}{2} \zeta_3 ( Q_{ik} Q_{kl} A_{lj} - Q_{jk} Q_{kl} A_{li} ) 
        + \zeta_4 ( Q_{ik} A_{kj} + Q_{jk} A_{ki} )
        \\
        &\hspace{1em}
        + \frac{1}{2} \zeta_5 ( Q_{ik} Q_{kl} A_{lj} + Q_{jk} Q_{kl} A_{li} )
        + \zeta_6 Q_{ij} Q_{kl} A_{lk}
        + \zeta_7 Q_{kl} Q_{lk} A_{ij}
        + \zeta_8 A_{ij}
        \\
        &\hspace{1em}
        + \zeta_{10} (\epsilon_{ikm} Q_{kj} + \epsilon_{jkm} Q_{ki} ) q_m
        + \zeta_9 \left(- \epsilon_{ikm} Q_{kl} Q_{lj} + \epsilon_{jkm} Q_{kl} Q_{li} + 2 \epsilon_{klm} Q_{ki} Q_{lj} \right) q_m
        .
    \end{align}
    \end{subequations}
    This stress reduces to the \EricksenLeslie, and a set of relations of the coefficients $\zeta_i\,(i=1,\dots,8)$ is obtained~\cite{Tovkach2017}:
    \begin{subequations}
    \begin{align}
        \alpha_1
        &=
        \zeta_6 S^2
        \\
        \alpha_2 
        &=
        \zeta_2 S + \left( - \zeta_1 + \frac{1}{6} \zeta_3 \right) S^2
        \\
        \alpha_3
        &=
        \zeta_2 S + \left( + \zeta_1 + \frac{1}{6} \zeta_3 \right) S^2
        \\
        \alpha_4
        &=
        \zeta_8 - \frac{1}{3} \zeta_4 S + \frac{1}{3} \left( \frac{1}{3} \zeta_5 + 2 \zeta_7 \right) S^2
        \\
        \alpha_5 
        &=
        \left( - \zeta_2 + \frac{1}{2} \zeta_4 \right) S + \left( - \frac{1}{6} \zeta_3 + \frac{1}{4} \zeta_5 \right) S^2
        \\
        \alpha_6
        &=
        \left( + \zeta_2 + \frac{1}{2} \zeta_4 \right) S + \left( + \frac{1}{6} \zeta_3 + \frac{1}{4} \zeta_5 \right) S^2
        .
    \end{align}
    \end{subequations}
    In the uniaxial phase, the Tovkach model is encompassed by the \ImuraOkano but imposes constraints:
    \begin{subequations}
    \begin{align}
        c_1
        &=
        0
        \\
        \frac{1}{4} \left( 6 c_2 + 5 c_3 - 2 c_4 \right)
        &=
        \zeta_1
        \\
        -\frac{3}{2} b_1
        &=
        \zeta_2
        \\
        -\frac{9}{2} b_2
        &=
        \zeta_3
        \\
        3 a_1
        &=
        \zeta_4
        \\
        3 \left( a_4 - a_5 \right)
        &=
        \zeta_5
        \\
        \frac{9}{4} \left( a_3 + a_5 \right)
        &=
        \zeta_6
        \\
        \frac{1}{8} \left( 18 a_2 + 2 a_4 + 7 a_5 \right)
        &=
        \zeta_7
        \\
        \eta
        &=
        \zeta_8
        .
    \end{align}
    \end{subequations}
    Here $c_1 =0$ indicates that $\alpha_3 - \alpha_2=O(S^2)$, which is also mentioned by the Beris--Edwards model~\cite{BerisEdwards1994}. The \ImuraOkano provides a general framework of nematodynamics but \OnsagerVariational reduces extra degrees of freedom, which is applicable to the Leslie effects: 
    \begin{subequations}
    \begin{align}
        \mu^{\text{A}}_{12}
        &=
        \zeta_{10}
        ,\\
        \mu^{\text{N}}_{22}
        = - \frac{1}{2} \mu^{\text{N}}_{23}
        &=
        \zeta_9
        ,\\
        \mu^{\text{A}}_{22} = \mu^{\text{A}}_{24}
        = \mu^{\text{N}}_{01} = \mu^{\text{N}}_{11} = \mu^{\text{N}}_{12} = \mu^{\text{N}}_{21} = \mu^{\text{N}}_{24} = \mu^{\text{N}}_{25}
        &=
        0
        .
    \end{align}
    \end{subequations}
    Consequently, the Leslie cross-coupling coefficients are given by
    \begin{subequations}
    \begin{align}
        \mu^{\text{A}}
        &=
        \frac{3}{2} \zeta_{10} S
        ,\\
        \mu^{\text{N}}
        &=
        - \frac{9}{4} \zeta_9 S^2
        ,\\
        \mu'^{\text{N}}
        &=
        O(S^3)
        .
    \end{align}
    \end{subequations}
    $\mu^{\text{A}}$ and $\mu^{\text{N}}$ are polynomials with terms of $S^1$ and $S^2$ or higher, respectively. The $\mu'^{\text{N}}$-term does not appear when considering up to quadratic contributions of $\tsr{Q}$ and/or $\corot{\tsr{Q}}$. The finite $\mu'^{\text{N}}$ emerges in higher terms such as $\left( \epsilon_{ikl} Q_{km} Q_{lj} + \epsilon_{jkl} Q_{km} Q_{li} \right) q_m \corot{Q}_{ij}$:
    \begin{equation}
        \frac{\partial}{\partial ( \nabla_j v_i)}\left( \epsilon_{ikl} Q_{km} Q_{lj} + \epsilon_{jkl} Q_{km} Q_{li} \right) q_m \corot{Q}_{ij}
        =
        \left[
        \frac{9}{8} S^3 \left( n_i \epsilon_{jkm} n_k - n_j \epsilon_{ikm} n_k \right) + \frac{3}{4} S^3 \epsilon_{ijm}
        \right] q_m
        .
    \end{equation}
    Other cubic terms can also be argued. \OnsagerVariational successfully reproduced the Leslie effects, as derived in the \ImuraOkano. These include the $\mu'^{\text{N}}$-term that was not derived in previous frameworks. Furthermore, an insightful suggestion was reached: The lowest-degree term of the Leslie cross-coupling coefficients is not necessarily $S^1$. \OnsagerVariational derives the $S$-dependence: $\alpha_4 = O(S^0),\ \alpha_2, \alpha_3, \alpha_5, \alpha_6 = O(S^1),\ \alpha_1 = O(S^2)$. These relations are led by also the \ImuraOkano~\cite{Imura1972} and explain the smallness of $\alpha_1$~\cite{de_Gennes}. Our achievement $\mu^{\text{A}} = O(S^1),\ \mu^{\text{N}} = O(S^2),\ \mu'^{\text{N}} = O(S^3)$ will expect their magnitude relationship and its shift driven by temperature. 

    \textit{Heat conduction}---
    The heat current $\vec{q}$ and the temperature gradient $- T^{-1} \vec{\nabla} T$ are conjugate variables, as in Eq.~\eqref{eq:2R_sigma_A_sigma_N_T_q}. The variational derivative of $\Rayleighian$ with $\vec{q}$ yields 
    \begin{subequations}
    \begin{align}
        - T^{-1} \nabla_i T
        &=
        \frac{\delta \Rayleighian}{\delta q_i}
        =
        \frac{\partial R}{\partial q_i}
        \\
        &=
        \left(
            \epsilon_{mki} Q_{kn} + \epsilon_{nki} Q_{km}
        \right) 
        \left(
            \zeta_9 \corot{Q}_{mn}
            + \zeta_{10} A_{mn}
        \right)
        + \left( \zeta_{11} \delta_{im} + \zeta_{12} Q_{im} \right) q_m
        .
    \end{align}
    \end{subequations}
    The first term shows the temperature gradient induced by the thermal Leslie effect. The second term shows pure heat conduction. Eq.~\eqref{eq:LQQ_Q} indicates the correspondence of the coefficients:
    \begin{subequations}
    \begin{align}
        \zeta_{11} &= \rho_0
        ,\\
        \zeta_{12} &= \rho_1
        .
    \end{align}
    \end{subequations}

    Only $R$ is responsible for heat conduction, and the corresponding reversible part is absent in $\Rayleighian$. The temperature is always uniform in equilibrium. Any systems with the temperature gradient are in nonequilibrium even if a heat current is canceled out. Equilibrium with a nonuniform temperature is only achieved when gravity, in the context of general relativity, is significant~\cite{Shimizu2021ThermoII}. However, diffusion and electric conduction may have reversible terms, and equilibrium is possible when the chemical potential gradient and the electric field are applied, respectively, but the conjugate currents are absent.

    In the present model, the hydrodynamic variables are $\vec{v}$, $\dot{Q}$ and $\vec{q}$, which are all $\mathcal{T}$-odd. This time reversal symmetry results in the symmetric transport matrix due to \Reciproc in Eq.~\eqref{eq:Onsager_Reciprocal_Relations}. Since \OnsagerVariational is consistent with the dynamics characterized by the symmetric transport matrix~\cite{Doi2011}, our choice of variables is appropriate for formulation with this framework. On the other hand, the conventional model~\cite{Leslie1968II, de_Gennes} adopts the $\mathcal{T}$-even $- T^{-1} \vec{\nabla} T$ instead of $\mathcal{T}$-odd $\vec{q}$. The thermal Leslie effect is represented by the antisymmetric part of the transport matrix, and thus cannot be described by \OnsagerVariational. MD simulation by Sarman\etal has revealed that the thermal Leslie torque is induced so that dissipation is minimized~\cite{Sarman2025PCCP}. This is consistent with \OnsagerVariational , where the sum of the free energy and the dissipation functional is minimized.

%==========================================================================
\section{Derivation of thermodynamic bounds}\label{sec:derivation_bounds}
    Thermodynamic bounds in Eqs.~\eqref{eq:bound_alpha_4}--\eqref{eq:bound_|Z|} are derived from the positive semidefiniteness of the transport matrix $\tsr{L}$. We have determined the conditions under which all eigenvalues of $\tsr{L}$ are non-negative. For the uniaxial phase, dissipation function $R$ is given by
    \begin{align}
        2R
        &=
        \sigma^{\text{sym}}_{ij} A_{ij} + \sigma^{\text{ant}}_{ij} N_{ij} + (- T^{-1} \nabla_i T) q_i
        \nonumber\\
        &=
        \alpha_1 n_i n_j n_k n_l A_{ij} A_{kl}
        + \alpha_4 A_{ij} A_{ij}
        + (\alpha_5 + \alpha_6) n_i n_j A_{ik} A_{jk}
        + 2 (\alpha_2 + \alpha_3) n_i A_{ij} \corot{n}_j
        + (\alpha_3 - \alpha_2) \corot{n}_i \corot{n}_i 
        \nonumber\\
        &\hspace{1.1em}
        + 4 \mu^{\text{A}} \epsilon_{ijk} n_j n_l A_{il} q_k
        - 4 \left( \mu^{\text{N}} - \mu'^{\text{N}} \right) \epsilon_{ijk} n_j \corot{n}_i q_k
        + \rho_\perp q_i q_i
        + (\rho_\parallel - \rho_\perp) n_i n_j q_i q_j
        .
    \end{align}
    The transport matrix $\tsr{L}$ relates the affinity with $21=9+9+3$ components and the current with $21=9+9+3$ and has $441=21\times21$ components, we obtain a meaningful $10\times 10$ matrix $\tilde{\tsr{L}}$ by selecting an appropriate coordinate system such that the director $\vec{n}$ lies along the $x$-axis
    \begin{equation}
        \vec{n} = \hat{\vec{x}}
        .
    \end{equation}
    Since the coordinate system is arbitrary, this selection preserves generality. The affinity has $10$ components that vary independently:
    \begin{equation}
        \tilde{\vec{F}} \coloneqq (A_{yy},\ A_{zz},\ A_{yz},\ A_{xy},\ A_{zx},\ \corot{n}_y,\ \corot{n}_z,\ q_y,\ q_z,\ q_x)^{\text{T}}
        .
    \end{equation}
    The positive semidefiniteness of $\tsr{L}$ is equivalent to that of $\tilde{\tsr{L}}$
    \begin{equation}
        \tilde{L}_{\mu\nu} \tilde{F}_\mu \tilde{F}_\nu \ge 0
        .
    \end{equation}
    Upon rearranging the affinity $\tilde{\vec{F}}$ into 
    \begin{equation}
        \bar{\vec{F}} \coloneqq (A_{yz},\ A_{yy},\ A_{zz},\ A_{xy},\ \corot{n}_y,\ q_z,\ A_{zx},\ \corot{n}_z,\ q_y,\ q_x)^{\text{T}}
        ,
    \end{equation}
    the matrix is block-diagonalized in
    \begin{subequations}
    \begin{align}
        \bar{\tsr{L}}
        &=
        \begin{pmatrix}
            \bar{\tsr{L}}^{(1)}  &   &   &   &   \\
            &   \bar{\tsr{L}}^{(2)}  &   &   &   \\
            &   &   \bar{\tsr{L}}^{(3)}  &   &   \\
            &   &   &   \bar{\tsr{L}}^{(4)}  &   \\
            &   &   &   &   \bar{\tsr{L}}^{(5)}
        \end{pmatrix}
        ,\\
        \bar{\tsr{L}}^{(1)}
        &=
        \begin{pmatrix}
            2 \alpha_4
        \end{pmatrix}
        ,\\
        \bar{\tsr{L}}^{(2)}
        &=
        \begin{pmatrix}
            \alpha_1 + 2 \alpha_4 + (\alpha_5 + \alpha_6 )  & \alpha_1 + \alpha_4 + (\alpha_5 + \alpha_6 )\\
            \alpha_1 + \alpha_4 +( \alpha_5 + \alpha_6 )    & \alpha_1 + 2\alpha_4 + ( \alpha_5 + \alpha_6  )
        \end{pmatrix}
        ,\\
        \bar{\tsr{L}}^{(3)}
        &=
        \begin{pmatrix}
            2 \alpha_4 + ( \alpha_5 + \alpha_6 )    & \alpha_2 + \alpha_3   & -2 \mu^{\text{A}}  \\
            \alpha_2 + \alpha_3 &  \alpha_3 - \alpha_2  & 2 \left( \mu^{\text{N}} - \mu'^{\text{N}} \right)\\
            -2 \mu^{\text{A}}    & 2 \left( \mu^{\text{N}} - \mu'^{\text{N}} \right)  & \rho_\perp
        \end{pmatrix}
        ,\\
        \bar{\tsr{L}}^{(4)}
        &=
        \begin{pmatrix}
            2 \alpha_4 + ( \alpha_5 + \alpha_6 )    & \alpha_2 + \alpha_3   & 2 \mu^{\text{A}} \\
            \alpha_2 + \alpha_3 & \alpha_3 - \alpha_2   & -2 \left( \mu^{\text{N}} - \mu'^{\text{N}} \right)\\
            2 \mu^{\text{A}} & -2 \left( \mu^{\text{N}} - \mu'^{\text{N}} \right)   & \rho_\perp
        \end{pmatrix}
        ,\\
        \bar{\tsr{L}}^{(5)}
        &=
        (\rho_\parallel)
        .
    \end{align}
    \end{subequations}
    All eigenvalues of each block matrix $\bar{\tsr{L}}^{(i)}\ (i=1,2,\dots,5)$ are non-negative.

    For $\bar{\tsr{L}}^{(1)}$, 
    \begin{equation}
        \alpha_4 \ge 0
    \end{equation}
    holds, and Eq.~\eqref{eq:bound_alpha_4} is obtained.

    For $\bar{\tsr{L}}^{(2)}$, the eigenvalues are positive
    \begin{equation}
        \alpha_1 + 2 \alpha_4 + ( \alpha_5 + \alpha_6 ) \pm ( \alpha_1 + \alpha_4 + ( \alpha_5 + \alpha_6 )) \ge 0
        ,
    \end{equation}
    or the equivalent representation is possible as
    \begin{subequations}
    \begin{align}
        2 \alpha_1 + 3 \alpha_4 + 2 ( \alpha_5 + \alpha_6 ) 
        &\ge 0
        ,\\
        \alpha_4
        &\ge 0
        .
    \end{align}
    \end{subequations}
    The former inequality is Eq.~\eqref{eq:bound_2alpha_1}, and the latter is Eq.~\eqref{eq:bound_2alpha_4}.

    For $\bar{\tsr{L}}^{(3)}$, its positive semidefiniteness is examined not by explicitly computing all eigenvalues, but through the necessary and sufficient condition that all its principal minors be non-negative. Since a $n\times n$ matrix has $2^n - 1$ principal minors, $\bar{\tsr{L}}^{(3)}$ provides $2^3 -1 = 7$ inequalities. The principal minors of order $1$ are to be non-negative:
    \begin{subequations}
    \begin{align}
            2 \alpha_4 + ( \alpha_5 + \alpha_6 )
            &\ge 0
            ,\\
            \alpha_3 - \alpha_2
            &\ge 0
            ,\\
            \rho_\perp
            &\ge 0
            ,
    \end{align}
    \end{subequations}
    and these inequalities derived Eqs.~\eqref{eq:bound_2alpha_4}, \eqref{eq:bound_alpha_3-alpha_2} and \eqref{eq:bound_rho_perp}. The non-negativity of the principal minors of order $2$ provides
    \begin{subequations}
    \begin{align}
        \begin{vmatrix}
            2 \alpha_4 + ( \alpha_5 + \alpha_6 )    &   \alpha_3 + \alpha_2 \\
            \alpha_3 + \alpha_2 &   \alpha_3 - \alpha_2
        \end{vmatrix}
        &=
        (2 \alpha_4 + ( \alpha_5 + \alpha_6 )) (\alpha_3 - \alpha_2) - (\alpha_3 + \alpha_2)^2
        \ge 0
        ,\\
        %&\Leftrightarrow |Z| \le 1 ,\quad Z \coloneqq \frac{\alpha_3 + \alpha_2}{\sqrt{(2 \alpha_4 + ( \alpha_5 + \alpha_6 )) (\alpha_3 - \alpha_2)}} ,\\
        \begin{vmatrix}
            \alpha_3 - \alpha_2   & 2 \left( \mu^{\text{N}} - \mu'^{\text{N}} \right)\\
            2 \left( \mu^{\text{N}} - \mu'^{\text{N}} \right)   & \rho_\perp
        \end{vmatrix}
        &=
        (\alpha_3 - \alpha_2) \rho_\perp - \left( 2 \left( \mu^{\text{N}} - \mu'^{\text{N}} \right) \right)^2
        \ge 0
        ,\\
        %&\Leftrightarrow |Y| \le 1 ,\quad Y \coloneqq \frac{2 \left( - \mu^{\text{N}} + \mu'^{\text{N}} \right)}{\sqrt{(\alpha_3 - \alpha_2) \rho_\perp}} ,\\
        \begin{vmatrix}
            2 \alpha_4 + ( \alpha_5 + \alpha_6 )   & 2 \mu^{\text{A}}   \\
            2 \mu^{\text{A}}   & \rho_\perp
        \end{vmatrix}
        &=
        (2 \alpha_4 + ( \alpha_5 + \alpha_6 )) \rho_\perp - \left( 2 \mu^{\text{A}} \right)^2
        \ge 0
        ,
        %&\Leftrightarrow |X| \le 1 ,\quad X \coloneqq \frac{2 \mu^{\text{A}}}{\sqrt{(2 \alpha_4 + ( \alpha_5 + \alpha_6 )) \rho_\perp}} ,
    \end{align}
    \end{subequations}
    and these inequalities derived Eqs.~\eqref{eq:bound_|Z|}, \eqref{eq:bound_|Y|} and \eqref{eq:bound_|X|}. The principal minor of order $3$ is non-negative
    \begin{align}
        &\hspace{1.1em}
        \begin{vmatrix}
            2 \alpha_4 + ( \alpha_5 + \alpha_6 )    & \alpha_2 + \alpha_3   & -2 \mu^{\text{A}} \\
            \alpha_2 + \alpha_3 & \alpha_3 - \alpha_2   & 2 \left( \mu^{\text{N}} - \mu'^{\text{N}} \right)\\
            -2 \mu^{\text{A}} & 2 \left( \mu^{\text{N}} - \mu'^{\text{N}} \right)   & \rho_\perp
        \end{vmatrix}
        \nonumber\\
        &=
        (2 \alpha_4 + ( \alpha_5 + \alpha_6 ))( \alpha_3 - \alpha_2 ) \rho_\perp - 2 ( \alpha_2 + \alpha_3 ) 2 \left(\mu^{\text{N}} - \mu'^{\text{N}} \right) 2\mu^{\text{A}} - \rho_\perp ( \alpha_2 + \alpha_3 )^2 
        \nonumber\\
        &\hspace{1.1em}
        - (2 \alpha_4 + ( \alpha_5 + \alpha_6 )) \left(2 \left(\mu^{\text{N}} - \mu'^{\text{N}} \right) \right)^2 - ( \alpha_3 - \alpha_2 ) \left(-2 \mu^{\text{A}} \right)^2
        \nonumber\\
        &\ge 0
        ,
    \end{align}
    which gives Eq.~\eqref{eq:bound_X2+Y2+Z2-2XYZ}. These $7$ inequalities are also derived from $\bar{\tsr{L}}^{(4)}$.
    
    Dimensionless quantities $X$ and $Y$ show the conversion efficiency between a heat current and stresses, and correspond to the dimensionless figure of merit $ZT$ in the thermoelectric effect. Rigorously, $X$ and $Y$ are equivalent to $\sqrt{ ZT / ( 1 + ZT )}$. The maximum efficiency at the reversible regime, where the entropy production vanishes, is called the Carnot efficiency. In the thermoelectric effect, the Carnot efficiency~\cite{Shiraishi2023} is attained when $\sqrt{ ZT / ( 1 + ZT )} \to 1\ (ZT \to +\infty)$. So, probable thermoelectric materials possess $\sqrt{ ZT / ( 1 + ZT )} \le 1$, which corresponds to Eqs.\eqref{eq:bound_|X|} and \eqref{eq:bound_|Y|}. Note that the Carnot efficiency is attained only if the power is absent~\cite{Shiraishi2023}, which is confirmed by the TUR since any currents are zero when $R=0$ [see Eq.~\eqref{eq:TUR_linear}]. In the Leslie effects, the heat current $q_z$ induces a stress
    \begin{subequations}
    \begin{align}
        \frac{ \sigma'_{xy} + \sigma'_{yx} }{2}
        &=
        + \frac{2 \alpha_4 + ( \alpha_5 + \alpha_6 )}{2} A_{xy} + \frac{\alpha_2 + \alpha_3}{2} \corot{n}_y - \mu^{\text{A}} q_z
        ,\\
        \frac{ \sigma'_{xy} - \sigma'_{yx} }{2} 
        &=
        - \frac{\alpha_2 + \alpha_3}{2} A_{xy} - \frac{\alpha_3 - \alpha_2}{2} \corot{n}_y - \left( \mu^{\text{N}} - \mu'^{\text{N}} \right) q_z
        ,
    \end{align}
    \end{subequations}
    and the flow $A_{xy},\ \corot{n}_y$ against the stress is work. The power $\dot{w}_1 (A_{xy},\, \corot{n}_y\, ;\, q_z)$ of this work is represented as
    \begin{align}
        \dot{w}_1 (A_{xy},\, \corot{n}_y\, ;\, q_z)
        &=
        -2 A_{xy} \frac{ \sigma'_{xy} + \sigma'_{yx} }{2} 
        -2 \left( - \corot{n}_y \right) \frac{ \sigma'_{xy} - \sigma'_{yx} }{2} 
        \\
        &=
        - A_{xy}
        \left[
            (2 \alpha_4 + ( \alpha_5 + \alpha_6 )) A_{xy} + (\alpha_2 + \alpha_3) \corot{n}_y -2 \mu^{\text{A}} q_z
        \right]
        \nonumber\\
        &\hspace{1.1em}
        - \corot{n}_y
        \left[
            (\alpha_2 + \alpha_3) A_{xy} + (\alpha_3 - \alpha_2) \corot{n}_y + 2 \left( \mu^{\text{N}} - \mu'^{\text{N}} \right) q_z
        \right]
        .
    \end{align}
    Assume that all eigenvalues of $\bar{\tsr{L}}^{(3)}$ are positive (never zero), and then $\bar{\tsr{L}}^{(3)}$ is invertible: $|\bar{\tsr{L}}^{(3)}| > 0$. Let $R^{(3)}$ be the dissipation due to $\bar{\tsr{L}}^{(3)}$, and it is represented as
    \begin{equation}
        R^{(3)}
        =
        \bar{\vec{F}}^{(3)\text{T}} \bar{\tsr{L}}^{(3)} \bar{\vec{F}}^{(3)} \ge 0
        ,\quad
        \bar{\vec{F}}^{(3)} = (A_{xy},\ \corot{n}_y,\ q_z)^{\text{T}}
        .
    \end{equation}
    The equality holds if and only if the affinity is trivial $\bar{\vec{F}}^{(3)} = (0,0,0)^{\text{T}}$, which is the kernel of $\bar{\tsr{L}}^{(3)}$. So, the Carnot efficiency is attained when $R^{(3)} = 0$, where the affinity $\bar{\vec{F}}^{(3)}$ is absent and the system never possesses finite power: $\dot{w}_1 = 0$. %Three coupling effects incorporate in Leslie effects, and thus the additional inequality Eq.~\eqref{eq:bound_X2+Y2+Z2-2XYZ} is imposed. 
    We next examine the maximum of the power. The power $\dot{w}_1$ is bounded as
    \begin{equation}
        \dot{w}_1 (A_{xy},\, \corot{n}_y\, ;\, q_z)
        \le
        \frac{1}{4} \rho_\perp {q_z}^2 \frac{X^2 + 2XYZ + Y^2}{1 - Z^2}
        \le
        \frac{1}{4} \rho_\perp {q_z}^2
        .
    \end{equation}
    Here
    \begin{equation}
        \rho_\perp \frac{X^2 + 2XYZ + Y^2}{1 - Z^2}
    \end{equation}
    is an intrinsic factor of the material, namely independent of size of the sample and the externally imposed heat current $q_z$. This factor characterizes the maximum power.
    The first inequality shows the maximum power for fixed $q_z$ and $\bar{\tsr{L}}^{(3)}$, and the (first) equality holds when
    \begin{subequations}
    \begin{align}
        A_{xy}^\ast
        &=
        \left( 1 - \frac{(\alpha_2 + \alpha_3)^2}{(2 \alpha_4 + ( \alpha_5 + \alpha_6 )) (\alpha_3 - \alpha_2)} \right)^{-1}
        %\frac{ (2 \alpha_4 + ( \alpha_5 + \alpha_6 )) (\alpha_3 - \alpha_2) }{(2 \alpha_4 + ( \alpha_5 + \alpha_6 )) (\alpha_3 - \alpha_2) - (\alpha_2 + \alpha_3)^2}
        \left[
            A_{xy}^0 
            - \frac{\alpha_2 + \alpha_3}{2 \alpha_4 + ( \alpha_5 + \alpha_6 )} \corot{n}_y^0
        \right]
        ,\\
        \corot{n}_y^\ast
        &=
        \left( 1 - \frac{(\alpha_2 + \alpha_3)^2}{(2 \alpha_4 + ( \alpha_5 + \alpha_6 )) (\alpha_3 - \alpha_2)} \right)^{-1}
        %\frac{ (2 \alpha_4 + ( \alpha_5 + \alpha_6 )) (\alpha_3 - \alpha_2) }{(2 \alpha_4 + ( \alpha_5 + \alpha_6 )) (\alpha_3 - \alpha_2) - (\alpha_2 + \alpha_3)^2}
        \left[
            \corot{n}_y^0
            - \frac{ \alpha_2 + \alpha_3 }{\alpha_3 - \alpha_2} A_{xy}^0 
        \right]
        ,
    \end{align}
    \end{subequations}
    where $A_{xy}^0,\ \corot{n}_y^0$ are flow in the absence of $\alpha_2 + \alpha_3$, 
    \begin{subequations}
    \begin{align}
        A_{xy}^0
        &=
        + \frac{ \mu^{\text{A}} }{2 \alpha_4 + ( \alpha_5 + \alpha_6 )}
        q_z
        ,\\
        \corot{n}_y^0
        &=
        - \frac{ \mu^{\text{N}} - \mu'^{\text{N}} }{ \alpha_3 - \alpha_2 }
        q_z
        .
    \end{align}
    \end{subequations}
    When $\alpha_2 + \alpha_3 = 0$, the Leslie thermohydrodynamic and thermomechanical effects induce the irrotational flow $A_{xy}^0$ and rotational flow $\corot{n}_y^0$, respectively. The flows are mixed by $\alpha_2 + \alpha_3$ to produce $ A_{xy}^\ast$ and $\corot{n}_y^\ast$. 
    The second inequality is obtained when the cross-coupling coefficients are optimized as 
    \begin{equation}
    \label{eq:X2-2XYZ+Y2=1-Z2}
        X^2 + 2XYZ + Y^2 = 1 - Z^2
        .
    \end{equation}
    That provides the maximum power. The hydrodynamic coupling of $A_{xy}$ and $\corot{n}_y$ brings dissipation $(\alpha_2 + \alpha_3) A_{xy} \corot{n}_y$. This dissipation is obviously absent when $\alpha_2 + \alpha_3$. At the optimal $A_{xy}^\ast$ and $\corot{n}_y^\ast$, the contribution of $\alpha_2 + \alpha_3$ (namely $Z$) is estimated as
    \begin{equation}
        (\alpha_2 + \alpha_3) A_{xy}^\ast \corot{n}_y^\ast
        =
        - XYZ \rho_\perp {q_z}^2 + O(Z^2)
        .
    \end{equation}
    When $XYZ > 0$, the coupling of $A_{xy}$ and $\corot{n}_y$ contributes to the work rather than dissipation and improves efficiency. Since nonequilibrium processes, in particular the Leslie effects, suppress dissipation~\cite{Sarman2025PCCP}, $XY$ tends to have the same sign as $Z$. For example, $X/Y < 0$ is preferred when $Z < 0$ so that the Leslie effects cooperatively reduce dissipation.
    The temperature gradient $-T^{-1} \nabla_z T$ is controlled in actual experiments. The power for the fixed temperature gradient $\dot{w}_2$ is bounded as
    \begin{align}
        \dot{w}_2
        &\le
        \frac{1}{4} (2 \alpha_4 + ( \alpha_5 + \alpha_6 )) (\alpha_3 - \alpha_2) (-T^{-1} \nabla_z T)^2 ( X^2 + 2XYZ + Y^2 )
        %\\
        %&\le
        %\frac{1}{4} (2 \alpha_4 + ( \alpha_5 + \alpha_6 )) (\alpha_3 - \alpha_2) (-T^{-1} \nabla_z T)^2 ( 1 - Z^2 )
        .
    \end{align}
    The temperature gradient drives the heat current that supplies thermal energy to the system of concern. The maximum work relative to the heat current is given by
    
    \begin{align}
        \frac{\dot{w}_2}{-q_z}
        &\le
        \frac{1}{2} \frac{X^2 + 2XYZ + Y^2}{1 - Z^2} \left( 2 - \frac{X^2 + 2XYZ + Y^2}{1 - Z^2} \right) \frac{ \nabla_z T }{T}
        \\
        &\le
        \frac{1}{2} \frac{ \nabla_z T }{T}
        %\\
        %&\le
        %\frac{1}{4} (2 \alpha_4 + ( \alpha_5 + \alpha_6 )) (\alpha_3 - \alpha_2) (-T^{-1} \nabla_z T)^2 ( 1 - Z^2 )
        ,
    \end{align}
    which represents the work done relative to the thermal energy supplied to the system. The second equality holds when the coefficients are optimized as in Eq.~\eqref{eq:X2-2XYZ+Y2=1-Z2}.
    Generally, the efficiency at the maximum power is half of the Carnot efficiency~\cite{Shiraishi2023}. With the temperature of the cold bath $T$ and the hot bath $T + \Delta T ,\ ( \Delta T > 0 )$, the Carnot efficiency is given by $ 1 - T / ( T + \Delta T ) \simeq \Delta T / T$.

    For $\bar{\tsr{L}}^{(5)}$, 
    \begin{equation}
        \rho_\parallel \ge 0
    \end{equation}
    gives Eq.~\eqref{eq:bound_rho_parallel}.

%==========================================================================
\section{Stress in some director fields}\label{sec:some_director_fields}
    This section shows representations of a stress in some director fields that are often observed in experiments.
    We follow previous studies, namely we chose the temperature gradient as the affinity and adopt Eq.~\eqref{eq:constitutive2} as the constitutive equations.
    Assume a uniform temperature gradient
    \begin{equation}
        \vec{\nabla} T ( \vec{r}, t ) = ( \nabla_z T )\, \hat{\vec{z}}
        .
    \end{equation}
    
    \textit{Planar}---The helical axis is parallel to the $z$-axis:
    \begin{equation}
        \vec{n} (\vec{r}, t) = (\cos{q_0 z}) \, \hat{\vec{x}} + ( \sin{q_0 z} ) \, \hat{\vec{y}}
        ,
    \end{equation}
    where $2 \pi / q_0$ is the cholesteric pitch. This director field occupies the concentric droplets~\cite{Yoshioka2014, Nishiyama2021SoftMatter}, and is used in MD simulations~\cite{Sarman2016} and theoretical investigations~\cite{Leslie1968II,de_Gennes}.
    In this director field, the stress due to the thermal Leslie effect is
    \begin{subequations}
    \begin{align}
        \sigma^{\text{L}}_{xx}
        =
        - \sigma^{\text{L}}_{yy}
        &=
        2 \tilde{\mu}^{\text{A}} \cos{q_0 z} \sin{q_0 z} \, ( - T^{-1} \nabla_z T )
        ,\\
        \sigma^{\text{L}}_{xy}
        &=
        \left[
            \tilde{\mu}^{\text{A}} (\sin^2{q_0 z} - \cos^2{q_0 z}) 
            - \tilde{\mu}^{\text{N}} + \tilde{\mu}'^{\text{N}}
        \right]
        \, ( - T^{-1} \nabla_z T )
        ,\\
        \sigma^{\text{L}}_{yx}
        &=
        \left[
            \tilde{\mu}^{\text{A}} (\sin^2{q_0 z} - \cos^2{q_0 z}) 
            + \tilde{\mu}^{\text{N}} - \tilde{\mu}'^{\text{N}}
        \right]
        \, ( - T^{-1} \nabla_z T )
        ,\\
        \sigma^{\text{L}}_{xz} = \sigma^{\text{L}}_{yz} = \sigma^{\text{L}}_{zx} = \sigma^{\text{L}}_{zy} = \sigma^{\text{L}}_{zz}
        &= 0
        .
    \end{align}
    \end{subequations}
    The system has periodicity along the $z$-axis with wavelength $2\pi / q_0$. The stress is averaged within this interval:
    \begin{equation}
        \frac{q_0}{2\pi}
        \int_{-\pi/q_0}^{+\pi/q_0}
        dz \,
        \left(
        \begin{array}{c}
            \sigma^{\text{L}}_{xy} \\
            \sigma^{\text{L}}_{yx}
        \end{array}
        \right)
        =
        \left(
        \begin{array}{c}
            -1 \\
            +1
        \end{array}
        \right)
        \left( + \tilde{\mu}^{\text{N}} - \tilde{\mu}'^{\text{N}} \right)
        \, ( - T^{-1} \nabla_z T )
        ,
    \end{equation}
    and the other components vanish. The averaged torque has only the $z$-component
    \begin{equation}
        \vec{\Gamma}
        =
        %- \sigma^{\text{L}}_{xy} + \sigma^{\text{L}}_{yx}
        %=
        2\left( \tilde{\mu}^{\text{N}} - \tilde{\mu}'^{\text{N}} \right)
        \, ( - T^{-1} \nabla_z T ) \, \hat{\vec{z}}
        .
    \end{equation}

    \textit{Banded}---The helical axis is normal to the $z$-axis:
    \begin{equation}
        \vec{n} (\vec{r}, t) = (\cos{q_0 y}) \, \hat{\vec{x}} - ( \sin{q_0 y} ) \, \hat{\vec{z}}
        .
    \end{equation}
    This director field occupies the banded (stripe) droplets~\cite{Oswald2008PRL, Yoshioka2014, Bono2020} and used in MD simulations~\cite{Sarman2025PCCP}. The stress is
    \begin{subequations}
    \begin{align}
        \sigma^{\text{L}}_{xx}
        =
        \sigma^{\text{L}}_{yy}
        =
        \sigma^{\text{L}}_{zz}
        =
        \sigma^{\text{L}}_{zx}
        =
        \sigma^{\text{L}}_{xz}
        &=
        0
        ,\\
        \sigma^{\text{L}}_{xy}
        &=
        \left[
            \left( - \tilde{\mu}^{\text{A}} 
            - \tilde{\mu}^{\text{N}} \right) \cos^2{q_0 y} + \tilde{\mu}'^{\text{N}}
        \right]
        \, ( - T^{-1} \nabla_z T )
        ,\\
        \sigma^{\text{L}}_{yx}
        &=
        \left[
            \left( - \tilde{\mu}^{\text{A}}
            + \tilde{\mu}^{\text{N}} \right) \cos^2{q_0 y} - \tilde{\mu}'^{\text{N}}
        \right]
        \, ( - T^{-1} \nabla_z T )
        ,\\
        \sigma^{\text{L}}_{yz}
        &=
        \left( \tilde{\mu}^{\text{A}} - \tilde{\mu}^{\text{N}} \right) \sin{q_0 y} \cos{q_0 y} 
        \, ( - T^{-1} \nabla_z T )
        ,\\
        \sigma^{\text{L}}_{zy}
        &=
        \left( \tilde{\mu}^{\text{A}} + \tilde{\mu}^{\text{N}} \right) \sin{q_0 y} \cos{q_0 y} 
        \, ( - T^{-1} \nabla_z T )
        .
    \end{align}
    \end{subequations}
    These components of the stress tensor are averaged along the $y$-axis:
    \begin{equation}
        \frac{q_0}{2\pi}
        \int_{-\pi/q_0}^{+\pi/q_0}
        dy \,
        \left(
        \begin{array}{c}
            \sigma^{\text{L}}_{xy} \\
            \sigma^{\text{L}}_{yx}
        \end{array}
        \right)
        =
        \frac{1}{2}
        \left(
        \begin{array}{cc}
            - \tilde{\mu}^{\text{A}} - \tilde{\mu}^{\text{N}} + 2 \tilde{\mu}'^{\text{N}} \\
            - \tilde{\mu}^{\text{A}} + \tilde{\mu}^{\text{N}} - 2 \tilde{\mu}'^{\text{N}}
        \end{array}
        \right)
        \, ( - T^{-1} \nabla_z T )
        ,
    \end{equation}
    and other components are absent. The $z$-component of the averaged torque survives:
    \begin{equation}
        \vec{\Gamma}
        =
        \left(
            \tilde{\mu}^{\text{N}} - 2 \tilde{\mu}'^{\text{N}}
        \right)
        \, ( - T^{-1} \nabla_z T ) \, \hat{\vec{z}}
        .
    \end{equation}
    The deviatoric stress is also possible.

    Sarman\etal decomposed the torque into three parts by focusing on the cholesteric axis $\vec{n}_{\text{ch}}$ and the temperature gradient $\vec{F}_{\text{Q}}$. Let $\hat{\vec{F}}_{\text{Q}} \coloneqq \vec{F}_{\text{Q}} / | \vec{F}_{\text{Q}}|$ be the unit vector that indicates the direction of $\vec{F}_{\text{Q}}$. The torque $\vec{\Gamma}$ is decomposed as
    \begin{equation}
        \vec{\Gamma} 
        = \Gamma_1 \, \vec{n}_{\text{ch}} + \Gamma_2 \, \frac{\vec{n}_{\text{ch}} \times \hat{\vec{F}}_{\text{Q}}}{\left| \vec{n}_{\text{ch}} \times \hat{\vec{F}}_{\text{Q}} \right|} + \Gamma_3 \, \frac{\vec{n}_{\text{ch}} \times \left( \vec{n}_{\text{ch}} \times \hat{\vec{F}}_{\text{Q}} \right)}{\left| \vec{n}_{\text{ch}} \times \hat{\vec{F}}_{\text{Q}} \right|}
        ,
    \end{equation}
    where $\Gamma_i \ (i = 1,2,3)$ is the component of $\vec{\Gamma}$ in each direction, and the basis vectors are normalized. In the linear response regime, $\Gamma_2$ is omitted due to its proportionality to $|\vec{F}_{\text{Q}}|^2$~\cite{Sarman2025PCCP}. For \textit{planar} configuration $\vec{n}_{\text{ch}} = \hat{\vec{F}}_{\text{Q}}$, 
    \begin{equation}
        \vec{\Gamma} = \Gamma_1 \, \hat{\vec{F}}_{\text{Q}}
        .
    \end{equation}
    For \textit{banded} configuration $\vec{n}_{\text{ch}} \perp \hat{\vec{F}}_{\text{Q}}$, 
    \begin{equation}
        \vec{\Gamma} = \Gamma_1 \, \vec{n}_{\text{ch}} - \Gamma_3 \, \hat{\vec{F}}_{\text{Q}}
        .
    \end{equation}
    So, $\Gamma_1, \ \Gamma_3$ contain the $\mu'^{\text{N}}$-term.

    \textit{Double twist}---We suppose a double-twist director field~\cite{Bono2019}:
    \begin{equation}
        \vec{n} (\vec{r}, t) = \sin{\phi} \frac{-y}{\sqrt{x^2 + y^2}} \, \hat{\vec{x}} + \sin{\phi} \frac{x}{\sqrt{x^2 + y^2}} \, \hat{\vec{y}} + \cos{\phi} \, \hat{\vec{z}}
        ,
    \end{equation}
    with $\phi \coloneqq - q_0 \sqrt{x^2 + y^2}$. The center and its vicinity of the director field is a Bloch-type skyrmion that occupies blue phases.
    The stress is
    \begin{subequations}
    \begin{align}
        \sigma^{\text{L}}_{xx}
        =
        - \sigma^{\text{L}}_{yy}
        &=
        2 \tilde{\mu}^{\text{A}} \sin^2{\phi} \frac{-xy}{x^2 + y^2} \, ( - T^{-1} \nabla_z T )
        ,\\
        \sigma^{\text{L}}_{xy}
        &=
        \left[
            \left(
                \tilde{\mu}^{\text{A}} \frac{x^2 - y^2}{x^2 + y^2}
                - \tilde{\mu}^{\text{N}} 
            \right) \sin^2{\phi}
            + \tilde{\mu}'^{\text{N}}
        \right]
        \, ( - T^{-1} \nabla_z T )
        ,\\
        \sigma^{\text{L}}_{yx}
        &=
        \left[
            \left(
                \tilde{\mu}^{\text{A}} \frac{x^2 - y^2}{x^2 + y^2}
                + \tilde{\mu}^{\text{N}} 
            \right) \sin^2{\phi}
            - \tilde{\mu}'^{\text{N}}
        \right]
        \, ( - T^{-1} \nabla_z T )
        ,\\
        \sigma^{\text{L}}_{zz} 
        &= 0
        ,\\
        \sigma^{\text{L}}_{xz}
        &= 
        \left( \tilde{\mu}^{\text{A}} - \tilde{\mu}^{\text{N}} \right)
        \sin{\phi} \cos{\phi} \frac{x}{\sqrt{x^2 + y^2}} \, ( - T^{-1} \nabla_z T )
        ,\\
        \sigma^{\text{L}}_{zx}
        &= 
        \left( \tilde{\mu}^{\text{A}} + \tilde{\mu}^{\text{N}} \right)
        \sin{\phi} \cos{\phi} \frac{x}{\sqrt{x^2 + y^2}} \, ( - T^{-1} \nabla_z T )
        ,\\
        \sigma^{\text{L}}_{yz}
        &= 
        \left( - \tilde{\mu}^{\text{A}} + \tilde{\mu}^{\text{N}} \right)
        \sin{\phi} \cos{\phi} \frac{-y}{\sqrt{x^2 + y^2}} \, ( - T^{-1} \nabla_z T )
        ,\\
        \sigma^{\text{L}}_{zy}
        &= 
        \left( - \tilde{\mu}^{\text{A}} - \tilde{\mu}^{\text{N}} \right)
        \sin{\phi} \cos{\phi} \frac{-y}{\sqrt{x^2 + y^2}} \, ( - T^{-1} \nabla_z T )
        .
    \end{align}
    \end{subequations}
    The director field has $D_\infty$-symmetry. With $x = r \cos{\theta},\ y = r \sin{\theta}$, the summed stress over the azimuth $\theta$ is
    \begin{equation}
        \int_{-\pi}^{+\pi}
        r d\theta \,
        \left(
        \begin{array}{c}
            \sigma^{\text{L}}_{xy} \\
            \sigma^{\text{L}}_{yx}
        \end{array}
        \right)
        =
        \left(
        \begin{array}{c}
            -1 \\
            +1
        \end{array}
        \right)
        2 \pi r \left( \tilde{\mu}^{\text{N}} \sin^2{\phi} - \tilde{\mu}'^{\text{N}} \right)
        \, ( - T^{-1} \nabla_z T )
        ,
    \end{equation}
    and other components are annihilated. The survived torque is
    \begin{equation}
        4 \pi r \left( \tilde{\mu}^{\text{N}} \sin^2{\phi} - \tilde{\mu}'^{\text{N}} \right)
        \, ( - T^{-1} \nabla_z T )
        .
    \end{equation}
    For a disk with radius $r_{\text{d}}$, entire torque is
    \begin{equation}
        \int_0^{r_{\text{d}}} dr\ 4 \pi r \left( \tilde{\mu}^{\text{N}} \sin^2{\phi} - \tilde{\mu}'^{\text{N}} \right)
        \, ( - T^{-1} \nabla_z T )
        =
        \pi
        \left[
            \left( \tilde{\mu}^{\text{N}} - 2 \tilde{\mu}'^{\text{N}} \right) {r_{\text{d}}}^2 
            + \tilde{\mu}^{\text{N}} 
            \left( 
                \frac{\sin^2{q_0 r_{\text{d}}}}{{q_0}^2} 
                - \frac{r_{\text{d}} \sin{2q_0 r_{\text{d}}}}{q_0} 
            \right) 
        \right]
        \, ( - T^{-1} \nabla_z T )
        .
    \end{equation}

%==========================================================================
%\bibliography{clan}

\end{document}